\documentclass[12pt,a4paper]{article}

\usepackage{bm}
\usepackage{amsmath, latexsym}
\usepackage{amssymb}
\usepackage{amsfonts}

\usepackage{graphicx}
\usepackage[onehalfspacing]{setspace}
\usepackage{natbib}
\usepackage{geometry}                		
\usepackage[bottom]{footmisc}
\usepackage{fancyhdr}
\usepackage{parskip}
\usepackage{graphicx}	
\usepackage{enumitem}
\usepackage{afterpage}
\usepackage[tableposition=t]{caption}

\usepackage{multirow}
\usepackage{float}
\usepackage{setspace}
\usepackage{natbib}
\usepackage{url}
\usepackage{eurosym}
\usepackage{tablefootnote}
\usepackage{longtable}
\AtBeginEnvironment{longtable}{\singlespacing}
\usepackage{array}

\usepackage[T1]{fontenc}
\usepackage{lmodern}
\usepackage{csquotes}
\renewcommand{\arraystretch}{1.2}
 
\usepackage{pgfplots} 
\usetikzlibrary{patterns}
\usepackage{lscape}
\usepackage[justification=centering]{caption}
\usepackage{adjustbox}
\usepackage{booktabs}
\usepackage{dcolumn}
\newcolumntype{.}{D{.}{.}{-1}}
{\def\sym#1{\ifmmode^{#1}\else\(^{#1}\)\fi}

\newcolumntype{H}{>{\setbox0=\hbox\bgroup}c<{\egroup}@{}}
\usepackage{multirow}
\usepackage{tikz}
\usetikzlibrary{trees}

\usepackage{placeins}

\date{}

\begin{document}


\begin{titlepage}
\title{Peer Effects in Labor Market Training}
\author{Ulrike Unterhofer\thanks{ I am grateful to Thomas Kruppe  and Julia Lang at the IAB for their assistance with the data, to Simone Balestra, Manuel Buchmann, David Card, Andrea Ghisletta, Bryan Graham, Patrick Kline, Rafael Lalive, Robert Moffit, Fanny Puljic, Jesse Rothstein, Alois Stutzer, Conny Wunsch and Véra Zabrodina, as well as participants at various seminars and conferences for helpful discussions and comments. All errors are my own. \textbf{Correspondence to:} ulrike.unterhofer@unibas.ch. University of Basel, Faculty of Business and Economics, Peter-Merian Weg 6, 4002 Basel, Switzerland. \textbf{Data availability:} This research resulted from the project \textit{Langzeitevaluation der aktiven Arbeitsmarktpolitik in Deutschland (IAB-Projekt 3309)}, based on confidential data provided by the Institute for Employment Research: IAB Beschäftigtenhistorik (BeH) V10.02.00, IAB Integrierte Erwerbsbiografien (IEB) V13.00.00, IAB Arbeitsuchendenhistorik (ASU) V06.09.00, IAB Arbeitsuchendenhistorik aus XSozial-BA-SGB II (XASU) V02.01.00-201704, IAB Maßnahmeteilnahmehistorik (MTH) V07.00.00-201704, Nürnberg 2017. Access to the restricted data can be requested directly with the Institute for Employment Research (iab.fdz@iab.de). Models/codes are available upon request. \textbf{Disclosure Notice:} The author declares that she has no relevant or material financial interests that relate to the research described in this paper.
}\\
\normalsize \textit{University of Basel} }
\maketitle
\date{\vspace{-2ex}}

\setstretch{1.5} %

\begin{abstract} 
\noindent 
This paper shows that group composition shapes the effectiveness of labor market training programs for jobseekers. Using rich administrative data from Germany and a novel measure of employability, I find that participants benefit from greater average exposure to highly employable peers through increased long-term employment and earnings. The effects vary significantly by own employability:  jobseekers with a low employability experience larger long-term gains, whereas highly employable individuals benefit primarily in the short term through higher entry wages. An analysis of mechanisms suggests that within-group competition in job search attenuates  part of the positive effects that operate through knowledge spillovers.
\vspace{0.2in}
\end{abstract}
\setcounter{page}{0}
\thispagestyle{empty}
\end{titlepage}
\pagebreak


\setstretch{2} %
\pagenumbering{arabic}


\section{Introduction}

Technological and demographic change requires continuous (re)training of the workforce. This involves labor market training programs that are partly or fully publicly funded. If implemented efficiently, these programs may overcome structural mismatches and alleviate labor shortages by providing jobseekers with the skills that are in demand. One possible, but so far understudied determinant of the program effectiveness is the group composition. Labor market training for unemployed workers, which is the focus of this study, combines two contexts for which positive peer effects have been documented: education and job search.
A large literature shows that peer ability is a significant  driver  of  student achievement and labor market outcomes (\citealp{Sacerdote2011}; \citealp{Payolo2020}). 
In job search, there is evidence that social networks facilitate information transmission, reduce search frictions and improve the match quality between firms and workers (see e.g., \citealp{Ioannides2004}). 
However, in combination, these contexts might also give rise to negative peer effects, making the direction of the overall effect unclear. On the one hand, jobseekers might benefit from existing networks or skills of more employable peers and find jobs more easily. On the other hand, they might lose out against better peers and face more competition in job-search. Finally, peer effects might depend on the macroeconomic context within which the program takes place. The within-group competition might be fiercer if jobs are scarce and the outside competition plays a larger role.\footnote{The literature quanifying spillover effects of labour market policies shows that such spillovers to non-participants are larger if labour markets are weak (E.g. \citealp{Crepon2013}). \cite{Lazear2018} show that competition between jobseekers shapes their labour market success when job slots are fixed.}

Designing effective training programs is of great economic relevance, as public spending on vocational training remains high in many industrialized countries (up to 1\% of GDP in OECD countries in 2022 \citep{OECDstat2023}). 
This paper shows that peer quality is a key design feature of labor market programs, influencing individual employment outcomes through different mechanisms. The findings suggest that policy makers can improve program effectiveness by strategically shaping group composition.

The setting for my analysis is public sponsored training for jobseekers in Germany. 
Specifically, I investigate how the labor market outcomes of individual program participants are affected by the employability of their peers in the same course.  
Employability is defined as the predicted probability to find a stable employment within one year after program start. It is estimated 
using a lasso model and summarizes a rich set of individual and labor market characteristics in a single score. To establish causality, I rely on idiosyncratic variation in the group composition within courses of the same competence level, offered by the same training provider over time. I control for dynamic sorting into peer groups with aggregate time trends that vary at the local labour market level (occupational area). I test the sensitivity of the results to various alternative specifications.
The analysis is based on rich administrative data on the universe of job-seeking individuals who participate in public sponsored training programs in the years 2007 to 2012. 
I study three types of training programs which involve skills training of different intensity. While short and long classic vocational training extend particular occupational skills of jobseekers, retraining provides them with the competences for a new vocational degree. Germany is an ideal setting for studying peer effects in labor market training programs, as the programs there are widely used and similar to those offered in many other developed countries such as Austria, Switzerland, the Netherlands, and the United States.

I document the following set of results. First, I find that attending training with more employable peers has no effects on the job search duration for the average program participant but improves days spent in employment in the long run. A one standard deviation increase in the average peer employability increases employment by 16 to 20 days within five years after program start. The increase in employment stability goes hand in hand with a substantial increase in individual earnings in the short and long run (2 to 7 percent). The effects on earnings are largest and most robust for individuals in short vocational training programs. The effects for long vocational training programs and retraining are lower but less precisely estimated. 
Second, I document important effect heterogeneity with respect to own employability. I find that jobseekers with an individual employability below the median draw higher long-term benefits from exposure to highly employable peers compared to jobseekers with an employability above the median. Highly employable jobseekers benefit from a better peer group through higher earnings in their first job -- an effect that is accompanied by a longer job search duration. This suggests that they might be more selective in their job search. 
Third, I investigate possible mechanisms that may explain these effects. In particular, I disentangle the effect of peer employability through knowledge spillovers from the within-group competition in job search. Competition is proxied by participants' ordinal rank. The analysis suggests that competition attenuates part of the effects of peer employability that operate through knowledge spillovers. Competition effects have a larger relative impact on short term outcomes such as the job search duration and earnings in the first job. Finally, the results suggest that peer effects may depend on the tightness of the labor market. For short training programs, I show that both knowledge spillovers and competition effects are quantitatively more important in not-tight labor markets compared to tight labor markets. 


The paper makes a number of contributions. It is the first to systematically analyze peer effects on labor market outcomes in large scale training programs using rich administrative data. While there is a large literature studying the direct effects of such programs on their participants \citep{Card2010, Card2018}, the existing evidence on peer effects in active labor market policies (ALMP) is limited to a few experimental studies that come to different conclusions. \cite{Lafortune2018} analyze a training program for low-skilled Chilean women and find no robust evidence for peer effects. \cite{Vandenberg2019} evaluate a job search assistance program for young unemployed workers in France and find that being in a group with higher average employability negatively affects the program's effectiveness. \cite{Baird2023} analyze a job training program for disadvantaged adults in the United States, for which they find large positive peer effects on employment and earnings.\footnote{Peer effects have also been found to matter for the adoption of labor market programs \citep{Mora-Garcia2023}.} Compared to these papers, I study a more representative set of programs and workers which allows me to shed light on which contextual differences might drive outcomes. Furthermore, this paper is the first to systematically examine the mechanisms underlying peer effects in labor market training and how they depend on the macroeconomic environment.\footnote{The role of skill spillovers and competition effects between workers have been primarily studied at the workplace e.g. \cite{Cornelissen2023, Cornelissen2017, Mas2009}.} My findings highlight that the average peer effect in labor market training can mask potential counteracting effects. Negative effects from within group competition in job search can confound positive skill spillovers between peers. 

Second, my findings inform optimal treatment assignment in labor market training and suggest that the peer composition is an important design feature.  Most of the existing literature in this area investigates the efficient allocation of participants to programs based on individual characteristics, but does not take compositional aspects into account (\citealp{Eberts2002}, \citealp{Lechner2007b} \citealp{Staghoj2010}, \citealp{Cockx2023}). The evidence points to potential benefits from introducing statistical treatment rules to assist caseworkers in assigning unemployed workers to program types. Nevertheless, the practical implementation of a functioning targeting system is challenging (see e.g., \citealp{Behncke2009}).\footnote{ For Germany, \cite{Doerr2017a} and \cite{Huber2018} find that a voluntary assignment system using vouchers is less effective in the short term compared to a system with caseworker assignment. In the long run however, positive employment effects of voluntary assignment prevail.}
The only study to derive optimal allocation rules in the presence of peer effects is \cite{Baird2023}. They show that in a setting where peer effects are positive and linear-in-means and programs have a limited capacity, allocating seats to relatively more able participants can significantly increase the impact of the program. My findings suggests that these positive effects can potentially be even increased if programs are designed in a way as to maximize knowledge exchange between participants and reduce competition in job search.

Finally, I also contribute to the large literature on peer effects in education by considering educational programs for adults. For school-aged children, higher levels of peer ability have been shown to have a positive effect on individual learning outcomes while there appear to be important non-linearities. See \cite{Sacerdote2011} or \cite{Payolo2020} for an overview. For adolescents and university students, studies find at most small peer effects on performance from classmates and room-mates (e.g.,  
\citealp{Brodaty2016}; \citealp{Booij2017}) but large effects on social skills \citep{Zarate2023} and social behavior like drinking, cheating and fraternity membership (e.g., \citealp{Sacerdote2001} and \citealp{Gaviria2001}). My results suggest that skill spillovers are not limited to a young age. 

The paper proceeds as follows. Section \ref{backround} gives an overview of the institutional background, Section \ref{data} introduces the data and reports some descriptive statistics. I define the peer variables of interest, discuss my identification and estimation strategy in Section  \ref{identific_estim}. The results are presented in Section \ref{results}, Section \ref{conclusion} concludes. 
 
\section{Institutional Background}\label{backround}

Public sponsored training for the unemployed has the objective to update and increase the human capital of  jobseekers, to adjust their skills to technological changes, and facilitate a successful labor market integration. In Germany it is typically organised in the form of courses where participants interact on a daily basis. On the one hand, the setting is thus similar to an educational setting where individuals are provided with a set of skills in a group. On the other hand, participants in further vocational training courses are simultaneously looking for jobs, which means that their peers are also potential competitors on the labor market. Finally, they also compete for jobs with jobseekers outside of their training program. 

Since 2003, the assignment of unemployed individuals to courses has been regulated by a voucher system.\footnote{ See e.g., \cite{Kruppe2009} or \cite{Doerr2017a} for a detailed description of the institutional details.} 
Once an individual registers as unemployed, a caseworker reviews their labor market prospects in a profiling process. If a lack of qualifications is identified, she recommends participation in a training program and issues a  voucher. The voucher specifies the program's planned duration, its occupational target, its geographical validity, and the maximum course fee to be reimbursed. Vouchers are valid for a period up to three months from the date of issuance. Training providers are independently organized and offer different types of courses repeatedly in varying intervals. All certified providers and courses are listed in an online tool of the employment agency (\textit{Kursnet}). In addition, training providers may advertize their courses at local employment agencies. Caseworkers are instructed not to issue any course-specific recommendations. 
Once jobseekers obtain a voucher, they can choose between certified providers courses within the period and area of validity of the voucher. If they decide to take up a course\footnote{Jobseekers might obtain a new voucher if the old one expires but their is no legal claim for it. \cite{Doerr2017a} finds that roughly 22\% of vouchers are not redeemed and 25\% of jobseekers who do not redeem obtain a new voucher. Once a voucher is redeemed, non-attendance can lead to benefit sanctions.}, participation is mandatory. However, jobseekers can exit the program without further consequences before its planned end date in case they find a job. In theory, courses are also open to workers who are not registered at the employment agencies. I do not observe these individuals in my data but argue in Appendix \ref{appendix:validity_checks_me} if anything this should be a small number of people.

For my analysis, I aggregate different public sponsored training programs into groups according to their homogeneity with respect to educational contents and organisation. I distinguish between short and long classic vocational training and retraining.  All considered types of trainings are offered for a wide range of fields, require full-time participation and combine classroom training with practical elements such as on-the-job training. Short vocational training programs  (short training)  are defined as programs with a maximum planned duration of 6 months. They have an average planned duration of about 3.7 months  and offer minor improvements of skills. An example for such a course is "financial accounting with SAP". Long vocational training programs (long training) have a planned duration of over 6 months and generally last up to one year. They provide more intensive updating and extending of occupational skills. Such courses involve e.g., training on software skills, operating construction machines or in marketing and sales strategies. Some of the courses offer the possibility to obtain partial degrees. Retraining or degree courses (retraining) have the longest duration of 2 to 3 years and provide a highly standardized training for a new vocational degree according to the German system of vocational education. They focus on jobseekers who have never completed any vocational training or have not worked in the occupation they are trained for a minimum of four years.\footnote{For those workers, retraining courses provide a first occupational qualification. The full terminology is ``programs with a qualification in a recognized apprenticeship occupation".} 

\section{Data and Sample Selection}\label{data}
The analysis is based on administrative data provided by the Institute for Employment Research (IAB). It covers the universe of individuals participating in public sponsored labor market programs between 2007 and 2012 (Database of Program Participants, MTH) and is linked to the Database of Registered Job-Seekers (ASU and XASU) and the Integrated Employment Biographies (IEB). In combination, the data contain detailed information on the training programs attended (e.g., a course and provider identifier, the timing and planned duration, the target occupation\footnote{The target occupation is based on the German classification of occupations KldB 2010. Its granularity allows to identify the occupational specialisation and competence level associated with an occupation. I make use of both dimensions in the paper.} , information on course intensity and costs) as well as a wide range of characteristics of the jobseekers (demographic characteristics, labor market histories with daily accuracy and the region of residence). 
I identify peer groups as individuals who attend the same course together.

I focus my analysis on courses for which a peer effects analysis is sensible given the group size and the organization of the training.\footnote{ For the construction of the relevant peer variables (see Section \ref{peervars}) it is important that I  have information on everyone in the group. I exclude thus courses where some individuals have for example missing labor market histories.} I restrict the sample to courses where the number of participants lies between 5 and 30\footnote{Larger group sizes are rare.}. Furthermore, I focus on courses where individuals start within the same month, overlap for at least one day and exclude self-learning programs and special programs.\footnote{Special programs usually target a particular sub-population of the labor market, such as the program \textit{WeGebAU}, \textit{Perspektive 50plus}, \textit{Gute Arbeit für Alleinerziehende}. I exclude courses if all participants are funded via such special programs. The programs are too small in order to allow for a separate estimation of peer effects in my setting. A number of vocational courses offered in Germany allow for continuous entry. I do not consider those courses as they are characterized by partially overlapping peer groups. Identifying peer effects in these types of courses requires a different identification strategy.} 
 For reasons related to my identification strategy (see Section  \ref{identification}), I only consider training providers which offer courses once per month but multiple times in the observation period. Further, I exclude course where the target occupation is missing.\footnote{The target occupation is almost never missing in retraining courses which are usually occupation-specific and for around 10 percent of courses in short and long vocational training. It is likely that some of these courses cannot clearly be attributed to a single occupation. However, since I cannot be sure about the exact reason for which no target occupation is assigned, it is cleaner to exclude these courses from the estimation sample.} Once the peer variables are constructed, I confine the estimation sample to individuals doing their first vocational training program within my observation window.\footnote{This does not exclude that individuals attended other types of programs in their current unemployment spell.} 
 This yields a final sample of 61758 program participants. 

Table \ref{tab:descr_cl_il} summarizes selected course and individual characteristics as well as outcome variables by program type. The majority of individuals are in short vocational training (3723 courses), followed by long vocational training (1317 courses) and retraining programs (1407 courses). Panel A reports the course characteristics. The average number of courses a provider offers over time ranges from 5 to 7. The average number of courses per provider and occupational area\footnote{I distinguish between 10 occupational areas as defined in the KldB 2010. Agriculture and forestry as well as the military are not represented in my sample.} is only slightly lower suggesting that providers in my sample specialize on few occupations. On the one hand, this reflects the fact that the sample is constraint to providers who offer one course per month, i.e. small providers.  On the other hand, providers have an incentive to specialize in specific topics, in particular, if they offer long programs which require more infrastructure and know-how. The largest share of courses falls within the health care, social and educational sector, traffic, logistics, safety and security, manufacturing as well as business and administration. See Appendix Table \ref{app_tab:occ} for figures on the distribution of courses across occupations. The health care, social and educational sector is particularly important for retraining courses. Overall, the occupational mix is more or less comparable across program types.  In contrast, competence levels associated with the target occupation vary across training programs. I differentiate between low-skilled, skilled and high-skilled courses.\footnote{The competence level of courses is based on the German classification of occupations KldB 2010. There exist four competence levels which are closely tied to the formal vocational qualifications of an occupation \cite{Paulus2010}. I aggregate the competence levels to three categories, considering the two categories with the highest skill levels jointly.} Retraining programs primarily focus on courses in occupations involving skilled tasks that typically require no more than two to three years of vocational training. Short and long training programs also involve many courses with a medium skill level but courses in occupations with a higher (and lower) level of specialization and complexity are comparatively more frequent. 

On average there are about 11 to 12 individuals in one course who spend the majority of course time in class and only few hours in on-the-job training.\footnote{Appendix Figure \ref{fig:tn_anz} shows the distribution of individuals over different course sizes.} Most of the individuals start at the earliest entry date (above 92 percent). A small share ends the course early which is likely due to them finding a job before the course ends. This share is largest for retraining with 17 percent and smallest for short training with 9 percent. Thus peer groups are relatively stable throughout the course. In Appendix \ref{appendix:mechanisms_other} I provide evidence showing that compositional changes are unlikely to be an important behind peer effects.\footnote{I find qualitatively similar results between courses with a large share of dropouts and courses with a low share of dropouts. See Appendix Table \ref{tab:mech_dropouts}.} 
 
Generally, it is likely that program participants do not know each other when entering the program. Although, it cannot be fully ruled out.\footnote{A mass lay-off in a particular region might indeed drive workers that know each other into the same courses. Except for that, it is arguably unlikely that friends or co-workers get unemployed at the same time and select into the same programs. } I flag all courses in which program participants previously worked with more than 20 percent of their peers in the same firm. This affects around 5 percent of courses in my sample. I show that my results are not sensitive to excluding these courses in a robustness check. 

Panel B of Table \ref{tab:descr_cl_il} shows a selection of individual characteristics. While individuals are relatively comparable across program types in terms of their recent labor market history, there exist some differences in terms of demographic characteristics, education and training.
Individuals in short training have the highest level of employment and earnings in years before entering unemployment and start the program relatively early in the unemployment spell. Participants in retraining programs are on average three years younger, more likely to be female, less likely to have a high-school degree or vocational training and worked in lower-paying jobs compared to participants in classic vocational training programs. 
These differences can be explained by the nature and target group of the respective programs.

\begin{table}
 \centering
  \caption{ Summary Statistics for Selected Individual and Course Characteristics and Outcomes across Program Types }
  \label{tab:descr_cl_il}
\resizebox{\linewidth}{!}{%
\begin{tabular}{l*{3}{rr}}
\hline\hline
 &\multicolumn{2}{c}{Short training}&\multicolumn{2}{c}{Long training}&\multicolumn{2}{c}{Retraining}\\
                &     mean&       sd&     mean&       sd&     mean&       sd\\
\hline
\multicolumn{3}{l}{\textbf{Panel A - Course characteristics}}  \\[3pt]
Number of courses per provider&     6.74&     4.47&     5.17&     3.38&     4.67&     3.75\\
Number of courses per competence level and provider&     5.84&     3.85&     4.81&     3.13&     4.45&     3.62\\
Number of courses per occupational area and provider&     5.02&     3.57&     4.42&     3.19&     3.55&     2.20\\
Target occupation: low-skilled &     0.07&     0.25&     0.04&     0.18&     0.02&     0.14\\
Target occupation: skilled &     0.76&     0.43&     0.77&     0.42&     0.95&     0.21\\
Target occupation: high-skilled&     0.17&     0.38&     0.20&     0.40&     0.03&     0.16\\
Course size     &    11.94&     5.06&    11.81&     5.20&    10.92&     5.12\\
Average planned duration in months&     3.69&     1.63&     9.09&     3.22&    22.79&     7.52\\
Weekly hours    &    38.60&     3.45&    38.46&     3.02&    38.30&     2.94\\
Total hours in practice &     3.13&    44.75&    12.93&   125.40&    18.61&   255.54\\
Total hours in class&   441.36&   246.07&   986.05&   397.74&  2068.47&   851.82\\
Apprenticeship certification exam or similar&     0.02&     0.15&     0.07&     0.26&     0.67&     0.47\\
Share starting at earliest entry date&     0.93&     0.12&     0.94&     0.11&     0.92&     0.13\\
Share ending course early&     0.09&     0.13&     0.12&     0.13&     0.17&     0.19\\
Share of peers working at the same firm in last job&     0.02&     0.11&     0.01&     0.08&     0.01&     0.07\\
\multicolumn{3}{l}{\textbf{Panel B - Individual characteristics}}  \\[3pt]
Age in years    &    37.95&    10.75&    37.92&     9.82&    34.51&     8.16\\
Female&     0.43&     0.50&     0.40&     0.49&     0.50&     0.50\\
Non-German&      0.10&     0.30&     0.11&     0.31&     0.11&     0.31\\
High-school degree (Abitur)&     0.17&     0.38&     0.19&     0.39&     0.12&     0.32\\
Vocational training         &      0.64&     0.48&     0.61&     0.49&     0.53&     0.50\\
Academic degree          &     0.08&     0.27&     0.10&     0.29&     0.03&     0.16\\
Months unemployed at program start&   10.01&    20.57&    12.05&    22.51&    11.90&    22.17\\
Months employed (last 2 years)&    13.88&     8.44&    12.85&     8.55&    12.86&     8.49\\
Total earnings (last 2 years, 1000 euro) &     23.17&    22.56&    20.84&    20.91&    18.41&    17.53\\
Program in same unemployment spell & 0.22&     0.42&     0.26&     0.44&     0.29&     0.45\\[5pt]
\multicolumn{3}{l}{\textbf{Panel C - Outcomes}}  \\[3pt]
Search duration for first job (in days)&      751.64&     961.411&      848.23&     944.869&     1001.79&     839.279\\
Total employment by month 60 (in days)            &      1120.62&     587.890&     1061.41&     558.943&      904.20&     506.319\\
Positive earnings by month 60 &        0.92&       0.269&        0.92&       0.266&        0.93&       0.255\\
Log daily earnings in first job (if > 0)*&        3.32&       1.071&        3.29&       1.095&        3.13&       1.118\\
Log total earnings by month 60 (if > 0)*&       10.62&       1.288&       10.60&       1.298&       10.37&       1.210\\
\hline
Observations    &    36662&         &    12683&         &    12413&         \\
Number of courses&  3723&     &  1317&     &  1407&     \\
Number of providers&   858&     &   369&     &   425&     \\
\hline\hline
\end{tabular}}
     \begin{minipage}{\textwidth} \linespread{1.0}\selectfont
          \vspace{6pt}
          \footnotesize \textit{Notes:} Summary statistics (mean and standard deviation) calculated at the individual level. All amounts in euro are inflation adjusted (in prices of 2010). Individual characteristics are measured at program start. * are based on a smaller sample with positive earnings (92\% of the observations). 
     \end{minipage}
\end{table}

The key outcomes of my analysis are individual employment and earnings (Panel C of Table \ref{tab:descr_cl_il}).\footnote{As it is common in the literature on labor market training (e.g. \citealp{Lechner2011}), outcomes are measured at the beginning of program start because the end of the course is endogenous.}  I consider job search duration after program start and total employment\footnote{I consider all types of employment. This includes regular, part-time and so-called marginal employment. The term marginal employment refers to small-scale employment, so-called mini jobs - according to §8 SGB IV and §7 SGB V. The job search duration is censored at the end of the observation period.} up to 60 months after program start -- a proxy for employment stability. I also study the effects of peer quality on employment more dynamically by considering individual employment probabilities in each month after program start.  
Appendix Figure \ref{fig:avg_emp_alo} shows the average employment rate by program type, up to 60 months after program start. Average employment is below 20 percent directly after the program starts and increases to about 65 to 70 percent after around 40 months. The employment rate increases faster for shorter programs, partially explained by shorter lock-in effects\footnote{The literature evaluating the overall effectiveness of such programs, documents such lock-in effects. See e.g., \cite{McCall2016}.}  

With regard to earnings, I consider daily earnings in the first job and total earnings up to 60 months after program start. Both earnings outcomes are measured in logs and can represent earnings from part-time or full-time jobs. Earnings measures are based on the subset of jobseekers who are employed (92 percent).\footnote{For total employment, this includes individuals who are employed at least once in the first 60 months after program start. For earnings in the first job, it includes individuals who found a job until the end of the observation period. } I also present robustness checks based on the full sample with earnings measured in levels and in logs.

\section{Empirical Strategy}\label{identific_estim}
\subsection{Measuring Employability}\label{peervars}
The objective of this study is to quantify the impact of peer quality on individual post-program labor market outcomes. I focus on the average employment prospects of individual $i's$ peers at program start, referred to as (ex-ante) employability in what follows. I first define employability on the individual level and then construct a measure for peer employability as the leave-one-out sample average of individual $i's$ peers employability in group $g$ of size $n$: 
$$\bar{X}_{(-i)g} = \frac{1}{n_g} \sum^{n_g}_{j\neq i} X_{jg}$$

For further analysis I also consider the ordinal rank of the focal worker among her peers. It is measured as percentile rank ${R}_{ig}$, bounded between 0 and 1, with the lowest-ranked jobseeker in each group having $R=0$ and the highest ranked having $R=1$.\footnote{The percentile rank is measured as: $\bar{R}_{ig} = \frac{r_{ig}-1}{n_g-1} \in{\{0,1\}}$, with $r_{ig}$ being the ordinal rank of individual $i$ in group $g$ and $n_g-1$ being the course size.}

Since I do not observe a measure of employability directly in the data, I summarize individual characteristics which are likely to contain information on a jobseeker's employability in a single score. The approach has similarities with \cite{Vandenberg2019}.

I define employability as the probability of finding a contract  that last for more than 6 months within one year of program start. First, I draw a random sample of comparable jobseekers that do not participate in any program (1.5 million observations) to estimate the employability score.\footnote{It is a random sample of individuals entering unemployment between 1998 and 2012. Note that in around 27 percent of cases individuals do not participate in any program after entering unemployment.} The reason for doing so is that the employment status without program participation is unobserved for actual participants.  Second, I apply nearest neighbour matching based on the propensity score to adjust for observable differences  between the two groups at the moment of entry into unemployment (the procedure is described in detail in Appendix \ref{appendix:matching}). The matching deals with with non-random selection into training. Third, I estimate an employability model based on the matched sample of non-participants using different algorithms: A logit model, a lasso logit model with penalization based on crossvalidation and a lasso logit model with theory-driven penalization \citep{Tibshirani1996}.  I regress individual employability on a large set of variables including demographic characteristics, information on health, education, past labor market outcomes as well as information on the local labor market situation.\footnote{The regression output is  displayed in Appendix Table \ref{tab:predictors}. While the matching is done separately by program type, to allow for heterogeneous selection into training, I estimate the employability score jointly for all program types. The idea is that employability should be independent of which programs jobseekers participate in.} I adopt cross-validation and choose the model with the best goodness of fit in the sample of non-participants. All models perform similarly well with the cross-validated lasso slightly outperforming the other models.\footnote{With the cross-validated lasso I reach an accuracy of prediction of 67.35 percent in the holdout sample. 
The results are robust to the estimation of the employability model using a logit model. They are available on request.}  Fourth, I use the estimated coefficients to predict employability scores for the sample of participants at program start.\footnote{For the sample of program participants, I measure the input variables at program start, wherever possible. Information on education, training and health is elicited at entry into unemployment.} Finally, I construct the peer measures using the predicted employability. 

Choosing the timing of the construction of the employability score is not trivial. Jobseekers in the training sample and the sample of participants are most comparable at the moment of unemployment entry. At the same time,  employability at program start is most relevant for measuring peer effects -- it represents the more natural choice. However, since non-participants are only observed at unemployment entry, a potential concern is that the model might perform worse in predicting the employability of jobseekers who start a program late in their unemployment spell. Appendix \ref{appendix:modelselection} provides more details on the choice of the timing and possible sources of measurement error. A robustness analysis in Section \ref{sec:robustness1} shows, that the results are not sensitive to the timing of the construction of the employability score. 


The distributions of the predicted individual and average employability are shown in Appendix Figures \ref{fig:pred_ind_emp_by_type} and \ref{fig:pred_avg_emp_by_type} separately by program type. The individual employability ranges from 0.03 to 0.97 while the average employability ranges from 0.16 to 0.88. Overall, the distributions are largely overlapping across program types. The distribution of the average peer employability in short training is slightly shifted to the right compared to the other two program types. On average, jobseekers in short training have the highest predicted individual employability, jobseekers in retraining the lowest (see also Table \ref{tab:peervars_sel}).

How can the employability score be interpreted? The measure correlates both with past and future success on the labor market. The prediction models identify age, gender, education, training, past labor market outcomes as important drivers of employability. While the selected predictors of the lasso model should be interpreted with caution, they indicate that what drives employability is correlated with past labor market performance. Furthermore, the results in Section \ref{results} reveal that the predicted employability is strongly correlated with actual labor market success after program participation. The employability score thus seems to be a valid proxy for the labor market success of jobseekers and their peers. This is confirmed by a sensitivity analysis presented in Appendix Table \ref{tab:other_peervar} where I include different peer characteristics that are highly correlated with peer employability directly in the empirical model. I find positive peer effects on employment and earnings from being exposed to peers with successful labor market histories.



Several studies use predicted measures of ability to study peer effects, see e.g., \cite{Burke2013}, \cite{Vandenberg2019} or \cite{Thiemann2022}.\footnote{Comparable employability measures based on predictive algorithms have also  been used by  public employment services e.g. for profiling jobseekers. See \cite{Koertner2022} for an overview. So far, profiling has mostly aimed at identifying individuals at risk of long-term unemployment and has been shown to be effective in reducing the duration of unemployment.  Also targeting of jobseekers to the best intervention can be based on such predicted scores.} 
In my context, using a predicted employability measure entails a number of advantages compared to an analysis based on other, readily available peer characteristics. First, it is data-driven and does not rely on any prior knowledge of the strongest predictors of employability.  Second, using a single predicted score achieves a dimension reduction that allows for a more flexible analysis and an easier interpretation of peer effects without having to estimate a high-dimensional model. As pointed out by \cite{Graham2011}, it is not straightforward to study the effects of  multiple peer attributes simultaneously since a ceteris paribus interpretation of such effects is difficult.  


\subsection{Identification of Peer Effects and Empirical Model}\label{identification}

Two main methodological challenges complicate the empirical analysis of peer effects: the reflection and the selection problem (see e.g., \citealp{Moffit2001}).  The reflection problem \citep{Manski1993} arises because of the simultaneity in peer behavior meaning that an individual affects his peer's outcomes and the peers affect the individual's outcome. In the context of labor market training, program participants might for example affect each other through their job search behavior. This complicates the separate identification of exogenous peer effects, i.e., the influence of average peer characteristics, and endogenous peer effects, i.e., the influence of peer behavior. I do not attempt to separate these effects, but estimate a joint effect.\footnote{Estimating the effects jointly is still of great interest, since  policy makers who decide on how to optimally allocate individuals to a specific program would focus their attention on predetermined characteristics of unemployed that are actually observable. If peer characteristics like the ex-ante employability for example matter for the effectiveness of a program, it might not be relevant whether it is peer employability per se or the unobservable characteristics or behaviors correlated with employability. } This joint effect is captured by the average peer employability which might proxy peer behavior but is determined before program start.

The selection problem arises because of common unobserved shocks at the group level and endogenous peer group formation. In my setting peer groups might be endogenously formed if individuals select into specific courses based on unobserved preferences or abilities which are correlated among those belonging to the same peer group. In fact, jobseekers do generally not know who they will be grouped with but there might be endogenous sorting into specific providers or courses depending on the characteristics of the course (i.e., timing, content or location). 

My identification strategy builds on a strand of literature in the educational context that exploits idiosyncratic variation in the peer group composition controlling for selection into peer groups by including fixed effects at the school or grade level (e.g., \citealp{Bifulco2011}; \citealp{Lavy2012}; \citealp{Elsner2017}; \citealp{Carell2018}).\footnote{Other approaches identify peer effects by relying on random group assignment (e.g., \citealp{Sacerdote2001}; \citealp{Duflo2011}; \citealp{Carrell2013}; \citealp{Booij2017}), by using the underlying network structure to construct instrumental variables  \citep{Bramoulle2009, DeGiorgi2010} or by exploiting varying group sizes \citep{Lee2007, Boucher2014}.} Based on this idea, I control for provider choice, and exploit the variation in peer employability between comparable courses offered by the same training providers over time. Specifically, I only compare courses within the same competence level at the same provider. This addresses several concerns with regard to selection that is time constant. First, it controls for potential sorting with regard to location, since training providers are location-specific. Second, it controls for sorting based on occupational skills that are captured by the competence level. Third, since providers are usually specialized on specific target occupations (see Section \ref{data}), I generally only compare program participants in the same local labor market. 

There might also be dynamic sorting into peer groups due to labor market conditions that vary with the business cycle. The composition of unemployed is likely to be better during recessions than during booms \citep{Mueller2017}. 
Business cycles might vary systematically across occupations and local labor markets. 
 To address such sorting, I flexibly control for aggregate time trends with month x year x occupation dummies. In a robustness check I also include region-specific time trends.

The time trends also take care of potential endogenous selection into peer groups (based on time preferences) that varies over the calendar year and possibly across occupations. Individuals might choose their peer group based on expectations on the peer composition. This could happen if there is an option value in waiting until redeeming the voucher, e.g. due to seasonality in the quality of jobseekers that enter the pool of unemployed throughout the calendar year. In fact, due to many regular contracts ending at the end of the year, jobseekers becoming unemployed at the end of a calendar year might be systematically different from jobseekers entering unemployment in the rest of the year. 
Individuals obtaining a voucher in November might e.g. want to wait for January to have more employable peers. Such seasonal patterns are differenced out. At the same time, this sorting behaviour will be limited by the unpredictable timing of voucher issuance\footnote{The issuance date of vouchers can hardly be hardly influenced by jobseekers. Vouchers are issued in meetings with caseworkers, whose appointment time depends on their entry into unemployment and the efficiency of administrative processes.}, the limited voucher validity (jobseekers can select courses only up to three months after they obtain their voucher), course availabilities and capacity constraints. To alleviate further concerns regarding self-selection into courses, I restrict my analysis to  providers which offer courses exactly once per month. Thus, given their choice of provider and course month, individuals can enroll in only one course.\footnote{An alternative would be to use the group of training participants who receive a voucher in the same month as exogenous peer group. With this strategy the peer group would change from the entire course to a smaller fraction of jobseekers who obtain their voucher at the same time. Possibly spillovers are not confined to these groups. While such an analysis might thus be able to uncover interesting heterogeneities it is likely to underestimate the true peer effects. The approach could not be implemented since the date of voucher receipt is not observed in my data.}

\subsection{The Empirical Model}\label{model}

This identification strategy gives rise to the following empirical model which I estimate separately for the three types of training programs:\footnote{This model corresponds to individual best response functions derived from a theoretical framework assuming continuous actions, quadratic pay-off functions and strategic complementarities. The assumption of strategic complementarities is likely to hold in the context of labor market training. It implies that if $i$’s peers increase their effort e.g., by coming regularly to class, studying more or applying acquired skills to job search, $i$ will experience an increase in utility if she does the same. Equilibria have been derived for these types of games by \cite{CalvoA2009} assuming small complementarities and by \cite{Bramoulle2014} using the theory of potential games.}

\begin{equation} \label{eq:main}
y_{ipcot} = \alpha + \gamma X_{ipcot} + \theta \bar{X}_{(-i)pcot} + \pi W_{ipcot} + \lambda_{pc} + \delta_{ot} + \epsilon_{ipcot},
\end{equation}  
It regresses the outcome of an individual $i$ in a course at a provider $p$, with competence level $c$, occupational area $o$, in month $t$ on her own own employability $X_{ipcot}$ and the leave-one-out mean employability of individuals in the same course $\bar{X}_{(-i)pcot}$. 
The model controls for  provider x competence level, $\lambda_{pc}$ and month x year x occupation fixed effects, $\delta_{ot}$. Provider x competence fixed effects control for all observable and unobservable mean differences across provider-competence level combinations that are constant over time. Time fixed effects control for correlated effects which change over time and occupational areas but are the same across providers and competence level groups. I additionally control for a vector of course-level characteristics, $W_{pcot}$ which contains the course size, the average planned  duration, weekly hours and total hours spent in practice and class. 

The coefficient of interest in this linear-in-means model is  $\theta$, which represents the impact of a marginal increase in the average peer employability on $i$’s outcome. It can be interpreted as a social effect and is a combination of exogenous and endogenous peer effects. 

\subsection{Validity Checks }\label{validation} 
The empirical strategy will produce valid estimates if there is sufficient residual variation in the mean peer employability and if this variation is exogenous, i.e. there is no endogenous sorting into peer groups.

\begin{table}[htbp]\centering
\caption{Variation in Employability Measures}\label{tab:peervars_sel}
\resizebox{\linewidth}{!}{%
\begin{tabular}{l*{3}{rrrr}}
\hline\hline
                &\multicolumn{4}{c}{Short training} &\multicolumn{4}{c}{Long training}  &\multicolumn{4}{c}{Retraining}     \\
                &     mean&       sd&      min&      max&     mean&       sd&      min&      max&     mean&       sd&      min&      max\\
\hline
\multicolumn{13}{l}{\textbf{Panel A - Raw variation}}  \\[1pt]
$X_{ipct}$&     0.61&    0.170&     0.01&     0.96&     0.60&    0.173&     0.07&     0.96&     0.59&    0.171&     0.06&     0.95\\
$\bar{X}_{(-i)pct}$&      0.61&    0.092&     0.21&     0.88&     0.60&    0.089&     0.25&     0.83&     0.59&    0.091&     0.16&     0.83\\
\multicolumn{13}{l}{\textbf{Panel B - Variation net of fixed effects and course characteristics }}  \\[1pt]
$X_{ipct}$&     0.00&    0.152&    -0.65&     0.49&     0.00&    0.154&    -0.54&     0.48&    -0.00&    0.152&    -0.57&     0.49\\
$\bar{X}_{(-i)pct}$&     0.00&    0.051&    -0.29&     0.28&     0.00&    0.043&    -0.26&     0.23&     0.00&    0.046&    -0.24&     0.23\\
\hline
\hline
Observations    &    36662&         &         &         &    12683&         &         &         &    12413&         &         &         \\
\hline\hline
\end{tabular}}
      \begin{minipage}{\textwidth} \linespread{1.0}\selectfont
          \vspace{6pt}
          \footnotesize \textit{Notes:} This table shows summary statistics (mean, standard deviation (sd), minimum (min) and maximum (max)) of the employability variables. $X_{ipct}$ designates the own employablity and $\bar{X}_{(-i)pct}$ the leave-one-out mean peer employability.).
     \end{minipage}
\end{table}

I characterize the raw and residual variation in the average peer employability separately by program types in Table \ref{tab:peervars_sel}. The residual variation is obtained by netting out fixed effects and course controls. The distributions of both variations are plotted in Appendix Figure \ref{app_fig:dist_raw_net}. The raw standard deviation is 0.09 on average. It is slighlty larger for short vocational training compared to the other two program types. Netting out fixed effects and course controls reduces the variation to about half in all program types with the reduction being strongest for long training. At the same time, the residual variation is sufficient in order to identify the effects of interest, as shown by the results in Section \ref{results}. 

\textbf{Endogenous sorting into courses.} \quad I assume that variation in peer employability across courses of the same competence level and provider and after removing labor market-specific seasonality results from random fluctuations. It should not be driven by  endogenous sorting into specific courses. To assess the plausibility of this assumption, I present several checks. 

First, I visually check whether the residual variation in the average peer employability is in line with a variation resulting from random fluctuations. For this I perform a resampling exercise as in \cite{Bifulco2011} and compare the observed residual variation with a simulated variation based on a random group allocation (for more details, see Appendix \ref{appendix:validity_checks_res}). I find that the residual variation in the simulated peer employability is close to the observed one  (Appendix Figure \ref{app_fig:simulation}).
 The residual variation is slightly larger than the simulated one for short training and slightly smaller for long training and retraining.

Second, I test for a for systematic correlation between the predicted own and average peer employability as in \citealp{Guryan2009} (see Appendix Table \ref{tab:test_guryan}). I regress own employability on the leave-one-out peer employability and control for the variables on which randomization was conditioned. Additionally, I control for the average employability in the pool of jobseekers who are starting a course in the same month, year and competence level.\footnote{Controlling for jobseekers in the same pool controls for a possible mechanical effect of group assignment.} I find no correlation for short vocational training but some negative correlation for the other two course types. Together with the visual check above, this suggests that my estimates for long vocational training programs and retraining programs might be slightly attenuated.

I present another test for potential selection into courses based on own employability, following \cite{Chetty2011}  and \cite{Balestra2022}. If jobseekers were randomly assigned to courses, then conditional on the set of controls, course indicator variables should not predict the ex-ante employability of jobseekers. I regress own employability on provider-by-competence level, month-by-year-by-occupation fixed effects and course level controls and retrieve the residuals from this regression. Next, I regress the residuals on a full set of course fixed effects and test their joint significance in a F-test. The F-statistics are not statistically significant on the 1\% significance level for any of the three course types. This suggests that there is no systematic selection into courses based on own employability. The test-statistics are shown in Appendix Table \ref{app_tab:chetty}.\footnote{I also test whether peer employability is correlated with some predetermined individual characteristics (age, gender, nationality, education, past labor market performance). See Appendix \ref{appendix:validity_corr_chars} for more details. I find some small imbalances but I believe that this is not a major threat to my identification approach. It allows for imbalances in individual characteristics as long as peer employability is orthogonal to own employability.} 

Overall, I am confident that the variation in peer employability that I use, allows me to plausibly identify causal effects for short training programs. Possibly, my results are less precisely estimated and slightly attenuated for long vocational training and retraining.
In order to provide additional support to my identification strategy, I perform a series of robustness checks in Section \ref{sec:robustness} using alternative sets of controls.

\section{Results}\label{results}
This section presents and discusses the results of the analysis in three parts. First, I examine the effects of an increase in the average predicted peer employability on individual  labor market outcomes after program start.  Second, I investigate whether there is effect heterogeneity with respect to own employability. Third, I decompose peer effects into knowledge spillover and competition effects.   All of the analyses are run separately by program type. Standard errors are clustered at the course level.\footnote{I do not correct standard errors for the imprecision coming from the estimated employability scores. The reason is that there exists no theoretically recognized procedure for how to do it. Bootstrapping is not feasible for nearest neighbour matching \citep{Abadie2008}. Furthermore, I rely on two different samples for the estimation of the employability and the peer effects model. The results are robust to clustering at the level of providers. See Appendix figure \ref{fig:rob_se}.}

\subsection{Peer Effects on Individual Labor Market Outcomes}

\subsubsection{Effects on Job Search Duration and Employment}\label{sec:main1}

The effects of an increase in the average predicted peer employability on individual employment after program participation are presented in Table \ref{tab:res_LMout} separately by program type. Panel A shows the effects on job search duration, Panel B the effects on total employment up to five years after program start. Both outcomes are measured in days. The table presents the effects of a standard deviation increase in the average peer employability ($\bar{X}_{(-i)pct}$) as well as the coefficients on own employability ($X_{ipct}$). The latter are also scaled in order to represent a standard deviation increase.\footnote{ To get a meaningful effect size, I calculate the effect of a standard deviation increase multiplying the marginal effect with the residual standard deviation of the respective variable. See Panel B of Table \ref{tab:peervars_sel}.} Notice, that the latter should be interpreted with caution as they capture correlations and no causal effects. 


I find that an increase in the average peer employability has no statistically significant effect on job search duration. In contrast, a standard deviation increase in peer employability moderately increases total employment by around 15 to 20 days up until 5 years after program start. The effect is strongest for participants in short programs and smallest for participants in retraining. When comparing it to the average days in employment after program start (See Table \ref{tab:descr_cl_il}) it implies an increase of about 2 percent for all program types.

\begin{table}
\centering
  \caption{Effects of Peer Employability on Employment and Earnings}
  \label{tab:res_LMout}
  \resizebox{0.8\linewidth}{!}{%
\begin{tabular}[h!]{lccc} 
\hline\hline
 &\multicolumn{1}{c}{Short training}&\multicolumn{1}{c}{Long training}&\multicolumn{1}{c}{Retraining}\\
   &\multicolumn{1}{c}{(1)}&\multicolumn{1}{c}{(2)}&\multicolumn{1}{c}{(3)}\\
\hline
\multicolumn{4}{l}{\textbf{Panel A - Search duration first job (in days)}}\\ 
$\bar{X}_{(-i)pct}$&   -1.368&    8.752&    2.687\\
          &  (4.966)&  (7.382)&  (6.575)\\
$X_{ipct}$& -173.560***&-158.653***&-160.919***\\
          &  (5.216)&  (8.458)&  (7.489)\\
\textit{N}     &    \textit{36662}          &     \textit{12683}         &     \textit{12413}    \\[4pt]
\hline
\multicolumn{4}{l}{\textbf{Panel B - Total employment in month 60 (in days)}}\\ 
$\bar{X}_{(-i)pct}$ & 19.541***&18.041***&15.632***\\
          &  (2.717)&  (3.890)&  (3.920)\\
$X_{ipct}$ & 186.109***&184.271***&170.167***\\
          &  (2.980)&  (4.687)&  (4.486)\\
\textit{N}     &    \textit{36662}          &     \textit{12683}         &     \textit{12413}    \\
 \hline
 \multicolumn{4}{l}{\textbf{Panel C - Log earnings first job (if $>$ 0)}}\\ 
$\bar{X}_{(-i)pct}$& 0.041***&    0.014&   -0.005\\
          &  (0.006)&  (0.009)&  (0.010)\\
$X_{ipct}$& -0.014**&-0.067***&-0.088***\\
          &  (0.006)&  (0.011)&  (0.011)\\[4pt]   
\textit{N}     &    \textit{31355}         &     \textit{10779}         &     \textit{10618}         \\
\hline
\multicolumn{4}{l}{\textbf{Panel D - Log total earnings in month 60 (if $>$ 0)}}\\ 
$\bar{X}_{(-i)pct}$& 0.067***& 0.034***&   0.016*\\
          &  (0.007)&  (0.010)&  (0.009)\\
$X_{ipct}$& 0.247***& 0.264***& 0.226***\\
          &  (0.007)&  (0.012)&  (0.012)\\[4pt]
\textit{N}     &    \textit{33781}          &     \textit{11713}         &     \textit{11548}    \\
\hline
Provider x competence level FE&      \checkmark&      \checkmark&      \checkmark\\
Month x year x occupation FE&      \checkmark&      \checkmark&      \checkmark\\
Additional Controls&      \checkmark&      \checkmark&      \checkmark\\
N         &    36661&    12683&    12413\\
\hline\hline
\end{tabular}}
      \begin{minipage}{\textwidth}
          \vspace{6pt}
          \footnotesize \textit{Notes:} $X_{ipct}$ designates the own employablity and $\bar{X}_{(-i)pct}$ the leave-one-out mean peer employability. All specifications control for  course-level controls. Standard errors (in round brackets)  are clustered at the course level.  Effects are reported in terms of  SD increases (see Table \ref{tab:peervars_sel}). Earnings are in prices of 2010 and measured in log(euro).  \sym{*} \(p<0.05\), \sym{**} \(p<0.01\), \sym{***} \(p<0.001\)
     \end{minipage}
\end{table}

\begin{figure}[H]
 \begin{center}
    \caption{Monthly Effects of Peer Employability on Regular Employment} 
  \label{fig:effects_regemp}
 \includegraphics[width=0.9\textwidth]{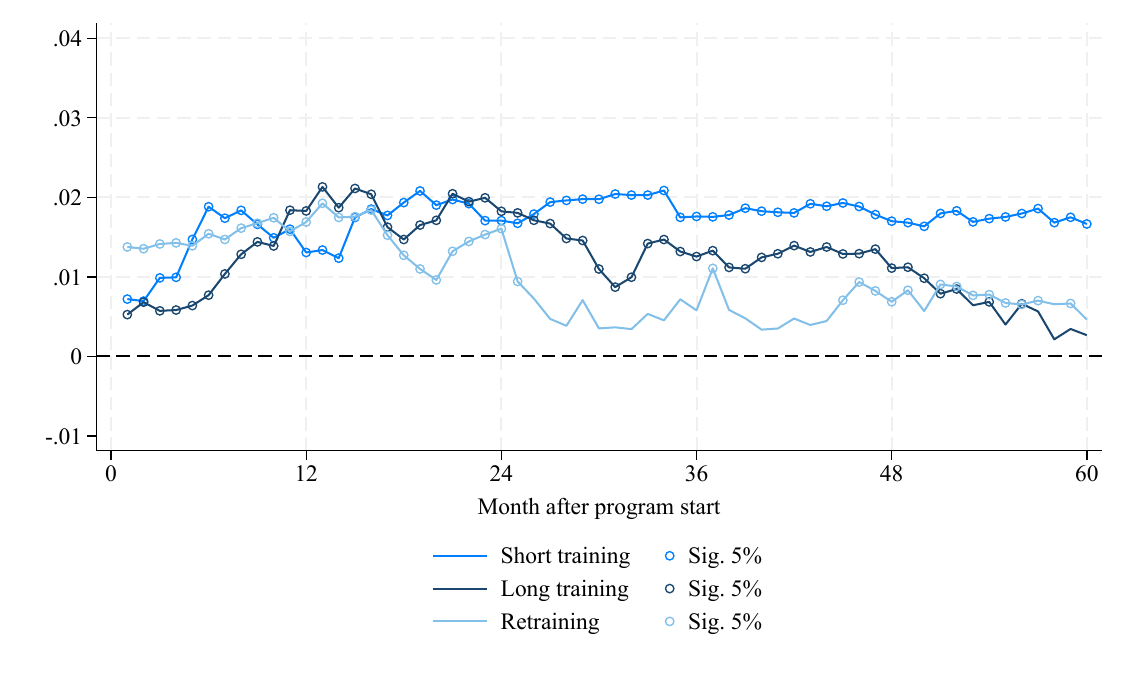}
 \end{center}
 \begin{minipage}{\textwidth}   \linespread{1.0}\selectfont
 \vspace{6pt}
     \footnotesize \textit{Notes:} The figure depicts the estimated effects (in percentage points/100) of a one standard deviation increase in the predicted mean peer employability on the individual probability to be employed in a regular job in the months 1-60 after program start. Significant effects at the 5 percent level are marked by circles. On top of the mean peer employability, the underlying model includes the individual ex-ante employability, a vector of course-level controls, provider x competence level and month x year x occupation fixed effects. Standard errors are clustered at the course level.
 \end{minipage}
\end{figure}

In order to investigate the dynamics behind the effects, I also estimate  monthly effects of an increase in the average predicted peer employability on individual employment up to 60 months after program start. Figure \ref{fig:effects_regemp} shows the effects on taking up a regular job (i.e. a job subject to social security contributions) by program type. It depicts the effect estimates (in percentage points) of a one standard deviation increase in the predicted mean peer employability for the months 1-60 after program start. Empty circles indicate that the effect is significant at the 5 percent level. Peer effects materialize directly after program start and increase in the months thereafter. This could be expected given that the program participants reduce their search efforts during the time of the program and that I find no effect on job search duration.\footnote{Several evaluation studies have found evidence for substantial lock-in effects of training programs in Germany. See e.g., \citealp{Lechner2011,Biewen2014}.} For short vocational training peer effects reach their first peak around six months after program start. For long vocational training and retraining they peak about one year after program start. At that time, a one standard deviation increase in the group's average employability increases  the  individual  employment  probability by around 2 percentage points. The effects are particularly stable for short training, persisting up to 60 months after program start. For long training, the effects decrease over time and fade out around five years after program start. For retraining, I find significant effects up until two years after program start, then 3 years and between four and five years after program start. When relating the effect sizes to the average employment rate about two years after program start (See Appendix Figure \ref{fig:avg_emp_alo}), they amount to about 3 percent for participants in vocational training programs and 6 percent for individuals in retraining who have a lower employment rate by that time.\footnote{I find similar patterns when considering a more global definition of employment that takes small jobs that are not subject to social insurance contributions into consideration (See Appendix Figure \ref{fig:effects_emp}). The effects are slightly lower amounting to around 1 percentage point per month. This raises the possibility that an exposure to high-quality peers does less likely lead to a take-up of employment in such small jobs compared to regular employment. However, the differences are small and mostly not statistically significant -- they should be interpreted with caution. 
}

In sum, the findings suggest that exposure to a more employable peer group does not necessarily lead to a faster integration of participants into the labor market but rather increases employment stability.  
Previous evaluation studies of the same types of programs in Germany, find that program participation increases the employment probability by 10 to 20 percentage points in the long run (see e.g., \citealp{McCall2016}, \citealp{Card2018}). In comparison, a peer effect of 1 to 2 percentage points seems like a moderate and realistic increase. However, a direct comparison of the effect sizes is difficult, since the peer effects analyzed in this study have to be interpreted conditional on participation.\footnote{The effect size is modest when compared to \cite{Baird2023} who find particularly large peer effects on employment. They find that 1 standard deviation increase in the average labor market history of peers increases the probability to be employed in the 6 quarters after training by 15 percentage points. This corresponds to about half of the average treatment effect and is thus exceptionally high.}


\subsubsection{Effects on Earnings}\label{sec:main2}

Table \ref{tab:res_LMout} also presents the effects of a more employable peer group on individual daily earnings in the first job and total earnings in the first five years after program participation. The sample is restricted to individuals with positive earnings. Specifically, I consider daily earnings in the first job in the sample of program participants who found a job up to December 2016 (Panel C) and total earnings in the sample of participants that were employed at least once in the first 60 months after program participation (Panel D). This is a possibly selected sample but I expect the earnings effect in the full sample to be similar given the small share of individuals that is excluded (roughly 8 percent) and the small employment effect that I find. I present some robustness checks with respect to the earnings results in Section \ref{sec:robustness3} which confirm this hypothesis. 



The results show that a one standard deviation increase in the average peer employability increases daily earnings in the first job by  4.1 percent for participants in short training (about 1 euro for the average participant). For participants in long training and retraining, I find no significant effect of a more employable peer group on earnings in the first job. However, in the long run, earnings across all program types are positively affected. My results point out a sizeable effect of 6.7 percent on total earnings in the first 5 years after program participation for participants in short training which corresponds to about 2700 euro for the average participant. The effects for participants in long training and retraining are lower amounting to 3.4 and 1.6 percent respectively.\footnote{ Sizeable peer effects on earnings were also found by \cite{Baird2023} for a job training program in the US and by \cite{Jarosch2021} for coworkers in firms.} The findings imply that being exposed to more employable peers allows jobseekers to secure higher-paying jobs and accumulate some earnings in the long run. In particular for short and long vocational training courses, these earnings effects are large and cannot be explained by more days in employment alone.

The results point to some differences in effects across program types in particular with respect to earnings in the first job. Some of these differences might be explained by a different composition of jobseekers across programs. As displayed in Appendix Figure \ref{fig:pred_avg_emp_by_type}, the distribution of peer employability at program start is shifted more to the right for participants in short vocational training compared to those in long vocational and retraining programs. Nevertheless, the differences should be interpreted with caution since confidence intervals are large.  Furthermore, as argued in Section \ref{validation}, due to possible selection effects the effects represent a lower bound for long vocational training and retraining programs.  


\subsection{Does the effect depend on one's own employability?}\label{sec:heterogeneity}

Findings from the education literature suggest that students of different ability levels benefit differently from peer ability (e.g., \citealp{Burke2013}, \citealp{Carrell2013}, \citealp{Feld2017}). Further, the program evaluation literature  has shown vocational training programs to be more effective for jobseekers with relatively bad employment prospects \citep{Card2018}. Therefore, I test whether individuals heterogeneously respond to the average employability in their group depending on their own employability. 

\begin{table}
\centering
  \caption{Heterogeneity in Effects of Peer Employability Depending on Own Employability}
  \label{tab:res_het_LMout_ownemp}
  \resizebox{0.8\linewidth}{!}{%
\begin{tabular}[htb]{lccc} 
\hline\hline
\multicolumn{4}{l}{\textbf{Panel A - Search duration first job (in days)}}\\ 
PE low employability&-16.252***&    0.285&    0.144\\
          &  (5.877)&  (8.470)&  (7.859)\\
PE high employability&23.072***&48.126***&31.647***\\
          &  (5.994)&  (9.594)&  (8.012)\\
P-value difference&        0&        0&        0\\
\textit{N}     &    \textit{36662}          &     \textit{12683}         &     \textit{12413}    \\[4pt]
\hline
\multicolumn{4}{l}{\textbf{Panel B - Total Employment (in days)}}\\ 
PE low employability&28.865***&16.948***&   7.942*\\
          &  (3.331)&  (4.944)&  (4.637)\\
PE high employability&    2.656&   -8.575&   -0.106\\
          &  (3.484)&  (5.342)&  (5.124)\\
P-value difference&        0&        0&     .132\\
\textit{N}     &    \textit{36662}          &     \textit{12683}         &     \textit{12413}    \\[4pt]
\hline
\multicolumn{4}{l}{\textbf{Panel C - Earnings in First Job (if > 0)}}\\ 
PE low employability& 0.021***&    0.009&   -0.016\\
          &  (0.006)&  (0.010)&  (0.011)\\
PE high employability& 0.068***& 0.032***&   0.024*\\
          &  (0.007)&  (0.012)&  (0.012)\\
P-value difference&        0&      .06&     .003\\
\textit{N}     &    \textit{31355}         &     \textit{10779}         &     \textit{10618}         \\[4pt]
\hline
\multicolumn{4}{l}{\textbf{Panel D - Total Earnings (if > 0)}}\\ 
PE low employability& 0.068***&  0.030**&   -0.006\\
          &  (0.008)&  (0.013)&  (0.012)\\
PE high employability& 0.059***&    0.003&    0.013\\
          &  (0.008)&  (0.013)&  (0.011)\\
P-value difference&      .32&     .035&     .173\\
\textit{N}     &    \textit{33781}          &     \textit{11713}         &     \textit{11548}    \\
\hline
Provider x competence level FE&      \checkmark&      \checkmark&      \checkmark\\
Month x year x occupation FE&      \checkmark&      \checkmark&      \checkmark\\
Additional Controls&      \checkmark&      \checkmark&      \checkmark\\
\hline\hline
\end{tabular}}
      \begin{minipage}{\textwidth}
          \vspace{6pt}
          \footnotesize \textit{Notes:} All specifications control for  course-level controls and individual employability. Standard errors (in round brackets)  are clustered at the course level.  Peer effects (PE) are reported in terms of  SD increases (see Table \ref{tab:peervars_sel}).  Earnings are in prices of 2010 and measured in log(euro). \sym{*} \(p<0.05\), \sym{**} \(p<0.01\), \sym{***} \(p<0.001\)
     \end{minipage}
\end{table}

I define program participants with a predicted employability below the sample median as participants with a low employability and participants with an employability greater or equal to the median as participants with a high employability. I then estimate the model similar to model (\ref{eq:main}) where I include a dummy variable for being a high-employability individual in place of the continuous measure for own employability ($X_{ipct}$) and an interaction term between this variable and the average peer employability.

The results are displayed in Table \ref{tab:res_het_LMout_ownemp} and show comparable patterns across program types. They suggest that jobseekers with a low employability draw higher long-term gains from being exposed to a better peer group compared to participants with a high employability. They accumulate more days in employment and in tendency higher earnings compared to highly employable jobseekers. In the short run, it is highly employable jobseekers who benefit relatively more. I find a larger positive effect on earnings in the first job which is accompanied by a positive effect on the job search duration for these types of individuals. This suggests that highly employable individuals might be more selective in their job search when exposed to a better peer group. They search for longer and possibly find better paid jobs. The positive effects on short-term earnings however, only materialize in significantly higher long-term earnings for participants in short training. Again, I do not over-interpreting these differences between programs, since the effects for long training and retraining programs are less precisely estimate.


\subsection{Mechanisms}\label{sec:mechanisms}

Different mechanisms could explain peer effects in the context of labor market training -- with possibly counteracting effects. Positive peer effects could arise from knowledge spillovers. Participants might learn from each other and benefit from the skills, knowledge and networks of more employable peers. These spillovers could improve the skill set and productivity of jobseekers and thereby positively affect their employment opportunities.\footnote{
 Knowledge spillovers have been discussed as main mechanism behind peer effects in education \citep{Booij2017, Kimbrough2022} and at the workplace \citep{Jarosch2021, Frakes2021, Cornelissen2023}. Such spillovers could also come in the form of referrals or knowledge transfers about effective job search strategies. Networks of friends, family or the neighbourhood have been shown to lead to more stable employment with higher wages (see e.g., \citealp{Cingano2012, Pelizzari2010, Brown2016, Dustmann2016}).}

Negative effects could result from an increased competition in job search. Jobseekers might loose out against highly employable peers when competing for jobs on the labor market.\footnote{Competition effects between peers have been analyzed at the workplace, e.g. by \cite{Cornelissen2023, Johnsen2024} and in job search, e.g. by \cite{Lazear2018}.}  Further, the importance of the within-group competition is likely to depend on the state of the labor market and the external competition for jobs. If the ratio of jobs to the number of unemployed in a labor market is low and the pool of unemployed is well qualified, then the outside competition might play a larger role.\footnote{The literature on macroeconomic effects of labor market programs suggests that jobseekers inside and outside these programs compete for the same jobs (\citealp{Ferracci2014, Gautier2018}). \cite{Crepon2013} for example found larger displacement effects of labor market programs in weak labor markets.} 

In the following, I disentangle peer effects into a knowledge spillover and competition effect. For this, I build on the literature on rank effects in education (See e.g. \citealp{Delaney2022}) and a strategy proposed by \cite{Cornelissen2023}. I identify knowledge spillovers from the average level of peer employability and competition from the ordinal rank of a jobseeker within the peer group. I estimate the following specification:

\begin{equation} \label{eq:rank}
y_{ipcot} = \alpha +  \sum_{j=2}^{30} \gamma_j X_{ipcot} + \theta \bar{X}_{(-i)pcot} + \psi R_{ipcot}  + \pi W_{ipcot} + \lambda_{pc} + \delta_{ot} + \epsilon_{ipcot},
\end{equation}  

which additionally includes $R_{ipcot}$, the percentile rank of jobseeker $i$ amongst the peers in her course. To isolate the effect of rank for a given level of employability, I control for own employability using percentile dummies. The rank effect ($\psi$) is thus identified from differences between courses in higher moments of the ability distribution than the mean.\footnote{Identifying rank effects is econometrically challenging since for a specific individual $i$, rank is the percentile of the employability distribution that corresponds to their own employability. Researchers thus usually rely on parametric assumptions to condition on rank and peer quality whose validity depend on the choice of correct functional form.} I choose a flexible specification with 30 percentile dummies to control for own employability. In order to interpret the coefficient on the average peer employability as pure knowledge spillover effect, I need to assume that rank perfectly controls for the within group competition and that that there are no other mechanisms that drive peer effects. In reality, rank will only capture specific aspects of how within-group competition affects labor market outcomes.\footnote{The education literature argues that rank effects operate primarily through self-confidence and less through competition. The competition channel seems more relevant in the job search context. However, shifts in self-confidence were also discussed as mechanism behind peer effect \citep{Vandenberg2019}.}  
Furthermore $\gamma$ might pick up other spillovers that do not constitute knowledge spillovers, if there are any.\footnote{Other possible mechanisms could be peer pressure \citep{Mas2009, Fu2019} or endogenous teacher behavior \citep{Brodaty2016}. However, they are unlikely to be the main drivers behind the effects. I further discuss them in Appendix \ref{appendix:mechanisms_other}.} Nevertheless, this excercise will inform us qualitatively about a possible split between these two most prominent mechanisms. 

As mechanisms are likely to depend on the state of the labor market, I additionally perform a heterogeneity analysis. I split courses depending on the tightness-level associated with their local labor market at the moment of program start.\footnote{ I measure labor market tightness as the ratio of vacancies registered at employment agencies to the number of unemployed in a specific occupation (2-digit level, KldB 2010) at the moment of program start.  The average tightness-level is about 11 percent, i.e. relatively low. Notice though, that the vacancy data does not cover the universe of jobs. Tightness is measured at program start, since the moment of exit is endogenous.} Courses are considered as tight if the ratio of vacancies to unemployed for their target occupation is higher than the median tightness across all courses starting in a given month. They are classified as not tight if the ratio is equal to or lower than the median. I then estimate a version of model (\ref{eq:rank}) where I include an indicator for whether a course is in a tight market, as well as an interaction term of this indicator with rank and average employability respectively. Notice, that this excercise is better suited to capture heterogeneity in rather short programs. The state of the labor market at program start is less relevant for jobseekers who exit a program only years after.

The results of the analysis are shown in Tables \ref{tab:mech_emp} and \ref{tab:mech_earn}. Columns (1), (4) and (7), measure a joint peer effect as in the main specification but control for own employability with percentile dummies instead of linearly. Columns (2), (5) and (8) show the decomposed effect. The coefficient on $\bar{X}_{(-i)pcot}$ captures knowledge spillovers, the coefficient on $R_{ipcot}$ an (opposite) competition effect. Since I control for own and average employability, the latter can be interpreted as pure ordinal rank effect. Columns (3), (6) and (9) show the estimated spillover and rank effects separately for not-tight markets (first two coefficients) and tight markets (second two coefficients). P-values from F-tests for the differences in effects between tight and non-tight markets are shown below.

The results for search duration (Panel A, Table \ref{tab:mech_emp}) suggest that the two mechanisms indeed have some counteracting effect. Once we condition on rank, the effects of average peer employability become smaller or more negative. The coefficients on rank point to lower search durations for higher ranked individuals. These workers have a better relative position among their peers and are able to find jobs faster. Note however, that we cannot precisely identify knowledge spillovers on search duration in any of the program types. Rank effects are only statistically significant for short training programs. 

I do not find statistically significant rank effects on long-term employment (Panel B, Table \ref{tab:mech_emp}). The coefficients on average peer employability are qualitatively similar to the joint effects. This suggests that the competition among jobseekers in the same course does not affect job stability in the long run.  

\begin{landscape}
\begin{table}[htbp]\centering
\def\sym#1{\ifmmode^{#1}\else\(^{#1}\)\fi}
\caption{Spillover and Competition Effects on Employment} \label{tab:mech_emp}
\footnotesize{
\begin{tabular}{l*{9}{c}} 
\hline\hline
          &\multicolumn{1}{c}{(1)}&\multicolumn{1}{c}{(2)}&\multicolumn{1}{c}{(3)}&\multicolumn{1}{c}{(4)}&\multicolumn{1}{c}{(5)}&\multicolumn{1}{c}{(6)}&\multicolumn{1}{c}{(7)}&\multicolumn{1}{c}{(8)}&\multicolumn{1}{c}{(9)}\\
          &\multicolumn{1}{c}{ST}&\multicolumn{1}{c}{ST}&\multicolumn{1}{c}{ST}&\multicolumn{1}{c}{LT}&\multicolumn{1}{c}{LT}&\multicolumn{1}{c}{LT}&\multicolumn{1}{c}{RT}&\multicolumn{1}{c}{RT}&\multicolumn{1}{c}{RT}\\
\hline
\multicolumn{10}{l}{\textbf{Panel A - Search duration first job (in days)}}\\ 
$\bar{X}_{(-i)pct}$&   -1.337&   -7.295&   -6.019&    8.581&    4.005&   -4.327&    1.523&   -1.955&    9.263\\
          &  (4.960)&  (6.074)&  (7.091)&  (7.359)&  (9.117)& (10.791)&  (6.517)&  (8.056)&  (9.354)\\
$R_{ipcot}$&         &  -8.247*&  -9.066*&         &   -7.343&   -9.222&         &   -5.282&  -10.606\\
          &         &  (4.781)&  (4.933)&         &  (8.148)&  (8.465)&         &  (7.161)&  (7.709)\\
$\bar{X}_{(-i)pct}$ (tight)&         &         &   -7.589&         &         &   14.472&         &         &  -11.700\\
          &         &         &  (8.107)&         &         & (11.424)&         &         & (10.050)\\
$R_{ipcot}$ (tight)&         &         &   -6.265&         &         &   -5.136&         &         &   -0.356\\
          &         &         &  (5.554)&         &         &  (9.121)&         &         &  (7.794)\\
P-value diff: $\bar{X}_{(-i)pct}$&         &         &     .861&         &         &     .136&         &         &     .055\\
P-value diff: $R_{ipcot}$&         &         &     .488&         &         &     .523&         &         &     .086\\
N         &    36662&    36662&    36662&    12683&    12683&    12683&    12413&    12413&    12413\\[4pt]
\hline
\multicolumn{10}{l}{\textbf{Panel B - Total Employment(in days)}}\\ 
$\bar{X}_{(-i)pct}$&19.497***&22.484***&23.601***&17.933***&14.874***&21.717***&15.906***&13.467***& 14.177**\\
          &  (2.709)&  (3.283)&  (3.982)&  (3.904)&  (4.746)&  (5.704)&  (3.893)&  (4.732)&  (5.526)\\
$R_{ipcot}$&         &    4.136&  5.508**&         &   -4.910&   -3.407&         &   -3.703&    1.720\\
          &         &  (2.684)&  (2.789)&         &  (4.436)&  (4.603)&         &  (4.096)&  (4.429)\\
$\bar{X}_{(-i)pct}$ (tight)&         &         &19.736***&         &         &    7.002&         &         & 12.667**\\
          &         &         &  (4.139)&         &         &  (5.988)&         &         &  (5.904)\\
$R_{ipcot}$ (tight)&         &         &    1.224&         &         &   -6.812&         &         & -8.847**\\
          &         &         &  (3.094)&         &         &  (5.019)&         &         &  (4.443)\\
P-value diff: $\bar{X}_{(-i)pct}$&         &         &     .417&         &         &     .031&         &         &     .816\\
P-value diff: $R_{ipcot}$&         &         &      .06&         &         &     .346&         &         &     .002\\
N         &    36662&    36662&    36662&    12683&    12683&    12683&    12413&    12413&    12413\\
\hline\hline
\end{tabular}}
      \begin{minipage}{\paperwidth}
          \vspace{6pt}
          \footnotesize \textit{Notes:} The results are displayed seperately by program type: short training (ST), long training (LT), retraining (RT). $\bar{X}_{(-i)pct}$ designates the leave-one-out mean peer employability and $R_{ipcot}$ the percentile rank. All specifications control for  course-level controls, provider x competence level FE and month x year x occupation FE. Standard errors (in round brackets)  are clustered at the course level.  Effects are reported in terms of  SD increases (see Table \ref{tab:peervars_sel}). \sym{*} \(p<0.05\), \sym{**} \(p<0.01\), \sym{***} \(p<0.001\)
     \end{minipage}
\end{table}

\begin{table}[htbp]\centering
\def\sym#1{\ifmmode^{#1}\else\(^{#1}\)\fi}
\caption{Spillover and Competition Effects on Earnings } \label{tab:mech_earn}
\footnotesize{
\begin{tabular}{l*{9}{c}}
\hline\hline
          &\multicolumn{1}{c}{(1)}&\multicolumn{1}{c}{(2)}&\multicolumn{1}{c}{(3)}&\multicolumn{1}{c}{(4)}&\multicolumn{1}{c}{(5)}&\multicolumn{1}{c}{(6)}&\multicolumn{1}{c}{(7)}&\multicolumn{1}{c}{(8)}&\multicolumn{1}{c}{(9)}\\
          &\multicolumn{1}{c}{ST}&\multicolumn{1}{c}{ST}&\multicolumn{1}{c}{ST}&\multicolumn{1}{c}{LT}&\multicolumn{1}{c}{LT}&\multicolumn{1}{c}{LT}&\multicolumn{1}{c}{RT}&\multicolumn{1}{c}{RT}&\multicolumn{1}{c}{RT}\\
\hline
\multicolumn{4}{l}{\textbf{Panel A - Earnings in First Job (if > 0)}}\\ 
$\bar{X}_{(-i)pct}$& 0.036***& 0.056***& 0.067***&    0.009&    0.009&    0.007&   -0.008&    0.000&    0.005\\
          &  (0.005)&  (0.007)&  (0.008)&  (0.009)&  (0.011)&  (0.014)&  (0.009)&  (0.012)&  (0.015)\\
          &         &         &         &         &         &         &         &         &         \\
$R_{ipcot}$&         & 0.027***& 0.029***&         &    0.001&   -0.003&         &    0.012&    0.004\\
          &         &  (0.006)&  (0.006)&         &  (0.010)&  (0.011)&         &  (0.011)&  (0.011)\\
$\bar{X}_{(-i)pct}$ (tight)&         &         & 0.040***&         &         &    0.014&         &         &   -0.003\\
          &         &         &  (0.009)&         &         &  (0.014)&         &         &  (0.014)\\
$R_{ipcot}$ (tight)&         &         & 0.023***&         &         &    0.005&         &         &   0.020*\\
          &         &         &  (0.007)&         &         &  (0.012)&         &         &  (0.011)\\
          &         &         &         &         &         &         &         &         &         \\
P-value diff: $\bar{X}_{(-i)pct}$&         &         &     .004&         &         &     .674&         &         &     .628\\
P-value diff: $R_{ipcot}$&         &         &     .256&         &         &     .398&         &         &     .051\\
N         &    31355&    31355&    31355&    10779&    10779&    10779&    10618&    10618&    10618\\[4pt]
\hline
\multicolumn{4}{l}{\textbf{Panel B - Total Earnings (if > 0)}}\\ 
$\bar{X}_{(-i)pct}$& 0.063***& 0.088***& 0.098***& 0.032***&  0.029**&   0.028*&    0.014&    0.016&  0.039**\\
          &  (0.007)&  (0.008)&  (0.010)&  (0.010)&  (0.012)&  (0.015)&  (0.009)&  (0.012)&  (0.015)\\
$R_{ipcot}$&         & 0.034***& 0.039***&         &   -0.004&   -0.003&         &    0.002&    0.014\\
          &         &  (0.006)&  (0.007)&         &  (0.011)&  (0.012)&         &  (0.011)&  (0.012)\\
$\bar{X}_{(-i)pct}$ (tight)&         &         & 0.070***&         &         &  0.031**&         &         &   -0.004\\
          &         &         &  (0.010)&         &         &  (0.016)&         &         &  (0.014)\\
$R_{ipcot}$ (tight)&         &         & 0.025***&         &         &   -0.007&         &         &   -0.009\\
          &         &         &  (0.007)&         &         &  (0.012)&         &         &  (0.012)\\
P-value diff: $\bar{X}_{(-i)pct}$&         &         &     .018&         &         &     .874&         &         &     .008\\
P-value diff: $R_{ipcot}$&         &         &      .01&         &         &     .675&         &         &     .014\\
N         &    33781&    33781&    33781&    11713&    11713&    11713&    11548&    11548&    11548\\
\hline\hline
\end{tabular}}
 \begin{minipage}{\paperwidth}
          \vspace{6pt}
          \footnotesize \textit{Notes:} The results are displayed seperately by program type: short training (ST), long training (LT), retraining (RT). $\bar{X}_{(-i)pct}$ designates the leave-one-out mean peer employability and $R_{ipcot}$ the percentile rank. All specifications control for  course-level controls, provider x competence level FE and month x year x occupation FE. Standard errors (in round brackets)  are clustered at the course level.  Effects are reported in terms of  SD increases (see Table \ref{tab:peervars_sel}). Earnings are in prices of 2010 and measured in log(euro).  \sym{*} \(p<0.05\), \sym{**} \(p<0.01\), \sym{***} \(p<0.001\)
     \end{minipage}
\end{table}

\end{landscape}

The decomposition for short and long-term earnings (Table \ref{tab:mech_earn}) points to important competition effects which attenuate positive spillovers, in particular for participants in short training. In these courses, knowledge spillover effects increase by 40 to 55 percent once we condition on rank. For individuals in long vocational training and retraining competition effects seem to play less of a role but cannot be precisely estimated.

For the discussion of the heterogeneity analysis by tightness level, I focus on short training programs.\footnote{The results point to a particularity for retraining programs. High-ranked participants in these programs have negative rank effects on employment and higher ones on first earnings in when starting their program in tight markets. This might suggest that high-ranked jobseekers are more selective in their job search. However, since retraining program usually last several years, there could be other drivers behind this result.}  With their duration of a few months, they are likely to provide the most sensible results. The analysis suggests that competition effects are quantitatively more important in not-tight labor markets compared to tight labor markets. This is pointed out by larger coefficients on rank, even though the difference is not statistically significant for all outcomes. Since the ratio of jobs to unemployed is comparatively low in not-tight markets, this results in a fiercer competition for jobs in these markets. At the same time, it seems to increase the within group competition. Knowledge spillover effects are also larger in these types of markets, in particular for earnings outcomes. They could be explained by a larger motivation and efforts of highly employable jobseekers in these programs from which other course participants might benefit. This result is also consistent with the literature finding smaller lock-in effects and larger employment effects of training programs when unemployment is high \citep{Lechner2009}. 

Overall, the findings suggest, that independent of the labor market situation, within group competition in job search attenuates  part of the effects of peer employability that operate through knowledge spillovers. Competition has a larger relative impact for short-term outcomes which could be expected. Finally, competition has the largest impact on earnings outcomes, implying that it primarily affects who obtains the better paid jobs.

\subsection{Robustness Checks}\label{sec:robustness}

This section presents robustness checks concerning the construction of the employability measure, the identification strategy and the measurement of earnings.

\subsubsection{Measurement of peer employability}\label{sec:robustness1}

Peer effects estimates might be biased if  the ``true'' peer employability to which jobseekers are exposed differs from the one that is observed. Below, I discuss two potential sources of error and argue that any resulting bias is likely to be small. 

\textbf{Timing of measurement.} \quad Measuring employability at program start, might result in some measurement error in the employability score in particular for individuals with a long unemployment duration at program start. 
See Appendix \ref{appendix:modelselection} for more details. Therefore, I estimate an alternative version of the employability score at the moment when jobseekers enter unemployment. This is also  where I can most precisely measure the inputs to the model and observe the sample of non-participants who are used to train the model. Appendix Figure \ref{app_fig:emp_pr_ue} shows that the employability distribution is clearly shifted to the right, when measured at unemployment entry. This suggests that some individuals' employability declines over the course of their unemployment spell, until they start a program.

Next, I test the sensitivity of the results to this alternative employability measure.\footnote{I estimate model (\ref{eq:main}) and additionally include the individual unemployment duration at program start as control variable.} The results are shown in Figure \ref{fig:rob_comb} by a diamond shaped marker in navy. They are qualitatively comparable to the results of the main specification ( shown by a black circle). I find somewhat lower effects on employment and earnings, in particular for jobseekers in long training but confidence intervals between the two models are largely overlapping. This suggests, that the timing of employability measurement has little effect on ultimate peer effects -- most likely, the shifts in employability over time are rank-preserving.

\textbf{Missing peers.} \quad I might not observe all peers in my data which could lead to a mismeasurement of peer variables. I discuss this possibility in detail in Appendix \ref{appendix:validity_checks_me}. In summary, I argue  for the fraction of missing peers to be small and for the missing data to be distributed independently of group assignment and the error term, conditional on fixed effects and course controls. 
Under these conditions a measurement error would merely cause a small attenuation bias and my results would represent a lower bound of the true peer effects (see \citealp{Ammermueller2009}, \citealp{Sojourner2013}, \citealp{Feld2017}).


\subsubsection{Threats to Identification}\label{sec:robustness2}

The tests presented in Section \ref{validation} suggest that the variation in peer emplyoability that I use to identify peer effects is similar to a variation resulting from random fluctuations. Nevertheless, the effects estimated for participants in long vocational training and retraining might suffer from some selection bias. In this section, I present additional checks that rule out possible sorting behaviors. 

\textbf{Sorting based on occupations.} \quad In the main specification I compare courses at the same provider, with the same competence level and difference out time trends specific to the occupational area of the target occupation. By doing so, I might still pick up some heterogeneity in peer quality that is due to sorting into specific occupations. For example, jobseekers in a course in construction might be systematically different from jobseekers in health care even though they train for occupations with the same competence level. To adress potential concerns with respect to such sorting, I estimate a model version where I compare courses within the same provider, occupational area and competence level. Specifically, I include provider x occupation x competence level fixed effects in model (\ref{eq:main}). I also control for aggregated time trends at the level of occupational area  (as in the main specification).\footnote{ This specification requires at least two courses of the same occupation and competence level per provider and at least two providers in an occupational labor market which reduces the sample slightly.} The results are shown in Figure \ref{fig:rob_comb}  by red triangles. For all program types, I find qualitatively similar results with slighlty attenuated effects. In particular, for long training and retraining the effects on long-term earnings are not significantly different from zero any more at the 95\% confidence level. 

\textbf{Regional business cycles.} \quad Business cycles might not only vary at the level of occupational areas but also across regions in which providers operate. To address potential concerns with respect to region-specific sorting over time, I estimate a version of the main model where I control for region-specific time trends instead of occupation-specific time trends. Specifically, I include month x year x region fixed effects in model (\ref{eq:main}) and separately control for occupational groups.\footnote{ I define 9 regions on the level of federal states: 1 "Hamburg and Bremen", 2 "Schleswig-Holstein und Niedersachsen"  3 "Nordrhein-Westfalen" 4 "Rheinland-Pfalz, Hessen and Saarland", 5 "Baden-Wuerttemberg", 6 "Freistaat Bayern", 7 "Berlin", 8 "Brandenburg und Mecklenburg-Vorpommern", 9 "Sachsen-Anhalt and Thüringen".}. The results are shown in Figure \ref{fig:rob_comb}  by a squared marker in purple. They are very close to the results of the main specification.

\textbf{Provider-specific mechanical sorting.}  \quad A dynamic self-selection into courses based on month and provider-specific preferences in combination with the three month redemption period of vouchers might cause some mechanical sorting. Consider the pool of jobseekers obtaining a voucher in the same month, e.g. February. If highly employable jobseekers within this pool select into  courses starting in a certain month (e.g. April), then the jobseekers attending the other months possible within their voucher validity (e.g. February, March or May) might "mechanically" be less employable.\footnote{This mechanical sorting is only problematic  if it is provider-specific. If not, it is already controlled for by the occupation-specific time trends.} I can potentially mitigate this mechanical sorting by differencing out sorting at the provider level that follows the same cyclicality in each quarter.\footnote{The seasonality of endogenous sorting into specific course months is in fact unknown.} I do so by comparing courses that are exactly 4 months apart at a specific provider. Then, the pool of individuals who select into a specific course month,  does not overlap with the pool of individuals who  select into courses starting exactly 4 and 8 months later. To take the same example: The feasible course options of jobseekers who start a course in the month of April do not overlap with those of jobseekers starting a course in August or December.  

In a robustness check, I thus compare only jobseekers at specific providers in courses that are 4 months apart, i.e. in a provider-specific \textit{month-group}. These are January-May-September, February-June-October, March-July-November and April-August-December.  Furthermore, I difference out time trends that are quarter-year-occupation specific and calendar month specific but constant across providers.\footnote{ A quarter refers to a 4-month divisions of the calendar year. Calendar month dummies control for potential seasonal sorting that is constant over the years. Controlling for time trends at the month x year level is not feasible in this approach. The competence level is included in the vector of course-level controls. The sample is restricted to providers offering at least two courses per \textit{month-group} and thus slighlty smaller.}  The intuition and execution of the approach is further illustrated in Appendix \ref{appendix:mech_sorting}. The results are reported in  Figure \ref{fig:rob_comb}  by an cross-shaped marker in orange. They are qualitatively similar to the results I obtain with my main identification strategy. For long training programs, the effects on long-term employment are lower and not statistically different from zero. I still find significant and positive earnings effects for individuals in this program type.

\textbf{Previous colleagues.} \quad Finally, I aim to rule out that the effects I estimate are endogenously driven by program participants that know each other from their previous job. This could bias the peer effects  if these jobseekers select into the same courses based on similar preferences or characteristics. In a robustness check, I thus exclude all courses from the sample where more than 20 percent of participants worked at the same firm before entering the program. The results are reported in  Figure \ref{fig:rob_comb}  by green arrows. They are very similar to the results of the main analysis with the effects on employment being slightly attenuated. In contrast to the main analysis, I find a significant reduction of the search duration for the first job for participants in short training. This difference might come from differences in the sample composition.
In sum, the check shows that my results are unlikely to be biased by the fact that jobseekers who know each other endogenously form peer groups.

\subsubsection{Earnings Effects}\label{sec:robustness3}
Unfortunately, I do not observe earnings for the entire sample since some individuals do not find employment in the observation window. The main results are thus based on a smaller sample (roughly 92\%) which is possibly selected. In the following, I thus examine the robustness of the results on earnings to the inclusion of zero earnings. Appendix Table \ref{tab:res_earnings} shows the results for the earnings outcomes in logs (Panels A and C)\footnote{For individuals who are not employed until the end of the observation period, their log earnings are set to 0. The same applies to individuals not having a first job by the end of the observation period. As a robustness check, I also measured zero earnings as log(euro+1). This has no impact on the results which are available on demand.} and in levels (Panels B and D) for the full sample. Both results in levels and logs are qualitatively similar to the results I obtain in the main estimation. The effects in logs are slightly larger compared to the effects in the sample that conditions on employment. Overall, I find a robust effect on earnings in the first job for individuals in short training and substantial positive earnings effects in the long run across all program types. Earnings effects are smallest for participants in retraining.

Overall, the effects are robust to different specifications, e.g. controlling for occupation-specific sorting and regional business cycles. Earnings effects are less precisely estimated for participants in long vocational training and retraining. 
They are more sensitive to fail the significance criteria in some of the robustness checks.

\section{Conclusion} \label{conclusion}

This paper investigates how labor market outcomes of participants in public sponsored training depend on the peer group composition as measured by the average employability of the peers. Using a number of predetermined characteristics, I construct a summary measure to proxy for employability. It is defined as the predicted probability to find a stable occupation within one year after program start. To identify a causal effect, I exploit the quasi-random variation in peer employability across courses of the same competence level offered by the same training provider over time. 

My results show that the composition of peers has a significant and positive impact on post program labor market outcomes. Effects on long-term employment are moderate and qualitatively comparable across program types. Effects on long-term earnings are sizeable amounting to 2 to 6 percent but can only be robustly identified for participants in short training programs.  I find substantial effect heterogeneity with respect to participants' own employability. Low-employability participants benefit relatively more from a better peer group than highly employable participants in the long run. The results further suggest that in the short run highly employable jobseekers are more selective in their job search when exposed to a highly employable peer group.

To understand the mechanisms behind the peer effects, I decompose them into a knowledge spillover  and competition component. For this, I proxy competition with a participant's ordinal rank in the group. While this analysis requires strong assumptions and effect sizes should be interpreted cautiously, it informs on the broad importance of both channels. The results suggest that knowledge spillovers are the main driver behind peer effects.  Jobseekers might  benefit from the skills and networks of their peers to obtain more stable and better paid jobs. At the same time, competition between participants attenuates this effect -- in particular for earnings outcomes. This implies that higher-ranked jobseekers tend to secure the better-paid jobs, while lower-ranked individuals are left with less attractive positions. Finally, a heterogeneity analysis suggests that peer effects and their drivers might depend on the state of the labor market. For short vocational training programs, I find that both knowledge spillover and competition effects are larger when jobseekers are exposed to strong external competition. 


My results bear important insights for policy makers. They suggest that peer effects could be used to increase the effectiveness of training programs through encouraging knowledge spillovers and reducing competition between participants. 
Accompanying measures could include greater promotion of knowledge sharing within the group and a targeted job search counselling for less employable participants, especially when conditions on the labor market are difficult. Furthermore, program effectiveness could be increased by optimal allocation policies. Recent findings by \cite{Baird2023} suggest, that in settings where programs target a specific subset of jobseekers, peer effects are linear-in-means and positive and where the social planner cares about the average outcome of individuals or about those most in need, assigning a higher fraction of highly employable jobseekers to programs is likely to increase program effectiveness. Similar effects could be expected for my setting. However, possible non-linearities in effects which might alter this recommendation cannot be ruled out.  Generally, policy recommendations regarding regrouping policies should be treated with caution since interventions that manipulate the peer composition have been shown to be confounded by endogenous sorting of individuals \citep{Carrell2013}. Finally, in the case of Germany, an assessment would be necessary to determine whether the net effects of a targeted allocation are greater than those of a voucher system, given that jobseekers are currently not assigned to courses but are free in their course choice.  

\setstretch{1.5} %
\vspace{12pt}
{\small
\bibliographystyle{econ}
\bibliography{library}
}


\newpage

\appendix
\addtocontents{toc}{\protect\setChapterprefix{Appendix }}
\setcounter{table}{0}
\setcounter{figure}{0}  
\renewcommand{\thetable}{A.\arabic{table}}
\renewcommand{\thefigure}{A.\arabic{figure}}

\section{Appendix}\label{appendix}

\subsection{Construction of the Employability Score} \label{appendix:employability}

\subsubsection{Matching Based on the Propensity Score} \label{appendix:matching}

I rely on a population of jobseekers that do not participate in any program to estimate the employability score. The reason for doing so is that the employment status without program participation is unobserved for actual participants. However, the sample of non-participant is likely to be selective which I try to mitigate with a matching approach. For the employability model to result in unbiased predictions, we need to assume that the relation between the observed predictors in the sample of non-participants and the employability outcome would be the same in the sample of program participants. This is more likely to hold the more comparable jobseekers in both samples are in terms of their characteristics.

I select the most comparable sample of jobseekers who do not enter any program during their unemployment spell applying exact nearest neighbour matching (with replacement) based on the propensity score (see e.g., \citealp{Huber2013}, \citealp{Abadie2016}). 
The propensity score is estimated as the conditional probability of being in the sample of program participants based on a set of variables that have been identified as sufficient to eliminate any selection bias in the context of program participation (see \citealp{Lechner2013}, \citealp{Huber2013}, \citealp{Caliendo2017}). It is estimated separately by program type to account for a possible heterogeneity in selection effects across program types. I include demographic characteristics, information on education, training, characteristics of the last job, the recent labor market history, as well as variables characterizing the local labor market situation in the model. For this step of the analysis, time sensitive variables are measured both for the sample of participants and non-participants at the moment of unemployment entry. 

Table \ref{tab:balancedness} 
gives an overview of the variables included in the propensity score estimation. It presents averages for the sample of participants (column 1) and non-participants (column 3) as well as the difference in means (column 5) and the standardized bias (column 7). Even though there remain small differences in means for single variables, the standardized bias lies mostly below 25\%, which is the reference value given by \cite{Rubin2001}. Note that the number of observations in the sample of participants is higher than in the final estimation sample. This is because some sample restrictions were imposed after the employability estimation. 

\begin{figure}[htbp]
      \caption{Predicted P-score (Program Participants and Non-participants)}\label{app_fig:pscore}
\begin{center}
  \includegraphics[width=0.55 \linewidth,trim=0 10 0 10, clip]{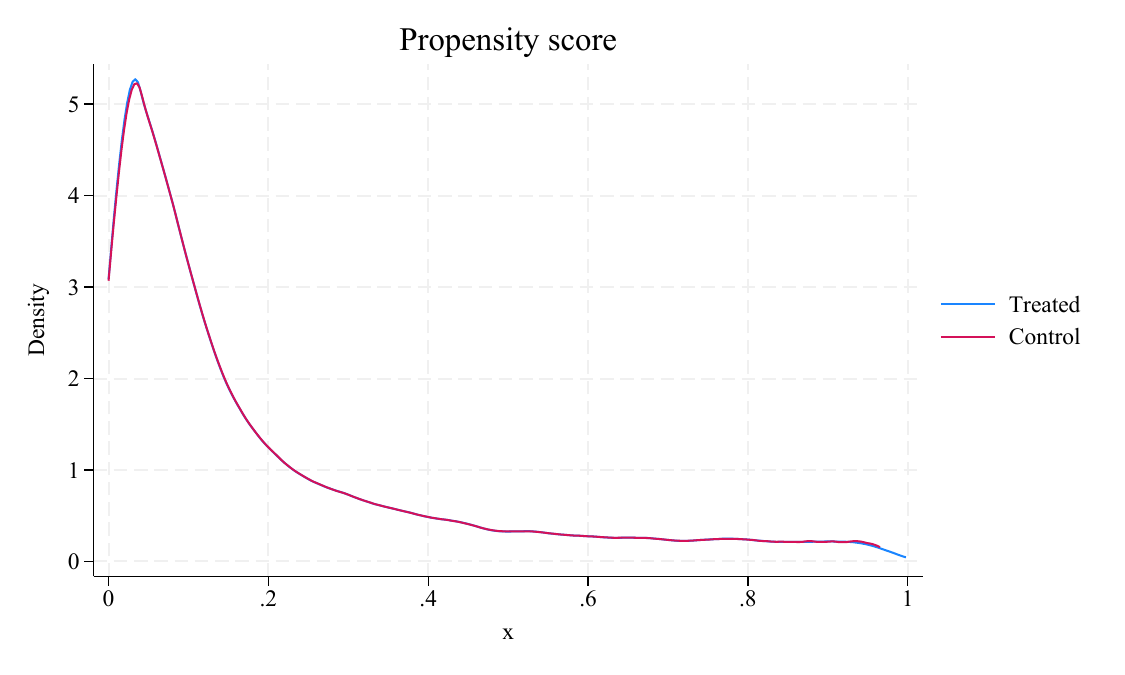}
 \caption*{(a) Short Training }
  \includegraphics[width=0.55\linewidth,trim=0 10 0 10, clip]{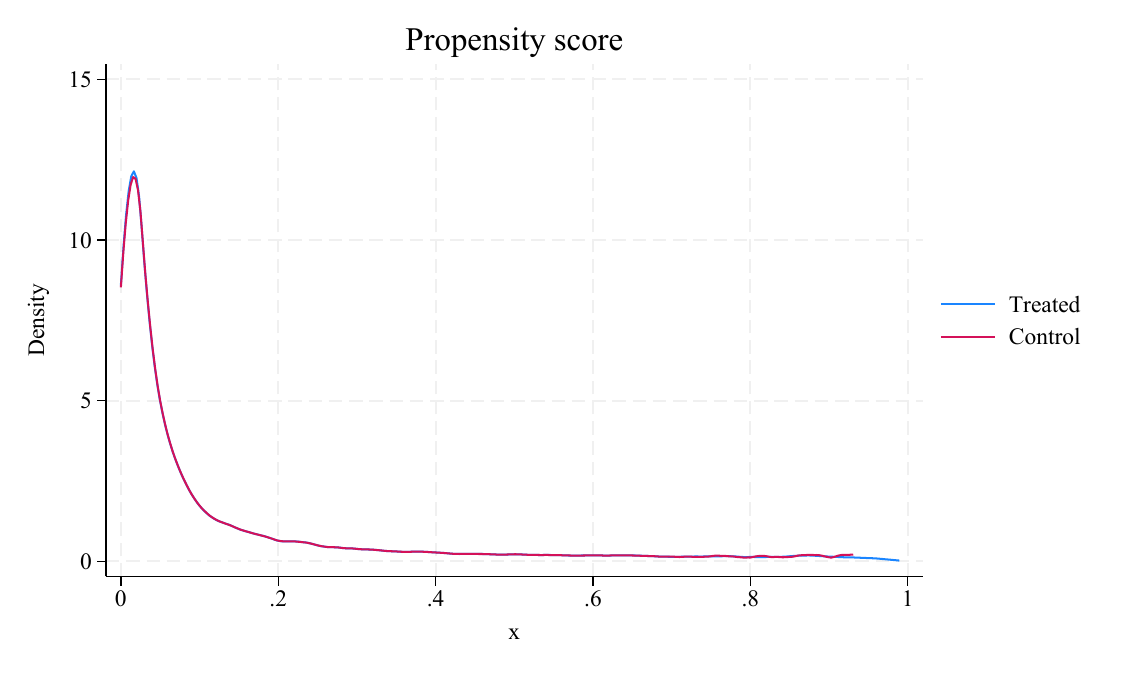} 
  \caption*{(b) Long Training}
   \includegraphics[width=0.55\linewidth,trim=0 10 0 10, clip]{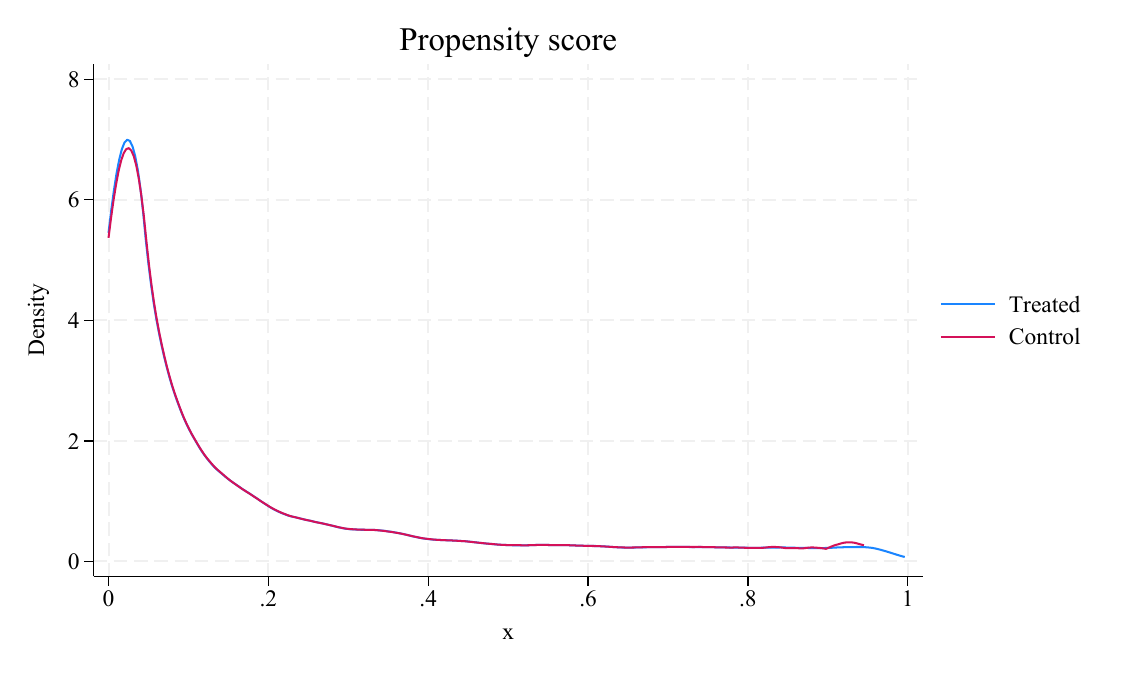} 
   \caption*{(c) Retraining}
   \end{center}
   \begin{minipage}{\textwidth}
          \vspace{2pt}
          \footnotesize \textit{Notes:} This figures display the distribution of the predicted p-score for program participants (PP) and matched non-participants (NP) by program type. 
     \end{minipage}
\end{figure}

Appendix Figure \ref{app_fig:pscore} shows the distribution of the predicted propensity scores for program participants and matched non-participants by program type. The curves largely overlap for both groups which indicates that the matching worked well.  Common support for individuals with very high propensity scores is violated. Nevertheless, this concerns only a small percentage of program participants. Overall, I am thus sufficiently able to balance the sample of program participants and non-participants based on their observable characteristics and therefore assume that they are also balanced in unobservable characteristics.

\FloatBarrier

\setlength\LTleft{10pt}
\setlength\LTright{10pt}
{\small
\begin{longtable}[!ht]{lrrrrr}
    \caption{Average Characteristics of Participants and Non-participants}\\
    \hline\hline
         & (1) & (2) & (3) & (4) & (5) \\ 
         & PP & NP & Diff & pval & SB \\ 
         \hline
               \multicolumn{2}{l}{\textit{Month of entry into unemployment}}\\
        January & 0.12 & 0.13 & 0.00 & 0.66 & 0.31 \\ 
        February & 0.09 & 0.08 & 0.00 & 0.10 & 1.36 \\ 
        March & 0.09 & 0.08 & 0.01 & 0.00 & 2.78 \\ 
        April & 0.09 & 0.09 & 0.00 & 0.86 & 0.13 \\ 
        May & 0.07 & 0.07 & 0.00 & 0.12 & 1.02 \\ 
        June & 0.07 & 0.07 & 0.00 & 0.31 & 0.67 \\ 
        July & 0.08 & 0.08 & 0.00 & 0.33 & 0.77 \\ 
        August & 0.08 & 0.08 & 0.00 & 0.18 & 1.22 \\ 
        September & 0.09 & 0.10 & -0.01 & 0.04 & 2.40 \\ 
        October & 0.09 & 0.09 & 0.00 & 0.32 & 1.37 \\ 
        November & 0.07 & 0.07 & 0.00 & 0.16 & 0.93 \\ 
        December & 0.07 & 0.07 & 0.00 & 0.56 & 0.44 \\ 
     \multicolumn{2}{l}{\textit{Year of entry into unemployment}}\\
        2000 & 0.01 & 0.02 & 0.00 & 0.01 & 1.95 \\ 
        2001 & 0.01 & 0.01 & 0.00 & 0.15 & 0.90 \\ 
        2002 & 0.01 & 0.01 & 0.00 & 0.18 & 0.80 \\ 
        2003 & 0.01 & 0.01 & 0.00 & 0.56 & 0.33 \\ 
        2004 & 0.02 & 0.02 & 0.00 & 0.25 & 0.68 \\ 
        2005 & 0.03 & 0.03 & 0.00 & 0.47 & 0.43 \\ 
        2006 & 0.06 & 0.06 & 0.00 & 0.05 & 1.40 \\ 
        2007 & 0.15 & 0.14 & 0.01 & 0.00 & 2.53 \\ 
        2008 & 0.21 & 0.22 & 0.00 & 0.56 & 0.51 \\ 
        2009 & 0.27 & 0.27 & -0.01 & 0.22 & 1.21 \\ 
        2010 & 0.15 & 0.15 & 0.00 & 0.21 & 1.19 \\ 
        2011 & 0.07 & 0.07 & 0.00 & 0.15 & 0.91 \\ 
      \multicolumn{2}{l}{\textit{Federal state}}\\
        Schleswig-Holstein & 0.04 & 0.04 & 0.00 & 0.13 & 1.09 \\ 
        Freie und Hansestadt Hamburg & 0.04 & 0.04 & 0.00 & 0.37 & 0.98 \\ 
        Niedersachsen & 0.08 & 0.09 & 0.00 & 0.71 & 0.30 \\ 
        Freie Hansestadt Bremen  & 0.02 & 0.02 & 0.00 & 0.42 & 0.96 \\ 
        Nordrhein-Westfalen & 0.24 & 0.23 & 0.00 & 0.40 & 0.68 \\ 
        Hessen & 0.05 & 0.05 & 0.00 & 0.70 & 0.26 \\ 
        Rheinland-Pfalz & 0.04 & 0.03 & 0.00 & 0.24 & 0.87 \\ 
        Baden-Wuerttemberg & 0.10 & 0.10 & 0.00 & 0.71 & 0.27 \\ 
        Bayern & 0.16 & 0.16 & 0.00 & 0.16 & 1.12 \\ 
        Saarland & 0.01 & 0.01 & 0.00 & 0.16 & 0.92 \\ 
        Berlin & 0.02 & 0.02 & 0.00 & 0.43 & 0.56 \\ 
        Brandenburg & 0.03 & 0.03 & 0.00 & 0.31 & 0.66 \\ 
        Mecklenburg-Vorpommern & 0.05 & 0.05 & 0.00 & 0.78 & 0.45 \\ 
        Freistaat Sachsen & 0.07 & 0.06 & 0.00 & 0.08 & 1.16 \\ 
        Sachsen-Anhalt & 0.02 & 0.02 & 0.00 & 0.29 & 0.69 \\ 
        Freistaat Thüringen & 0.04 & 0.04 & 0.00 & 0.96 & 0.04 \\ 
       \multicolumn{2}{l}{\textit{Demographic characteristics}}\\
        Younger than 25 years & 0.14 & 0.15 & -0.01 & 0.00 & 3.58 \\ 
        25-29 years & 0.17 & 0.17 & 0.00 & 0.77 & 0.22 \\ 
        30-34 years & 0.15 & 0.15 & 0.00 & 0.29 & 0.88 \\ 
        35-39 years  & 0.14 & 0.14 & 0.00 & 0.71 & 0.40 \\ 
        40-44 years  & 0.15 & 0.14 & 0.00 & 0.25 & 1.13 \\ 
        45-49 years & 0.13 & 0.12 & 0.00 & 0.07 & 1.40 \\ 
        50-54 years & 0.09 & 0.09 & 0.00 & 0.77 & 0.24 \\ 
        Older than 54 years  & 0.03 & 0.04 & 0.00 & 0.67 & 0.29 \\ 
        Female & 0.44 & 0.44 & 0.00 & 0.74 & 0.29 \\ 
        German & 0.90 & 0.90 & 0.00 & 0.79 & 0.21 \\ 
        Single & 0.46 & 0.46 & 0.00 & 0.55 & 0.52 \\ 
        Married & 0.37 & 0.37 & 0.00 & 0.42 & 0.65 \\ 
        Marital status: missing & 0.16 & 0.16 & 0.00 & 0.89 & 0.14 \\ 
        Children < 15 & 0.23 & 0.23 & 0.00 & 0.87 & 0.15 \\ 
        Children < 3 & 0.15 & 0.16 & 0.00 & 0.25 & 1.21 \\ 
        Single parent & 0.07 & 0.06 & 0.01 & 0.03 & 3.14 \\ 
        Restrictions or disability  & 0.10 & 0.10 & 0.00 & 0.26 & 1.02 \\ 
        No schooling degree & 0.06 & 0.06 & 0.00 & 0.05 & 1.36 \\ 
        Schooling degree without Abitur & 0.73 & 0.74 & -0.01 & 0.08 & 1.37 \\ 
        Schooling degree with Abitur & 0.17 & 0.16 & 0.00 & 0.06 & 1.34 \\ 
        Schooling: missing & 0.04 & 0.04 & 0.00 & 0.22 & 1.14 \\ 
        Without vocational training & 0.27 & 0.27 & 0.00 & 0.46 & 0.75 \\ 
        Vocational training & 0.61 & 0.61 & 0.00 & 0.69 & 0.37 \\ 
        Academic degree & 0.07 & 0.07 & 0.00 & 0.88 & 0.12 \\ 
        Vocational Training: missing & 0.05 & 0.04 & 0.00 & 0.18 & 0.89 \\ 
        Job searched: part time  & 0.05 & 0.05 & 0.00 & 0.06 & 1.25 \\ 
        Job searched: full time  & 0.74 & 0.73 & 0.01 & 0.16 & 1.40 \\ 
        Job searched: missing  & 0.21 & 0.22 & -0.01 & 0.04 & 2.19 \\ 
        Job returner & 0.03 & 0.02 & 0.00 & 0.00 & 2.21 \\ 
       \multicolumn{2}{l}{\textit{Characteristics of the last job}}\\
        No last job & 0.01 & 0.02 & -0.01 & 0.00 & 4.83 \\ 
        Last job: unskilled/semiskilled tasks & 0.09 & 0.09 & 0.00 & 0.25 & 0.91 \\ 
        Last job: skilled tasks & 0.66 & 0.66 & 0.00 & 0.68 & 0.34 \\ 
        Last job: complex tasks & 0.04 & 0.04 & 0.00 & 0.61 & 0.32 \\ 
        Last job: highly complex tasks & 0.07 & 0.07 & 0.00 & 0.66 & 0.34 \\ 
        Last job: skills missing & 0.14 & 0.14 & 0.00 & 0.14 & 1.30 \\ 
        Last job: part time & 0.67 & 0.69 & -0.01 & 0.00 & 3.00 \\ 
        Last job: full time & 0.31 & 0.29 & 0.02 & 0.00 & 4.49 \\ 
        Last job: working time missing & 0.01 & 0.02 & -0.01 & 0.00 & 4.97 \\ 
        \multicolumn{2}{l}{\textit{LM history relative to unemployment start}}\\
        Regular EMP 3 months before & 0.61 & 0.63 & -0.02 & 0.00 & 3.65 \\ 
        Marginal EMP 3 months before & 0.14 & 0.12 & 0.02 & 0.00 & 5.62 \\ 
        Part-time 3 months before & 0.10 & 0.11 & 0.00 & 0.31 & 0.67 \\ 
        Program 3 months before & 0.09 & 0.09 & -0.01 & 0.01 & 3.04 \\ 
        OLF 3 months before& 0.02 & 0.02 & 0.00 & 0.16 & 0.84 \\ 
        Regular EMP 6 months before & 0.61 & 0.63 & -0.02 & 0.00 & 3.76 \\ 
        Marginal EMP 6 months before & 0.13 & 0.11 & 0.02 & 0.00 & 5.89 \\ 
        Part-time 6 months before & 0.10 & 0.11 & 0.00 & 0.18 & 0.90 \\ 
        Program 6 months before & 0.09 & 0.10 & -0.01 & 0.01 & 3.73 \\ 
        OLF 6 months before& 0.03 & 0.04 & -0.01 & 0.06 & 2.85 \\ 
        Regular EMP 9 months before & 0.60 & 0.62 & -0.02 & 0.00 & 3.30 \\ 
        Marginal EMP 9 months before & 0.13 & 0.11 & 0.02 & 0.00 & 6.29 \\ 
        Part-time 9 months before & 0.10 & 0.10 & 0.00 & 0.11 & 1.06 \\ 
        Program 9 months before & 0.09 & 0.09 & 0.00 & 0.35 & 0.94 \\ 
        OLF 9 months before& 0.04 & 0.05 & -0.01 & 0.07 & 2.52 \\ 
        Regular EMP 12 months before & 0.61 & 0.62 & -0.02 & 0.00 & 3.27 \\ 
        Marginal EMP 12 months before & 0.12 & 0.11 & 0.02 & 0.00 & 4.83 \\ 
        Part-time 12 months before & 0.10 & 0.10 & 0.00 & 0.21 & 0.83 \\ 
        Program 12 months before & 0.09 & 0.09 & 0.00 & 0.88 & 0.13 \\ 
        OLF 12 months before & 0.04 & 0.05 & -0.01 & 0.05 & 2.70 \\ 
        Regular EMP 18 months before & 0.57 & 0.58 & -0.01 & 0.03 & 1.94 \\ 
        Marginal EMP 18 months before & 0.12 & 0.11 & 0.01 & 0.01 & 2.51 \\ 
        Part-time 18 months before & 0.09 & 0.09 & 0.00 & 0.37 & 0.60 \\ 
        Program 18 months before & 0.09 & 0.09 & -0.01 & 0.13 & 2.08 \\ 
        OLF 18 months before & 0.05 & 0.06 & -0.01 & 0.03 & 2.76 \\ 
        Regular EMP 24 months before & 0.53 & 0.54 & -0.01 & 0.15 & 1.29 \\ 
        Marginal EMP 24 months before4 & 0.12 & 0.11 & 0.01 & 0.00 & 2.54 \\ 
        Part-time 24 months before & 0.09 & 0.09 & 0.00 & 0.78 & 0.19 \\ 
        Program 24 months before & 0.09 & 0.09 & 0.00 & 0.69 & 0.38 \\ 
        OLF 24 months before & 0.06 & 0.06 & 0.00 & 0.02 & 1.86 \\ 
        Regular EMP 36 months before & 0.48 & 0.49 & -0.01 & 0.18 & 1.15 \\ 
        Marginal EMP 36 months before & 0.12 & 0.12 & 0.00 & 0.92 & 0.13 \\ 
        Part-time 36 months before & 0.08 & 0.08 & 0.00 & 0.67 & 0.31 \\ 
        Program 36 before & 0.09 & 0.09 & 0.00 & 0.71 & 0.33 \\ 
        OLF 36 before & 0.07 & 0.07 & 0.00 & 0.09 & 1.36 \\ 
        Total days in EMP 6 months before & 132.98 & 133.17 & -0.19 & 0.78 & 0.29 \\ 
        Total days in ALO 6 months & 38.50 & 36.07 & 2.43 & 0.00 & 4.04 \\ 
        Total days in program 6 months before & 15.26 & 17.18 & -1.93 & 0.00 & 4.39 \\ 
        Total days in EMP 12 months before & 263.39 & 263.99 & -0.60 & 0.62 & 0.50 \\ 
        Total days in ALO 12 months before & 76.72 & 70.67 & 6.05 & 0.00 & 5.66 \\ 
        Total days in program 12 months before & 30.60 & 33.23 & -2.63 & 0.00 & 3.34 \\ 
        Total days in EMP 24 months before & 512.65 & 513.89 & -1.24 & 0.61 & 0.55 \\ 
        Total days in ALO 24 months before& 154.84 & 142.11 & 12.72 & 0.00 & 6.91 \\ 
        Total days in program 24 months before & 61.81 & 65.54 & -3.73 & 0.02 & 2.75 \\ 
        Total days in EMP 60 months before & 1177.16 & 1178.72 & -1.57 & 0.79 & 0.30 \\ 
        Total days in ALO 60 & 415.91 & 386.40 & 29.51 & 0.00 & 7.15 \\ 
        Total days in program 60 months & 156.67 & 162.16 & -5.49 & 0.06 & 2.09 \\ 
        Total days in EMP 120 months before & 2129.20 & 2128.06 & 1.14 & 0.91 & 0.11 \\ 
        Total days in ALO 120 & 739.56 & 683.61 & 55.95 & 0.00 & 7.73 \\ 
        Total days in program 120 months before & 273.21 & 281.18 & -7.97 & 0.13 & 1.98 \\ 
        Number of ALMP 24 months before & 0.61 & 0.63 & -0.02 & 0.23 & 1.67 \\ 
        Number of ALMP 60 months before & 1.26 & 1.25 & 0.01 & 0.67 & 0.48 \\ 
        Log. total earnings 12 months before & 8.51 & 8.47 & 0.04 & 0.21 & 1.55 \\ 
        Log. total benefits 12 months before & 1.87 & 1.82 & 0.05 & 0.09 & 1.39 \\ 
        Log. total earnings 24 months before & 9.37 & 9.30 & 0.06 & 0.07 & 2.61 \\ 
        Log. total benefits 24 months before & 2.69 & 2.63 & 0.06 & 0.06 & 1.52 \\ 
        Log. total earnings 48 months before & 10.12 & 10.04 & 0.08 & 0.03 & 3.59 \\ 
        Log. total benefits 48 months before & 3.95 & 3.87 & 0.08 & 0.02 & 1.91 \\ 
        Log. total earnings 60 months before & 10.35 & 10.26 & 0.09 & 0.02 & 4.11 \\ 
        Log. total benefits 60 months before & 4.47 & 4.43 & 0.04 & 0.24 & 1.02 \\ 
        Log. total earnings 120 months before & 11.03 & 10.91 & 0.12 & 0.00 & 5.86 \\ 
        Log. total benefits 120 months before & 5.97 & 5.80 & 0.17 & 0.00 & 4.20 \\ 
        \multicolumn{2}{l}{\textit{Regional characteristics (at district level)}}\\ 
        Local unemployment rate & 8.62 & 8.54 & 0.08 & 0.00 & 2.38 \\ 
        Population density & 792.50 & 785.65 & 6.85 & 0.33 & 0.77 \\ 
        Employment share primary sector & 1.84 & 1.86 & -0.01 & 0.29 & 0.77 \\ 
        Employment share secondary sector  & 26.25 & 26.37 & -0.12 & 0.06 & 1.39 \\ 
        Employment share third sector  & 71.91 & 71.77 & 0.14 & 0.06 & 1.42 \\ 
        GDP per capita  & 30.68 & 30.69 & 0.00 & 0.97 & 0.03 \\ 
        No regional information & 0.09 & 0.10 & -0.01 & 0.07 & 2.74 \\ 
        \hline\hline
        Observations & 62909  &  52486\\ 
        \hline
    \label{tab:balancedness}
\end{longtable}
 \begin{minipage}{0.9\textwidth} \linespread{1.0}\selectfont
          \vspace{-30pt}
          \footnotesize \textit{Notes:} This table displays the average characteristics of the sample of program participants (PP) and the average characteristics of the sample of matched non-participants (NP) in columns (1) and (2), respectively. Column (3) displays the difference in means and column (4) the standardized bias (SB) calculated for each characteristic $X_k$ as $SB_k = \frac{|E_{cs}(X_{k})-E_{p}(X_{k})|}{\sqrt{(V_{cs}(X_{k})+ V_{p}(X_{k}))/2}}\times100$.
     \end{minipage}
}

\subsubsection{Estimation of the Employability Model} \label{appendix:modelselection}

For the estimation of the employability model I compare the predictive performance of three models. A logit model without regularization, a lasso logit model where the regularization parameter (lambda) is chosen by cross-validation (10-folds) and a lasso logit based on a theory driven regularization parameter. To test the performance of the model, I split the sample of matched non-participants and keep one fourth as a holdout sample. The rest of the observations are used to train the model. 

All models yield a similar predictive performance of around 67-68 percent (ROC area) in the training and holdout sample. The lasso based on cross-validation slightly outperforms the other models. See Table \ref{tab:ROC}. Compared to the so called rigorous lasso that is based on the theory driven penalization, it selects a larger lambda (0.0001556) and 206 out of 220 covariates. The rigorous lasso, which has the lowest performance selects 62.

The set of variables included is the same as the one for the matching exercise with the addition of interactions between individual characteristics and gender. 
For my main analysis, I choose to measure employability at program start. This implies that I measure time sensitive variables for program participants at program start, wherever possible. Note, that I do not observe all time sensitive variable at program start but some only at unemployment entry. This concerns for example education, vocational training (with the exception of training in the context of ALMP) and the health and disability status.  For the sample of non-participants, who I use to train the model with, time sensitive variables are measured at entry into unemployment.  Measuring their employability at some hypothetical moment of program start is not possible, since for a part of the sample it would condition on the outcome. This is the case for individuals who start a program more than 12 months after they enter unemployment.

An employability score measured at program start thus suffers from the following potential sources of imprecision: First, I do not have true variation in all predictors with regard to timing. Second, the group who I train the model with is only observed once, at unemployment entry. If actual program participants are unemployed for a long time when entering a program, the model might perform worse in predicting their true employability at program start. This is because the matched training sample of non-participants might not be representative for jobseekers with long unemployment durations. Furthermore, if there is dynamic selection into programs, e.g. ex-ante non-employable individuals enter programs later than highly employable individuals or vice versa, there might be a systematic measurement error in the employability score. To test the sensitivity of the results to the timing of measurement, I perform a robustness check where I measure employability at unemployment entry.

Given the performance of the three models in the holdout sample (ROC area), I choose the cross-validated lasso model for my main specification. As Appendix Figure   \ref{fig:emp_lasso_logit}
 shows, the distribution of the employability measure (average peer employability) estimated by logit and cross-validated lasso are virtually the same. Not surprisingly, they also yield the same results (can be obtained by demand). Appendix Table \ref{tab:predictors} gives an overview of the estimated coefficients  in the preferred prediction model.

\begin{table}[ht]
\centering
\caption{Summary of Model Performance} \label{tab:ROC}
\resizebox{0.9\linewidth}{!}{%
\begin{tabular}{llccccc} 
\toprule
\textbf{Estimator} & \textbf{Sample}  & \textbf{ROC Area} & \textbf{Std. Err.} & \textbf{95\% CI Lower} & \textbf{95\% CI Upper} \\
\midrule
Logit               & Training   & 0.6860 & 0.0028 & 0.68054 & 0.69153 \\
Logit               & Hold-out   & 0.6721 & 0.0049 & 0.66242 & 0.68173 \\
Rigorous Lasso      & Training   & 0.6695 & 0.0029 & 0.66389 & 0.67510 \\
Rigorous Lasso      & Hold-out   & 0.6691 & 0.0050 & 0.65942 & 0.67884 \\
Lasso Logit (CV)    & Training   & 0.6858 & 0.0028 & 0.68027 & 0.69127 \\
Lasso Logit (CV)    & Hold-out   & 0.6735 & 0.0049 & 0.66382 & 0.68310 \\
\bottomrule
\end{tabular}}
\end{table}

\newpage
\renewcommand{\arraystretch}{1}
{\footnotesize
\begin{longtable}{lc} 
\caption{Predictors of Employability in Training Sample of Non-Participants (crossvalidated lasso logit)}\\
\hline \hline
 &\multicolumn{1}{c}{Coefficient}\\
\hline
\multicolumn{2}{l}{\textit{Demographics}}\\
Younger than 25 years&    -0.07         \\
30-34 years     &     0.01         \\
40-44 years     &    -0.04         \\
45-49 years     &    -0.00         \\
50-54 years     &    -0.11         \\
Older than 54 years  &    -0.57         \\
Female          &     0.72         \\
Non-German      &    -0.03         \\
Married         &     0.06         \\
Maritial status missing         &     0.09         \\
Children        &     0.10         \\
Children < 3    &    -0.05         \\
Restrictions or disability&    -0.35         \\
No schooling degree&    -0.08         \\
Schooling degree with Abitur&    -0.08         \\
Education missing         &    -0.31         \\
Without Vocational Training&    -0.20         \\
Vocational degree missing         &    -0.05         \\
Job searched:Part-time       &    -0.15         \\
Job searched: missing         &    -0.11         \\
Job returner    &    -0.13         \\
Welfare benefits&    -0.19         \\
\multicolumn{2}{l}{\textit{Month and year of entry into unemployment}}\\
Feb             &    -0.06         \\
Mar             &    -0.20         \\
Apr             &    -0.25         \\
May             &    -0.12         \\
Jun             &    -0.21         \\
Jul             &    -0.23         \\
Aug             &    -0.26         \\
Sep             &    -0.25         \\
Oct             &    -0.19         \\
Nov             &    -0.02         \\
Dec             &     0.07         \\
2006&    -0.11         \\
2007&     0.06         \\
2008&    -0.15         \\
2010&     0.27         \\
2011&     0.15         \\
\multicolumn{2}{l}{\textit{Residence - Federal state}}  \\              
Schleswig-Holstein&    -0.07         \\
Freie und Hansestadt Hamburg&     0.16         \\
Niedersachsen   &     0.09         \\
Freie Hansestadt Bremen&     0.29         \\
Hessen          &     0.02         \\
Rheinland-Pfalz &     0.08         \\
Freistaat Bayern&     0.14         \\
Saarland        &     0.02         \\
Berlin          &     0.10         \\
Brandenburg     &     0.17         \\
Mecklenburg-Vorpommern&     0.16         \\
Freistaat Sachsen&     0.18         \\
Sachsen-Anhalt  &    -0.05         \\
Freistaat Thüringen&     0.08         \\
\multicolumn{2}{l}{\textit{Characteristics of the last job}}    \\            
No last job         &    -3.03         \\
Unskilled/semiskilled tasks&    -0.00         \\
Complex tasks   &    -0.16         \\
Highly complex tasks&    -0.02         \\
Part-time       &    -0.01         \\
Time missing         &     0.07         \\
Log daily earnings in last job&     0.07         \\
Agriculture, forestry, farming, and gardening&     0.45         \\
Construction, architecture, surveying and technical building services&     0.17         \\
Natural sciences, geography and informatics&    -0.22         \\
Traffic, logistics, safety and security&     0.08         \\
Commercial services, trading, sales, the hotel business and tourism&     0.11         \\
Business organisation, accounting, law and administration&    -0.03         \\
Health care, the social sector, teaching and education&     0.04         \\
Philology, literature, humanities, social sciences, media, art&    -0.36         \\
Occupation: missing         &    -0.07         \\
\multicolumn{2}{l}{\textit{Labor market history}} \\              
Regular EMP 3 months before &    -0.02         \\
Marginal EMP 3 months before &     0.28         \\
Part-time EMP 3 months before &    -0.04         \\
In ALMP 3 months before &     0.05         \\
OLF 3 months before &    -0.23         \\
Regular EMP 6 months before &     0.04         \\
Marginal EMP 6 months before &     0.04         \\
In ALMP 6 months before &     0.07         \\
OLF 6 months before &    -0.12         \\
Regular EMP 9 months before &     0.01         \\
Marginal EMP 9 months before &     0.15         \\
In ALMP 9 months before &    -0.07         \\
OLF 9 months before &    -0.14         \\
Regular EMP 12 months before &    -0.02         \\
Marginal EMP 12 months before &     0.32         \\
Part-time EMP 12 months before &    -0.10         \\
In ALMP 12 months before &     0.05         \\
OLF 12 months before &    -0.12         \\
Regular EMP 18 months before &     0.18         \\
Marginal EMP 18 months before &     0.13         \\
Part-time EMP 18 months before &    -0.09         \\
In ALMP 18 months before &     0.21         \\
OLF 18 months before &     0.01         \\
Regular EMP 24 months before &    -0.17         \\
Marginal EMP 24 months before &     0.05         \\
Part-time EMP 24 months before &     0.18         \\
In ALMP 24 months before &    -0.03         \\
OLF 24 months before &     0.04         \\
Regular EMP 36 months before &    -0.05         \\
Marginal EMP 36 months before &     0.11         \\
Part-time EMP 36 months before &    -0.11         \\
In ALMP 36 months before &     0.11         \\
OLF 36 months before &    -0.04         \\
Cum. EMPL 6 months before  (days)&     0.00         \\
Cum. UE 6 months before  (days)&    -0.00         \\
Cum. EMPL 12 months before  (days)&    -0.00         \\
Cum. UE 12 months before  (days)&     0.00         \\
Cum. ALMP 12 months before  (days)&    -0.00         \\
Cum. EMPL 24 months before  (days)&     0.00         \\
Cum. UE 24 months before  (days)&     0.00         \\
Cum. EMPL 60 months before  (days)&     0.00         \\
Cum. UE 60 months before  (days)&    -0.00         \\
Cum. ALMP 60 months before  (days)&    -0.00         \\
Cum. EMPL 120 months before  (days)&     0.00         \\
Cum. UE 120 months before  (days)&    -0.00         \\
No of ALMP 24 months before &     0.02         \\
No of ALMP 60 months before &    -0.00         \\
Log cum. earnings 12 months before &     0.04         \\
Log cum. benefits 12 months before &    -0.02         \\
Log cum. earnings 24 months before &    -0.01         \\
Log cum. benefits 24 months before &     0.01         \\
Log cum. earnings 48 months before &     0.04         \\
Log cum. benefits 48 months before &    -0.01         \\
Log cum. earnings 60 months before &    -0.00         \\
Log cum. benefits 60 months before &     0.02         \\
Log cum. benefits 120 months before &     0.00         \\
\hline
Interactions with gender    &    Yes            \\
Regional labor market characteristics    &    Yes         \\
Observations    &    39364     \\
\hline\hline
\label{tab:predictors}
\end{longtable}
 \begin{minipage}{0.95\textwidth} \linespread{1.0}\selectfont
          \vspace{-30pt}
          \footnotesize \textit{Notes:} This table displays the coefficients of a lasso logit regression (lambda selected by cross-validation) of the employability measure on individual characteristics measured at the beginning of the unemployment spell in the training sample of non-participants. Coefficients based on interaction terms and for regional labor market characteristics are omitted in the table.
     \end{minipage}
\pagebreak
\FloatBarrier
\renewcommand{\arraystretch}{1.3}

\begin{figure}[H] 
  \centering
    \caption{Distribution of the Average Peer Employability estimated by Lasso and Logit} 
  \label{fig:emp_lasso_logit}
 \includegraphics[width=0.7\textwidth]{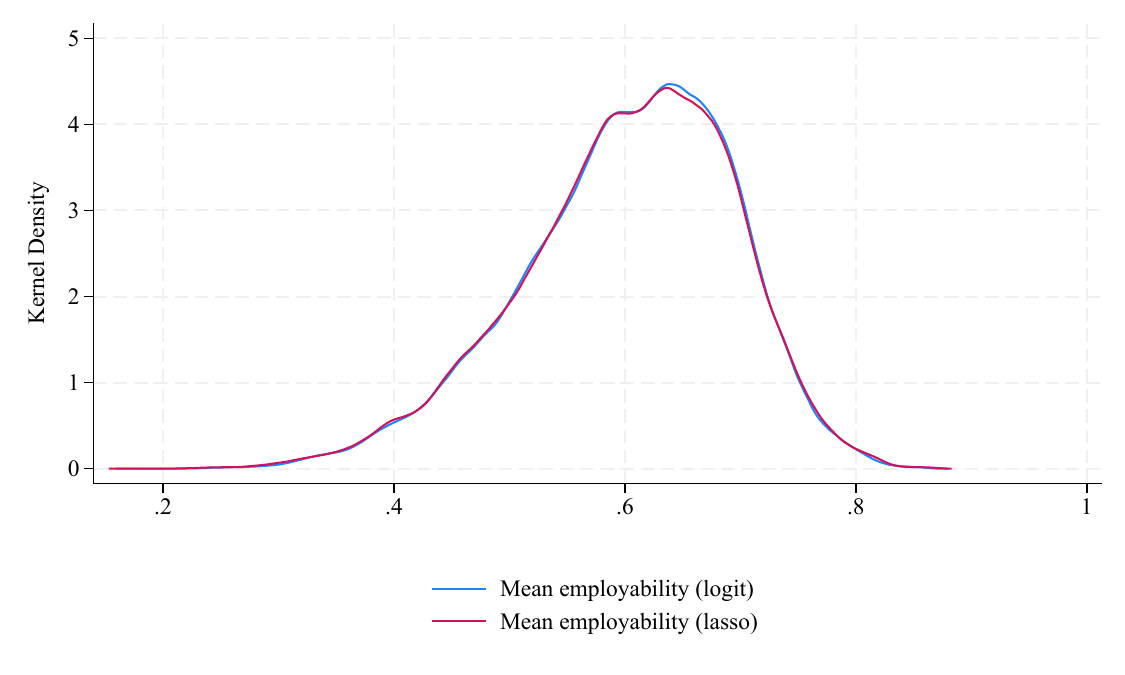}
 \begin{minipage}{\textwidth} \linespread{1.0}\selectfont
          \vspace{6pt}
          \footnotesize \textit{Notes:} This figure plots the distribution of the average peer employbility estimated by a lasso (main specification) and logit model.
     \end{minipage}
\end{figure}

\begin{figure}[H] 
  \centering
    \caption{Distribution of the Individual Predicted Employability by Program Type} 
  \label{fig:pred_ind_emp_by_type}
 \includegraphics[width=0.7\textwidth]{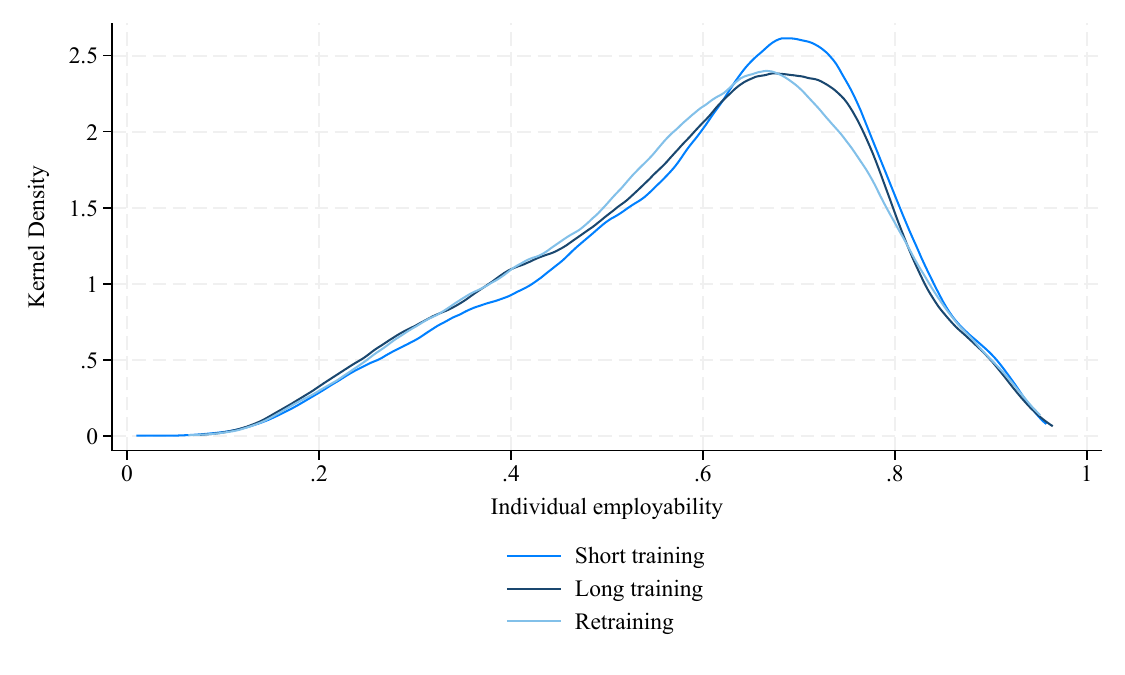}
 \begin{minipage}{\textwidth} \linespread{1.0}\selectfont
          \vspace{6pt}
          \footnotesize \textit{Notes:} This figure plots the distribution of the predicted individual employbility for the sample of program participants in short training, long training and retraining. It is based on a lasso model.
     \end{minipage}
\end{figure}

\begin{figure}[H] 
  \centering
    \caption{Distribution of the Average  Predicted Employability by Program Type} 
  \label{fig:pred_avg_emp_by_type}
 \includegraphics[width=0.7\textwidth]{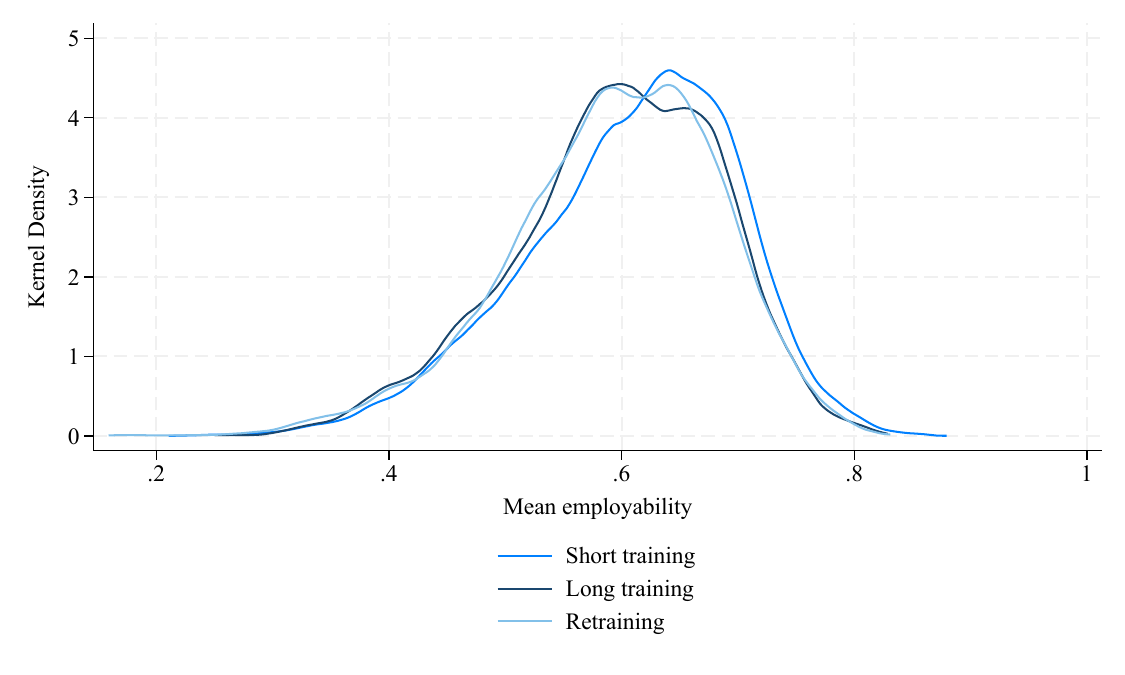}
 \begin{minipage}{\textwidth}\linespread{1.0}\selectfont
          \vspace{6pt}
          \footnotesize \textit{Notes:} This figure plots the distribution of the predicted average employbility for the sample of program participants in short training, long training and retraining. It is based on a lasso model.
     \end{minipage}
\end{figure}

\begin{figure}[htbp]
      \caption{Employability Estimated at Program Start and Unemployment Entry}\label{app_fig:emp_pr_ue}
\begin{center}
  \includegraphics[width=0.55\linewidth]{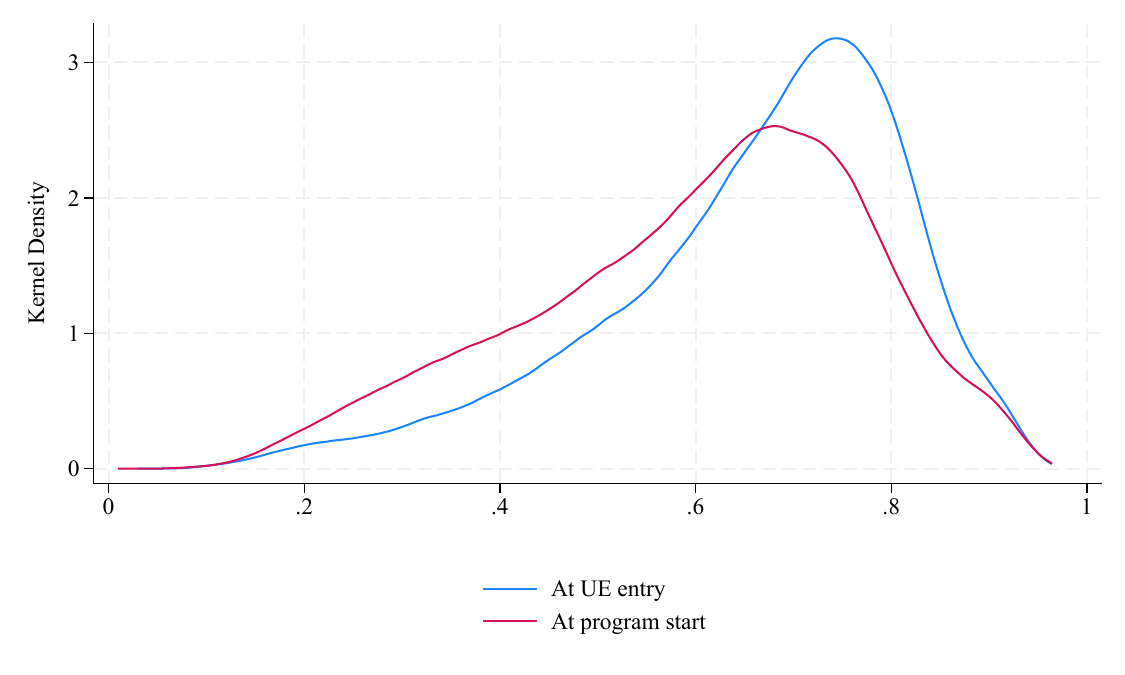}
 \caption*{(a) Individual Employability }
  \includegraphics[width=0.55\linewidth]{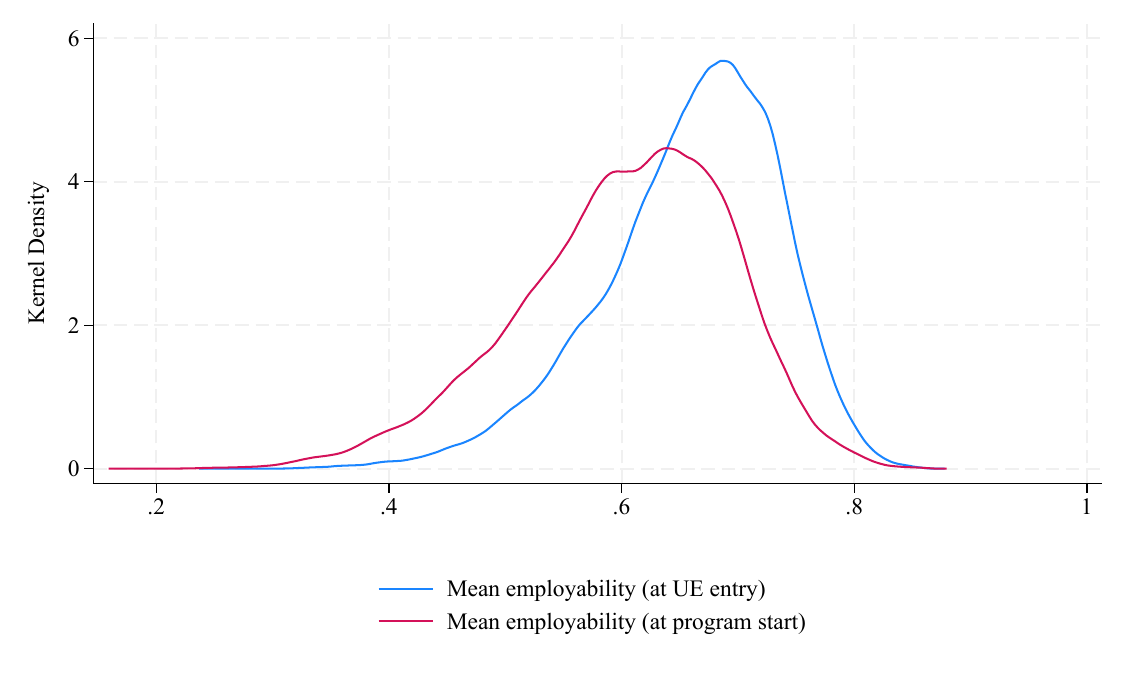} 
  \caption*{(b) Average Peer Employability}
\end{center}
   \begin{minipage}{\textwidth}
          \vspace{2pt}
          \footnotesize \textit{Notes:} The figure plots the kernel distribution of the individual employability and average peer employability estimated at program start (red) and at unemployment (UE) entry (blue). 
     \end{minipage}
\end{figure}

\pagebreak
\subsection{Descriptive Statistics} \label{appendix:descriptives}

\begin{figure}[H]
  \centering
    \caption{Distribution of Individuals over Different Course Sizes} 
  \label{fig:tn_anz}
 \includegraphics[width=0.7\textwidth]{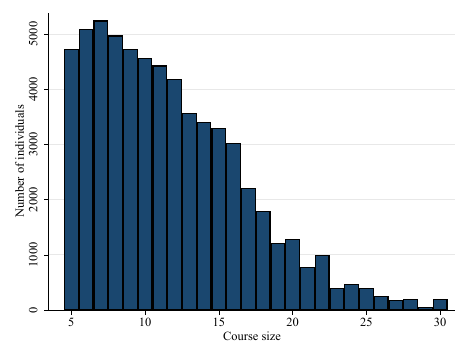}
  \begin{minipage}{\textwidth} \linespread{1.0}\selectfont
          \vspace{6pt}
          \footnotesize \textit{Notes:} This figure plots the distribution program participants in the estimation sample over different course sizes. It displays the total number of participants on the y-axis.
     \end{minipage}
\end{figure}

\begin{figure}[htb]
  \centering
      \caption{ Average Employment Rate for Different Program Types}
 \includegraphics[width=0.8\linewidth, clip=true, trim=0cm 0cm 0cm 0cm]{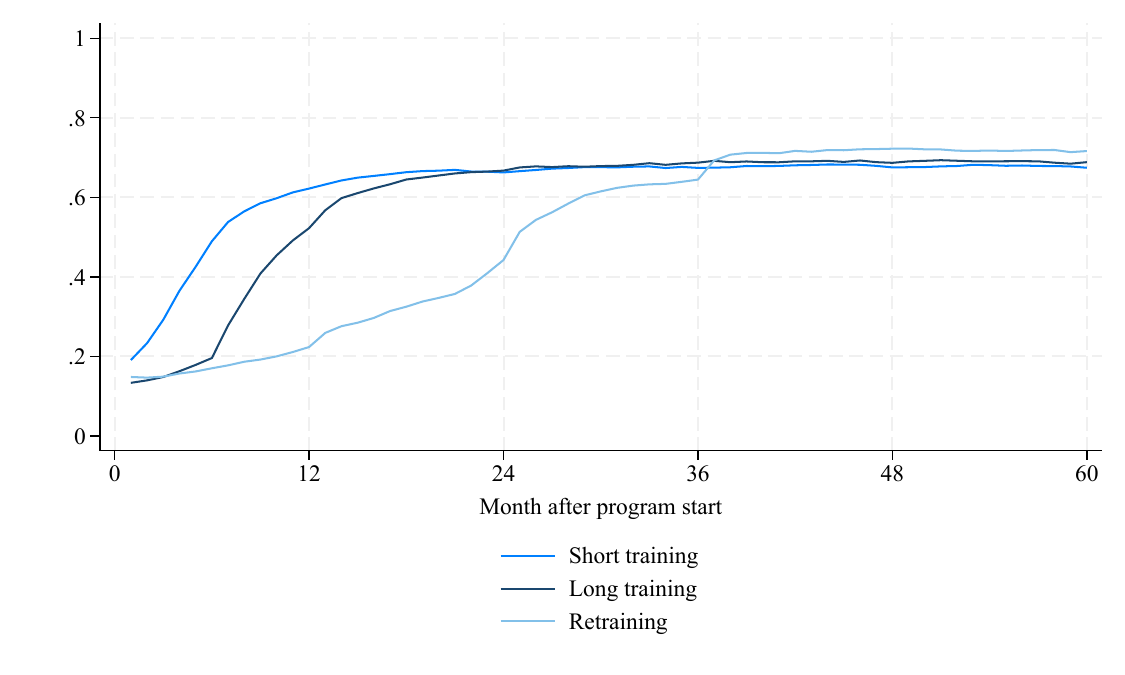} 
   \label{fig:avg_emp_alo}
    \begin{minipage}{\textwidth}   \linespread{1.0}\selectfont
 \vspace{6pt}
     \footnotesize \textit{Notes:} The figure depicts the average employment probability by month after program start separately for the three programs of interest.
 \end{minipage}
\end{figure}

\small
\begin{table}
 \centering
  \caption{ Occupational Areas across Programs }
  \label{app_tab:occ}
\resizebox{\linewidth}{!}{%
\begin{tabular}{l*{3}{rr}}
\hline\hline
 &\multicolumn{2}{c}{Short training}&\multicolumn{2}{c}{Long training}&\multicolumn{2}{c}{Retraining}\\
                &     mean&       sd&     mean&       sd&     mean&       sd\\
\hline
Production of raw materials and goods, and manufacturing&     0.19&     0.39&     0.21&     0.41&     0.24&     0.42\\
Construction, architecture, surveying and building services&     0.02&     0.14&     0.03&     0.17&     0.01&     0.07\\
Natural sciences, geography and informatics&     0.05&     0.21&     0.06&     0.24&     0.03&     0.17\\
Traffic, logistics, safety and security&     0.24&     0.43&     0.22&     0.42&     0.16&     0.37\\
Commercial services, trading, sales and tourism&     0.06&     0.24&     0.03&     0.18&     0.04&     0.20\\
Business organization, accounting, law and administration&     0.19&     0.40&     0.16&     0.37&     0.18&     0.38\\
Health care, social sector and education &     0.21&     0.41&     0.24&     0.43&     0.34&     0.47\\
Philology, literature, humanities, \& others&     0.03&     0.16&     0.03&     0.18&     0.01&     0.09\\[5pt]
\hline
Observations    &    36662&         &    12683&         &    12413&         \\
\hline\hline
\end{tabular}}
     \begin{minipage}{\textwidth} \linespread{1.0}\selectfont
          \vspace{6pt}
          \footnotesize \textit{Notes:} Summary statistics (mean and standard deviation) calculated at the individual level.
     \end{minipage}
\end{table}

\begin{figure}[htbp]
      \caption{Raw and Net Variation in the Average Peer Employability}\label{app_fig:dist_raw_net}
\begin{center}
  \includegraphics[width=0.55\linewidth]{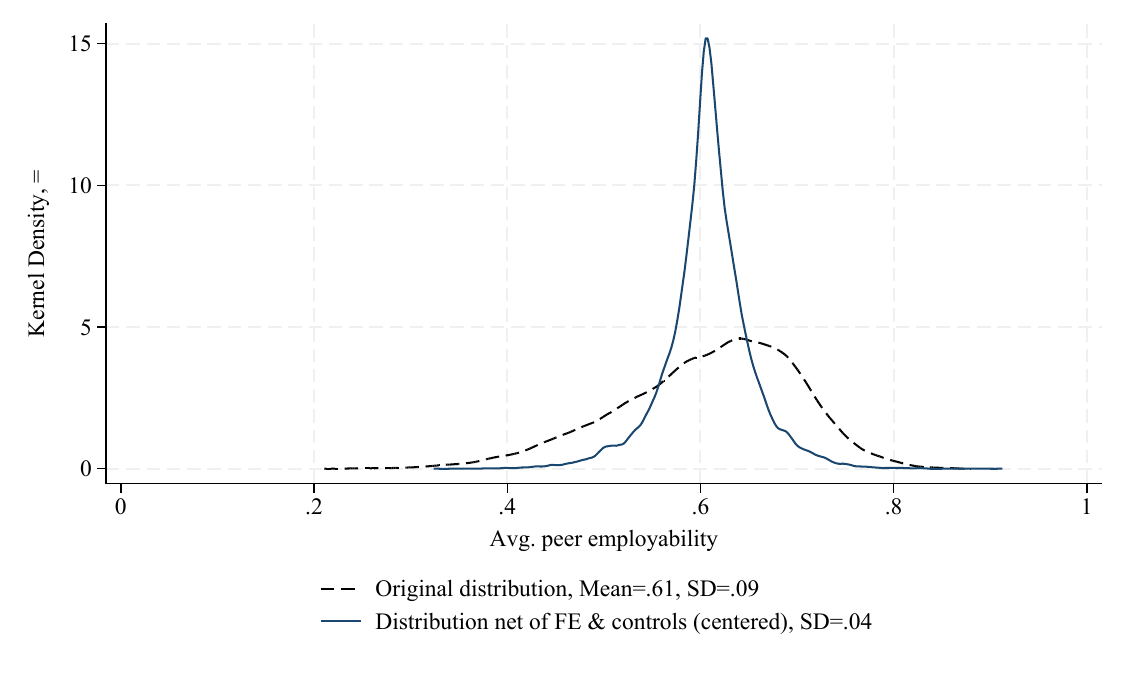}
 \caption*{(a) Short Training }
  \includegraphics[width=0.55\linewidth]{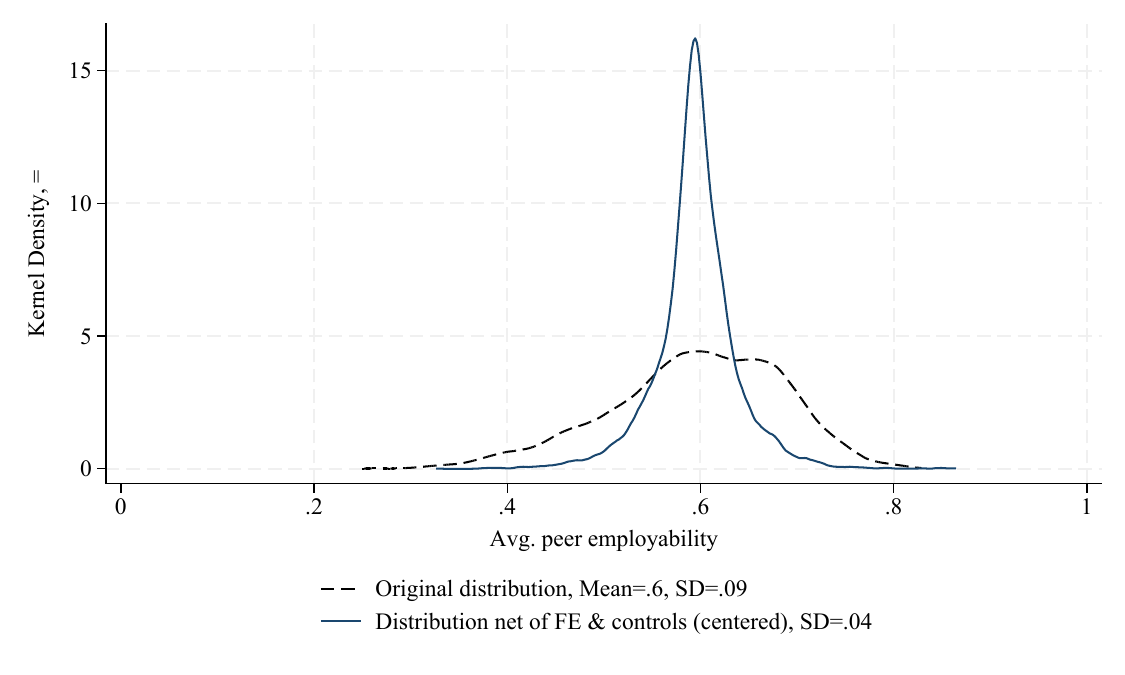} 
  \caption*{(b) Long Training}
   \includegraphics[width=0.55\linewidth]{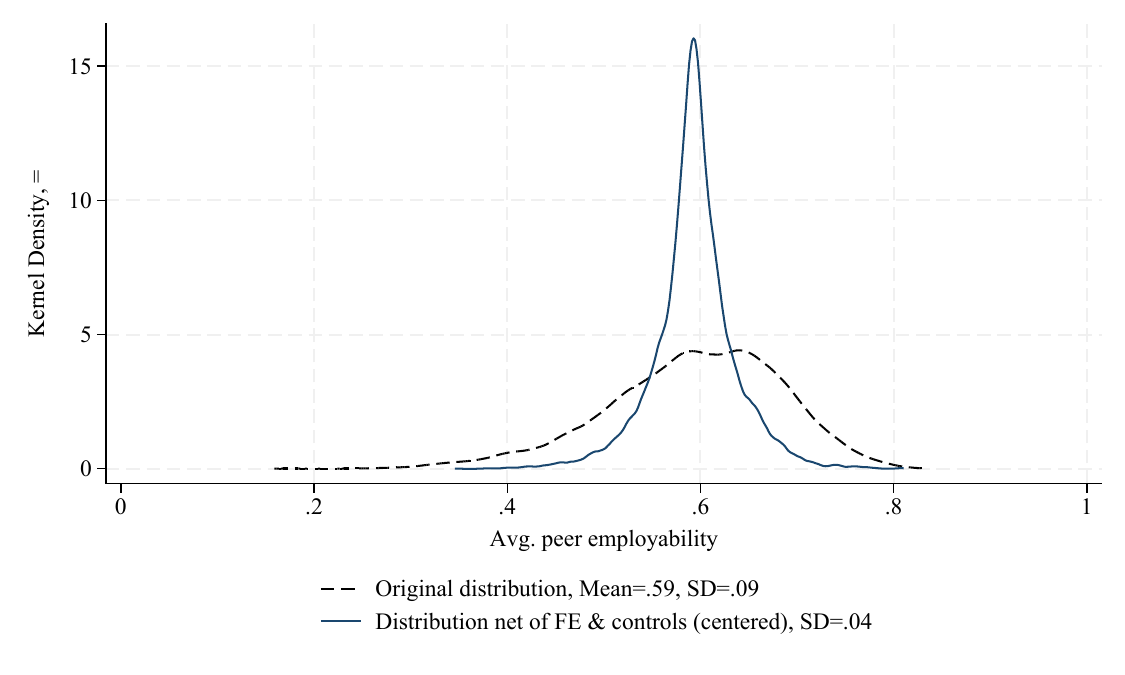} 
   \caption*{(c) Retraining}
   \end{center}
   \begin{minipage}{\textwidth}
          \vspace{2pt}
          \footnotesize \textit{Notes:} The figure plots the raw distribution of the average peer employability (solid line) and the distribution of the average peer employability net of provider x competence level, month x year x occupation fixed effects and course controls (dashed line) separately by program type. The residual distribution is centered at the mean. SD refers to the standard deviation.
     \end{minipage}
\end{figure}

\pagebreak
\FloatBarrier
\clearpage

\subsection{Identifying Assumptions and Validity Checks}\label{appendix:validity_checks}

\subsubsection{Resampling Test}\label{appendix:validity_checks_res}

To provide evidence that the residual variation in my imputed employability variables results from random fluctuation, I ran a series of simulations  (similar to \citealp{Bifulco2011}). In each simulation I randomly reallocate program participants to groups of the same size within the same provider-competence level group and use this simulated distribution to compute the variation of the predicted peer employability measures. Across 500 simulations and after removing fixed effects and course controls the average residual standard deviation in the simulated peer employability measures ranges from 0.048 to 0.05. The simulated residual variation is slightly smaller than the observed one for short training programs. It is somewhat larger for long training programs and retraining. Thus, there might be some excess variation in these types of programs. 

\FloatBarrier

\begin{figure}[htbp]
      \caption{Observed and Simulated Residual Variation in the Average Peer Employability}\label{app_fig:simulation}
\begin{center}
  \includegraphics[width=0.55\linewidth]{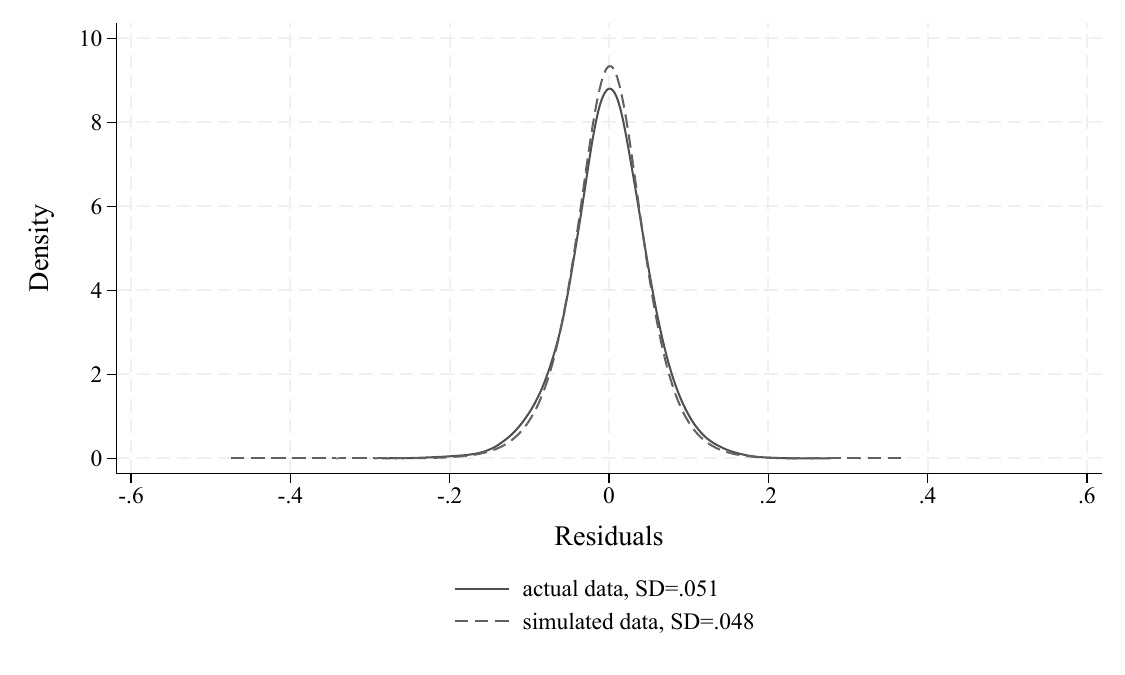}
 \caption*{(a) Short Training }
  \includegraphics[width=0.55\linewidth]{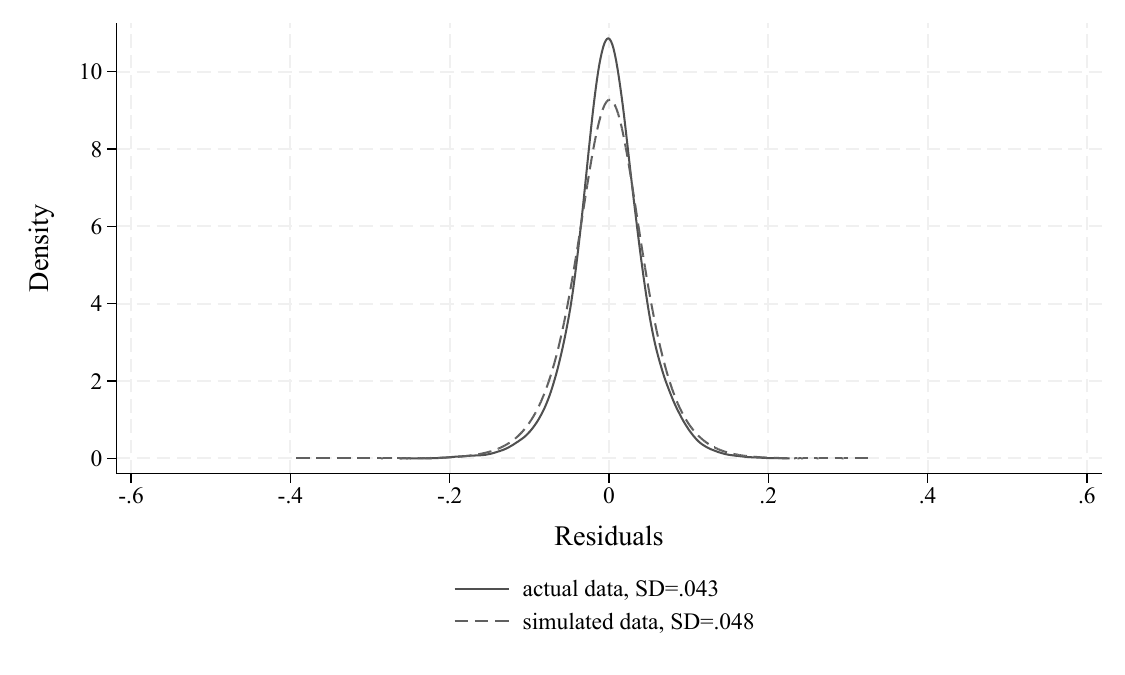} 
  \caption*{(b) Long Training}
   \includegraphics[width=0.55\linewidth]{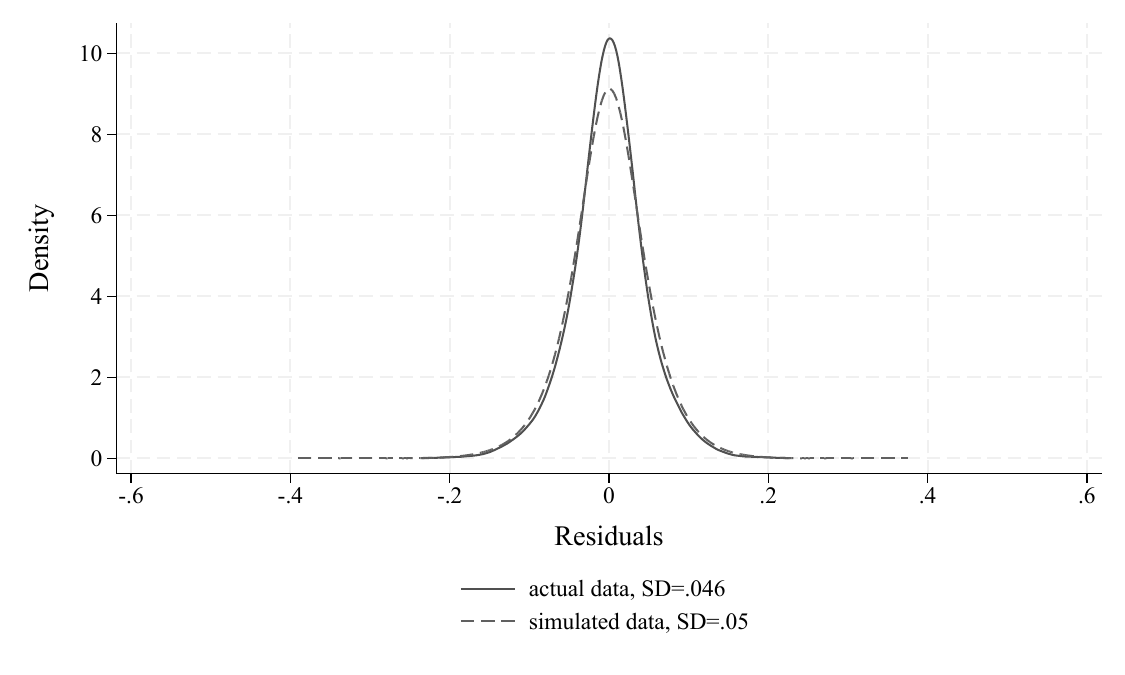} 
   \caption*{(c) Retraining}
   \end{center}
   \begin{minipage}{\textwidth}
          \vspace{2pt}
          \footnotesize \textit{Notes:} This figure plots the observed (solid line) simulated (dashed line) distribution of the average peer employability  (kernel density) by program type. SD refers to the standard deviation. 
     \end{minipage}
\end{figure}

\FloatBarrier

\subsubsection{Systematic Correlation of Own and Peer Employability}\label{appendix:validity_corr}

\begin{table}[htbp]\centering
\def\sym#1{\ifmmode^{#1}\else\(^{#1}\)\fi}
\caption{Exogeneity Test Controlling for the Average Employability in the Pool} \label{tab:test_guryan}
\resizebox{0.9\linewidth}{!}{%
\begin{tabular}{l*{3}{c}}
\hline\hline
          &\multicolumn{1}{c}{(1)}&\multicolumn{1}{c}{(2)}&\multicolumn{1}{c}{(3)}\\
          &\multicolumn{1}{c}{Short Training}&\multicolumn{1}{c}{Long Training}&\multicolumn{1}{c}{Retraining}\\
\hline
Mean peer employability ($\bar{X}_{ipct}$) &   -0.002&-0.025***&-0.024***\\
          &  (0.001)&  (0.003)&  (0.003)\\
Mean peer employability (pool)&-0.017***&-0.041***&-0.140***\\
          &  (0.006)&  (0.014)&  (0.033)\\
          \hline
Provider x competence level FE&     \checkmark &     \checkmark &     \checkmark \\
Month x year x occupation FE&     \checkmark &     \checkmark &     \checkmark \\
Additional Controls&     \checkmark &     \checkmark &     \checkmark \\
N         &    36662&    12683&    12413\\
\hline\hline
\end{tabular}}
   \begin{minipage}{\textwidth}
          \vspace{6pt}
          \footnotesize \textit{Notes:} This table displays the coefficients of a regression of the individual employability on the average employability in the course and in the pool of participants starting in the same month with the same competence level. Standard errors are clustered at the course level. \sym{*} \(p<0.05\), \sym{**} \(p<0.01\), \sym{***} \(p<0.001\)
     \end{minipage}
\end{table} 

\begin{table}[htbp]
 \centering
  \caption{Orthogonality of Own Employability and Course Dummies}
  \label{app_tab:chetty}
\resizebox{0.6\linewidth}{!}{%
\begin{tabular}{lrrr}
\hline\hline
 &\multicolumn{1}{c}{Short training}&\multicolumn{1}{c}{Long training}&\multicolumn{1}{c}{Retraining}\\
\hline
F-statistic& 0.982  & 0.593 & 0.619  \\
P-value &  0.766 & 1.000 & 1.000 \\
\hline
Observations    &    36662     &    12683&          12413        \\
\hline\hline
\end{tabular}}
     \begin{minipage}{\textwidth} \linespread{1.0}\selectfont
          \vspace{6pt}
          \footnotesize \textit{Notes:} F-tests and corresponding p-values from a regression of the residualized own employability (net of provider-by-competence level, month-by-year-by-occupation fixed effects and course level controls) on  course fixed effects.
     \end{minipage}
\end{table}

\FloatBarrier
\newpage
\subsubsection{Systematic Correlation of Peer Employability and other Predetermined Characteristics}\label{appendix:validity_corr_chars}

\begin{table}[htb]\centering
\def\sym#1{\ifmmode^{#1}\else\(^{#1}\)\fi}
\caption{Correlation of Peer Employability with Individual Characteristics} \label{tab:corr_ind}
\resizebox{0.8\linewidth}{!}{%
\begin{tabular}{l*{3}{c}}
\hline\hline
          &\multicolumn{1}{c}{(1)}&\multicolumn{1}{c}{(2)}&\multicolumn{1}{c}{(3)}\\
          &\multicolumn{1}{c}{Short Training}&\multicolumn{1}{c}{Long Training}&\multicolumn{1}{c}{Retraining}\\
\hline
Age       &-0.0000799\sym{*}  &-0.0000160         &0.0000153         \\
          &  (-2.07)         &  (-0.33)         &   (0.25)         \\
Female    &  0.00376\sym{***}& 0.000999         &-0.000588         \\
          &   (4.29)         &   (0.81)         &  (-0.52)         \\
Not German    & -0.00276\sym{**} &  0.00213         &  0.00205         \\
          &  (-2.78)         &   (1.67)         &   (1.62)         \\
High Education &-0.000425         &-0.000605         &  0.00140         \\
          &  (-0.45)         &  (-0.56)         &   (1.02)         \\
Vocational Degree&  0.00263         & -0.00127         & -0.00257         \\
          &   (1.85)         &  (-0.83)         &  (-1.05)         \\
High-skilled& -0.00192\sym{**} & 0.000397         &  0.00244\sym{**} \\
          &  (-2.91)         &   (0.47)         &   (2.60)         \\
Unemployment duration at program start    &0.0000529\sym{**} &0.0000773\sym{***}&0.0000792\sym{***}\\
          &   (3.00)         &   (4.12)         &   (3.71)         \\
Months employed in the last 2 years &0.0000999         &-0.000594\sym{***}&-0.000533\sym{***}\\
          &   (1.36)         &  (-6.73)         &  (-5.63)         \\
Months OLF in the last 2 years & 0.000790\sym{***}& 0.000611\sym{***}& 0.000529\sym{***}\\
          &   (8.92)         &   (5.88)         &   (4.34)         \\
Months employed in the last 10 years &-0.0000349\sym{*}  &-0.0000304         &-0.0000386         \\
          &  (-2.39)         &  (-1.61)         &  (-1.68)         \\
Months OLF in the last 10 years &-0.0000127         &0.0000108         &-0.0000159         \\
          &  (-0.54)         &   (0.38)         &  (-0.52)         \\
Earnings in  last 2 years (1000 EUR)& 0.000227\sym{***}& 0.000207\sym{***}& 0.000192\sym{***}\\
          &   (8.17)         &   (5.74)         &   (4.00)         \\
Earnings in  last 10 years (1000 EUR)&0.00000353         &-0.00000610         &-0.0000128         \\
          &   (0.68)         &  (-0.93)         &  (-1.33)         \\
\hline
P-val: joint F-test&        0         &        0         &        0         \\
Provider x competence level FE&      Yes         &      Yes         &      Yes         \\
Month x year x occupation FE&      Yes         &      Yes         &      Yes         \\
Additional Controls&      Yes         &      Yes         &      Yes         \\
N         &    36662         &    12683         &    12413         \\
\hline\hline
\multicolumn{4}{l}{\footnotesize \sym{*} \(p<0.05\), \sym{**} \(p<0.01\), \sym{***} \(p<0.001\)}\\
\end{tabular}}
\end{table}

\FloatBarrier
\newpage

\subsection{Further Results} \label{appendix:results}

\begin{figure}[htb]
 \begin{center}
    \caption{Monthly Effects of Peer Employability on All Employment} 
  \label{fig:effects_emp}
 \includegraphics[width=0.9\textwidth]{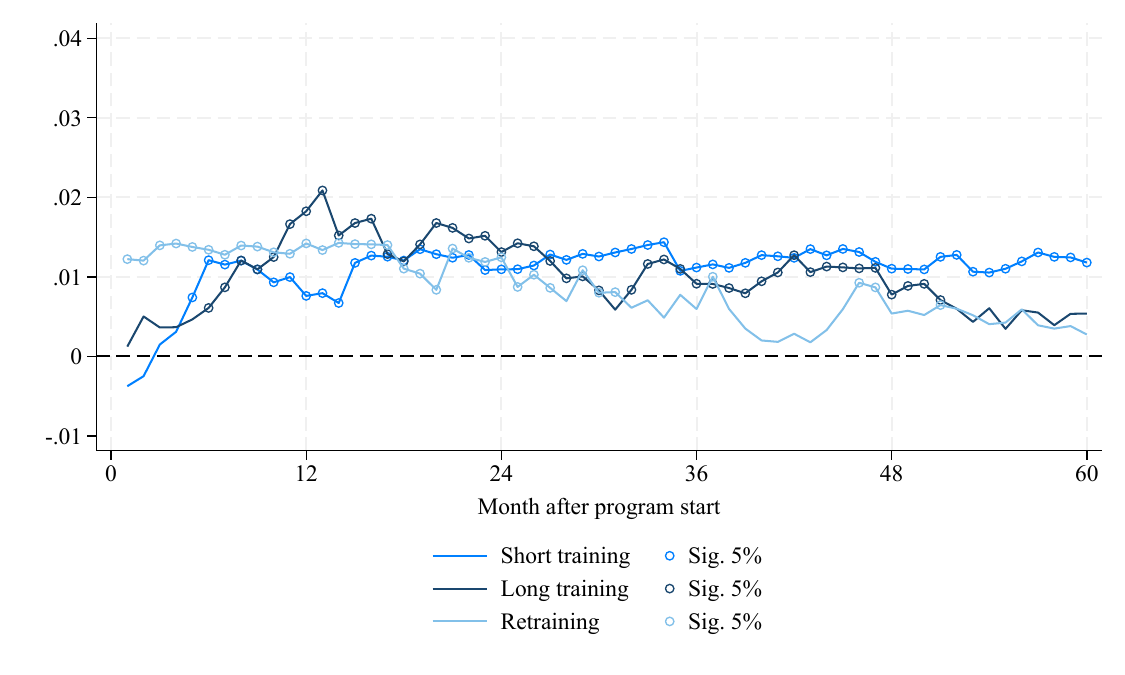}
 \end{center}
 \begin{minipage}{\textwidth}   \linespread{1.0}\selectfont
 \vspace{6pt}
     \footnotesize \textit{Notes:} The figure depicts the estimated effects (in percentage points/100) of a one standard deviation increase in the predicted mean peer employability on the individual probability to be employed in any job in the months 1-60 after program start. Significant effects at the 5 percent level are marked by circles. On top of the mean peer employability, the underlying model includes the individual ex-ante employability, a vector of course-level controls, occupation and provider-by-month group and seasonal fixed effects. Standard errors are clustered at the course level.
 \end{minipage}
\end{figure}

\begin{table}
 \centering
  \caption{\small The Effects of Other Average Peer Characteristics}
  \label{tab:other_peervar}
\def\sym#1{\ifmmode^{#1}\else\(^{#1}\)\fi}
\resizebox{\linewidth}{!}{%
\begin{tabular}{lcHccccc}
\hline\hline
                &\multicolumn{4}{c}{\textbf{Total employment in month 60}}&\multicolumn{3}{c}{\textbf{Log total earnings month 60}}\\
                &\multicolumn{2}{c}{(ST)}&\multicolumn{1}{c}{(LT)}&\multicolumn{1}{c}{(RT)}&\multicolumn{1}{c}{(ST)}&\multicolumn{1}{c}{(LT)}&\multicolumn{1}{c}{(RT)}\\
\hline                
Peer earnings in last 2 years (1000 euro)&     1.86***&     1.86***&     1.98** &     2.36** &     0.01** &     0.01*  &     0.01** \\
                &   (0.39)         &   (0.39)         &   (0.71)         &   (0.84)         &   (0.00)         &   (0.00)         &   (0.00)         \\[5pt]
                \hline
Peer months of employment in last 2 years&     6.59***&     6.59***&     5.34** &     5.51** &     0.03***&     0.02*  &     0.00         \\
                &   (0.98)         &   (0.98)         &   (1.71)         &   (1.67)         &   (0.01)         &   (0.01)         &   (0.01)         \\[5pt]
                \hline
Peer months of employment in last 10 years&     0.42         &     0.42         &     1.50***&     0.87*  &     0.00         &     0.01** &     0.00         \\
  &   (0.25)         &   (0.25)         &   (0.42)         &   (0.40)         &   (0.00)         &   (0.00)         &   (0.00)         \\
\hline
Individual-level controls&      \checkmark         &      \checkmark         &      \checkmark         &      \checkmark         &      \checkmark         &      \checkmark         &      \checkmark         \\
Provider x competence level FE&      \checkmark         &      \checkmark         &      \checkmark         &      \checkmark         &      \checkmark         &      \checkmark         &      \checkmark         \\
Month x year x occupation FE&      \checkmark         &      \checkmark         &      \checkmark         &      \checkmark         &      \checkmark         &      \checkmark         &      \checkmark         \\
Additional Controls&      \checkmark         &      \checkmark         &      \checkmark         &      \checkmark         &      \checkmark         &      \checkmark         &      \checkmark         \\
Observations    &    36661         &    36661         &    12683         &    12413         &    36661         &    12683         &    12413         \\
\hline\hline
\end{tabular}}
 \begin{minipage}{\textwidth} \linespread{1.0}\selectfont
          \vspace{6pt}
          \footnotesize \textit{Notes:} All specifications control for course-level controls, individual age, gender, nationality, education, vocational degree, unemployment duration at program start as well as the individual labor market attachment. Standard errors are clustered at the course level. \sym{*} \(p<0.05\), \sym{**} \(p<0.01\), \sym{***} \(p<0.001\).
     \end{minipage}
\end{table}

\FloatBarrier
\clearpage

\subsection{Robustness Checks }\label{appendix:rob_checks}
\begin{figure}[htbp]
          \vspace{10pt}
    \caption{Robustness Checks - Main}\label{fig:rob_comb}
    \centering
        \begin{minipage}{0.49\textwidth}
        \centering
        \includegraphics[width=\linewidth,trim=0 110 0 0,clip]{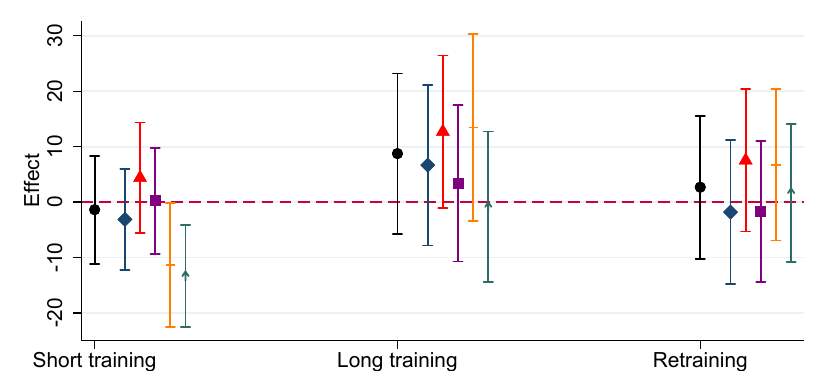}
        \caption*{(a) \small{Search Duration First Job (in days)}}
    \end{minipage}%
    \hfill
    \begin{minipage}{0.49\textwidth}
        \centering
        \includegraphics[width=\linewidth,trim=0 110 0 0,clip]{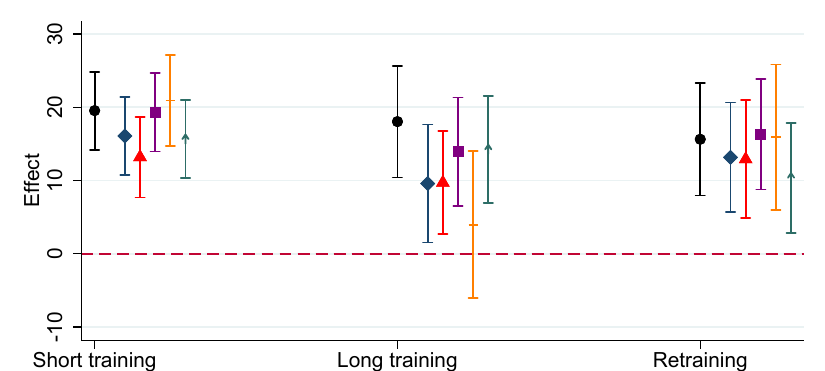}
        \caption*{(b) \small{Total Employment in month 60}}
    \end{minipage} \\[10pt]%
        \begin{minipage}{0.49\textwidth}
        \centering
        \includegraphics[width=\linewidth,trim=0 110 0 0,clip]{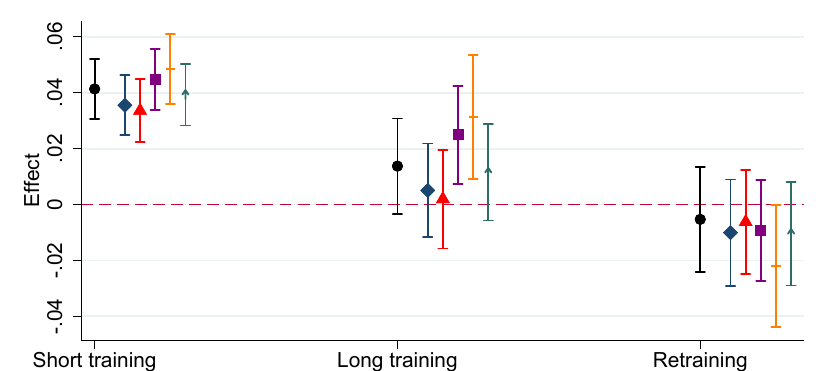}
        \caption*{(c) \small{Log Earnings in First Job}}
    \end{minipage}
    \hfill
    \begin{minipage}{0.49\textwidth}
        \centering
        \includegraphics[width=\linewidth,trim=0 110 0 0,clip]{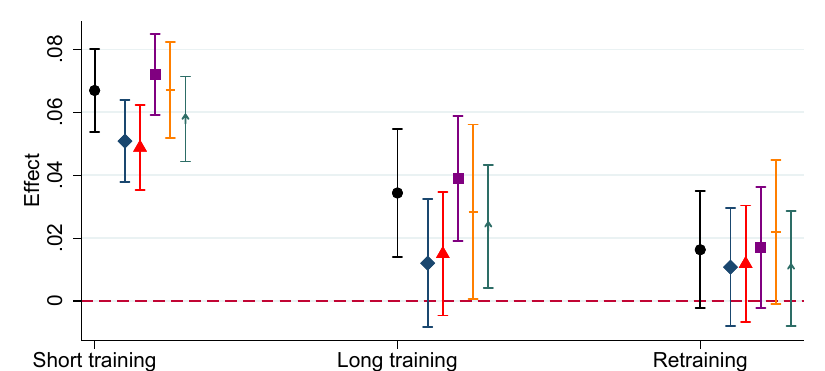}
        \caption*{(d) \small{Log Total Earnings in month 60}}
    \end{minipage}\\[10pt]
        \centering
         \begin{minipage}{0.70\textwidth}
        \includegraphics[width=\linewidth,trim=0 10 0 185,clip]{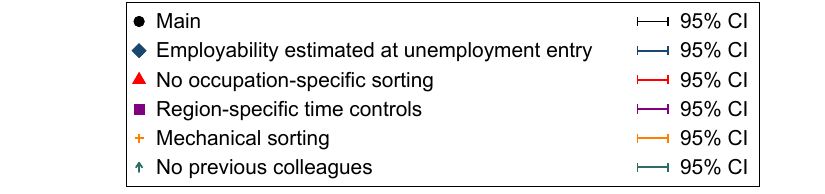}
    \end{minipage}
        \vspace{10pt}
    \begin{minipage}{\textwidth}
          \vspace{10pt}
          \footnotesize \textit{Notes:} This figure summarizes the results of the main specification ("Main") and various robustness checks described in Section \ref{sec:robustness2}. It shows the effect sizes by program type with the corresponding 95 percent confidence intervals. More detailed information about point estimates, standard errors and the number of observations underlying each estimation can be found in Appendix Table \ref{tab:rob_comb_tab}.
     \end{minipage}
\end{figure}

\FloatBarrier
\newpage
{\scriptsize
\begin{longtable}{lllrrr} 
\caption{Detailed Information for Robustness Checks in Figure \ref{fig:rob_comb}}\\
\hline
Specification & Outcome & Program & Effect & SE & N \\
\hline
Main&Total employment in month 60&Short training&19.541&2.717&36661\\
Main&Search duration first job&Short training&-1.368&4.966&36662\\
Main&Log earnings first job (if > 0)&Short training&.041&.006&31355\\
Main&Log total earnings in month 60 (if > 0)&Short training&.067&.007&33781\\
Employability estim. at UE entry&Total employment in month 60&Short training&16.082&2.725&36661\\
Employability estim. at UE entry&Search duration first job&Short training&-3.123&4.627&36662\\
Employability estim. at UE entry&Log earnings first job (if > 0)&Short training&.036&.005&31355\\
Employability estim. at UE entry&Log total earnings in month 60 (if > 0)&Short training&.051&.007&33781\\
No occupation-specific sorting&Total employment in month 60&Short training&13.194&2.806&33159\\
No occupation-specific sorting&Search duration first job&Short training&4.385&5.096&33160\\
No occupation-specific sorting&Log earnings first job (if > 0)&Short training&.034&.006&28373\\
No occupation-specific sorting&Log total earnings in month 60 (if > 0)&Short training&.049&.007&30634\\
Region-specific time controls&Total employment in month 60&Short training&19.347&2.741&36494\\
Region-specific time controls&Search duration first job&Short training&.238&4.873&36495\\
Region-specific time controls&Log earnings first job (if > 0)&Short training&.045&.006&31236\\
Region-specific time controls&Log total earnings in month 60 (if > 0)&Short training&.072&.007&33617\\
Mechanical sorting&Total employment in month 60&Short training&20.936&3.148&25031\\
Mechanical sorting&Search duration first job&Short training&-11.332&5.724&25032\\
Mechanical sorting&Log earnings first job (if > 0)&Short training&.049&.006&21387\\
Mechanical sorting&Log total earnings in month 60 (if > 0)&Short training&.067&.008&23076\\
No previous colleagues&Total employment in month 60&Short training&15.66&2.727&34512\\
No previous colleagues&Search duration first job&Short training&-13.355&4.687&34512\\
No previous colleagues&Log earnings first job (if > 0)&Short training&.039&.006&29708\\
No previous colleagues&Log total earnings in month 60 (if > 0)&Short training&.058&.007&31704\\
Main&Total employment in month 60&Long training&18.041&3.89&12683\\
Main&Search duration first job&Long training&8.752&7.382&12683\\
Main&Log earnings first job (if > 0)&Long training&.014&.009&10779\\
Main&Log total earnings in month 60 (if > 0)&Long training&.034&.01&11713\\
Employability estim. at UE entry&Total employment in month 60&Long training&9.593&4.106&12683\\
Employability estim. at UE entry&Search duration first job&Long training&6.68&7.402&12683\\
Employability estim. at UE entry&Log earnings first job (if > 0)&Long training&.005&.009&10779\\
Employability estim. at UE entry&Log total earnings in month 60 (if > 0)&Long training&.012&.01&11713\\
No occupation-specific sorting&Total employment in month 60&Long training&9.708&3.599&11513\\
No occupation-specific sorting&Search duration first job&Long training&12.71&7.003&11513\\
No occupation-specific sorting&Log earnings first job (if > 0)&Long training&.002&.009&9805\\
No occupation-specific sorting&Log total earnings in month 60 (if > 0)&Long training&.015&.01&10675\\
Region-specific time controls&Total employment in month 60&Long training&13.916&3.788&12238\\
Region-specific time controls&Search duration first job&Long training&3.417&7.216&12238\\
Region-specific time controls&Log earnings first job (if > 0)&Long training&.025&.009&10402\\
Region-specific time controls&Log total earnings in month 60 (if > 0)&Long training&.039&.01&11305\\
Mechanical sorting&Total employment in month 60&Long training&3.971&5.126&7827\\
Mechanical sorting&Search duration first job&Long training&13.447&8.615&7827\\
Mechanical sorting&Log earnings first job (if > 0)&Long training&.031&.011&6662\\
Mechanical sorting&Log total earnings in month 60 (if > 0)&Long training&.028&.014&7271\\
No previous colleagues&Total employment in month 60&Long training&14.259&3.738&12225\\
No previous colleagues&Search duration first job&Long training&-.848&6.937&12225\\
No previous colleagues&Log earnings first job (if > 0)&Long training&.011&.009&10449\\
No previous colleagues&Log total earnings in month 60 (if > 0)&Long training&.024&.01&11279\\
Main&Total employment in month 60&Retraining&15.632&3.92&12413\\
Main&Search duration first job&Retraining&2.687&6.575&12413\\
Main&Log earnings first job (if > 0)&Retraining&-.005&.01&10618\\
Main&Log total earnings in month 60 (if > 0)&Retraining&.016&.009&11548\\
Employability estim. at UE entry&Total employment in month 60&Retraining&13.164&3.814&12413\\
Employability estim. at UE entry&Search duration first job&Retraining&-1.809&6.644&12413\\
Employability estim. at UE entry&Log earnings first job (if > 0)&Retraining&-.01&.01&10618\\
Employability estim. at UE entry&Log total earnings in month 60 (if > 0)&Retraining&.011&.01&11548\\
No occupation-specific sorting&Total employment in month 60&Retraining&12.903&4.101&11460\\
No occupation-specific sorting&Search duration first job&Retraining&7.518&6.541&11460\\
No occupation-specific sorting&Log earnings first job (if > 0)&Retraining&-.006&.009&9778\\
No occupation-specific sorting&Log total earnings in month 60 (if > 0)&Retraining&.012&.009&10683\\
Region-specific time controls&Total employment in month 60&Retraining&16.29&3.836&11878\\
Region-specific time controls&Search duration first job&Retraining&-1.686&6.473&11878\\
Region-specific time controls&Log earnings first job (if > 0)&Retraining&-.009&.009&10151\\
Region-specific time controls&Log total earnings in month 60 (if > 0)&Retraining&.017&.01&11056\\
Mechanical sorting&Total employment in month 60&Retraining&15.918&5.06&8241\\
Mechanical sorting&Search duration first job&Retraining&6.7&6.963&8241\\
Mechanical sorting&Log earnings first job (if > 0)&Retraining&-.022&.011&7064\\
Mechanical sorting&Log total earnings in month 60 (if > 0)&Retraining&.022&.012&7702\\
No previous colleagues&Total employment in month 60&Retraining&10.383&3.831&11769\\
No previous colleagues&Search duration first job&Retraining&1.628&6.344&11769\\
No previous colleagues&Log earnings first job (if > 0)&Retraining&-.01&.009&10133\\
No previous colleagues&Log total earnings in month 60 (if > 0)&Retraining&.01&.009&10927\\
\hline 
\end{longtable}
  \label{tab:rob_comb_tab}
  \begin{minipage}{0.9\textwidth} \linespread{1.0}\selectfont
          \vspace{-30pt}
          \footnotesize \textit{Notes:} This table displays the point estimates (Effect), standard errors (SE), and numbers of observations (N) for the robustness checks described in Section \ref{sec:robustness2} for the main outcomes. UE: unemployment.
     \end{minipage}
}

\pagebreak
\FloatBarrier

\begin{figure}[htbp]
          \vspace{10pt}
    \caption{Robustness Checks - Clustering}\label{fig:rob_se}
    \centering
        \begin{minipage}{0.49\textwidth}
        \centering
        \includegraphics[width=\linewidth,trim=0 50 0 0,clip]{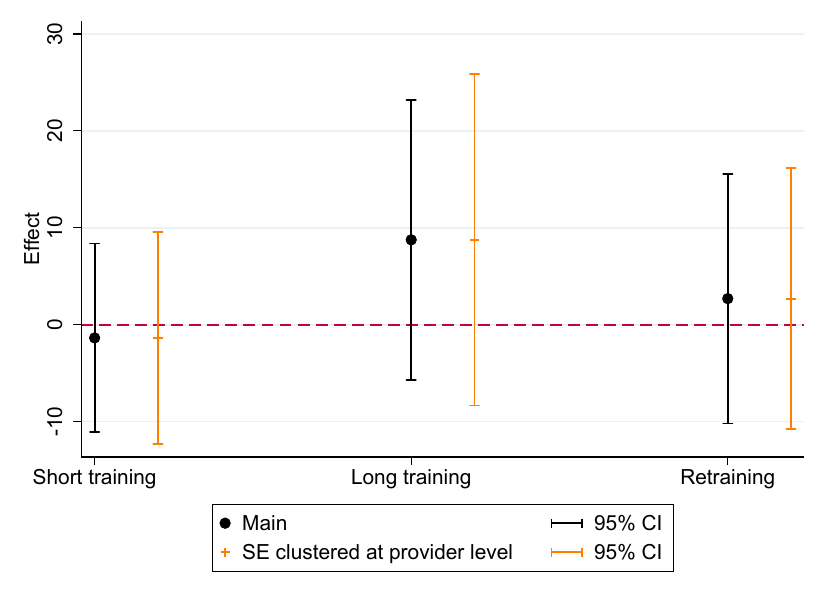}
        \caption*{(a) \small{Search Duration First Job (in days)}}
    \end{minipage}%
    \hfill
    \begin{minipage}{0.49\textwidth}
        \centering
        \includegraphics[width=\linewidth,trim=0 50 0 0,clip]{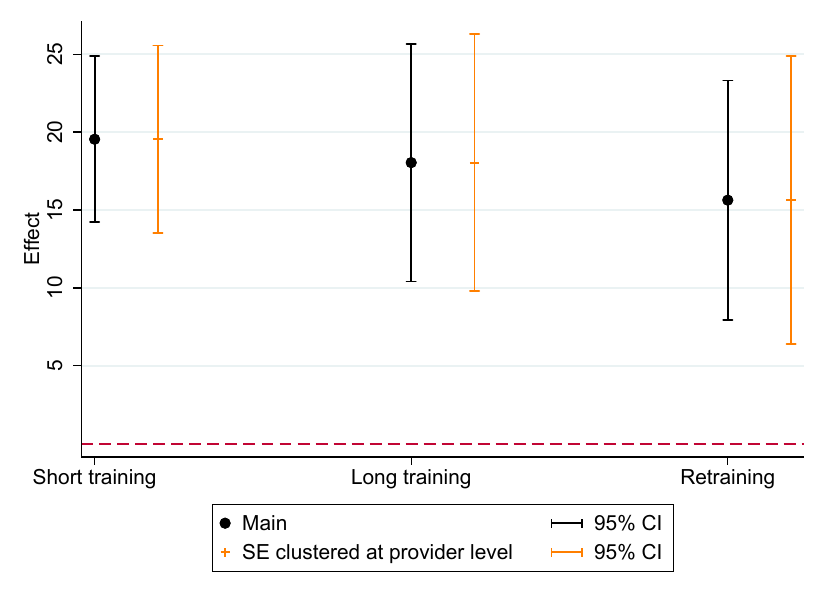}
        \caption*{(b) \small{Total Employment in month 60}}
    \end{minipage} \\[10pt]%
        \begin{minipage}{0.49\textwidth}
        \centering
        \includegraphics[width=\linewidth,trim=0 50 0 0,clip]{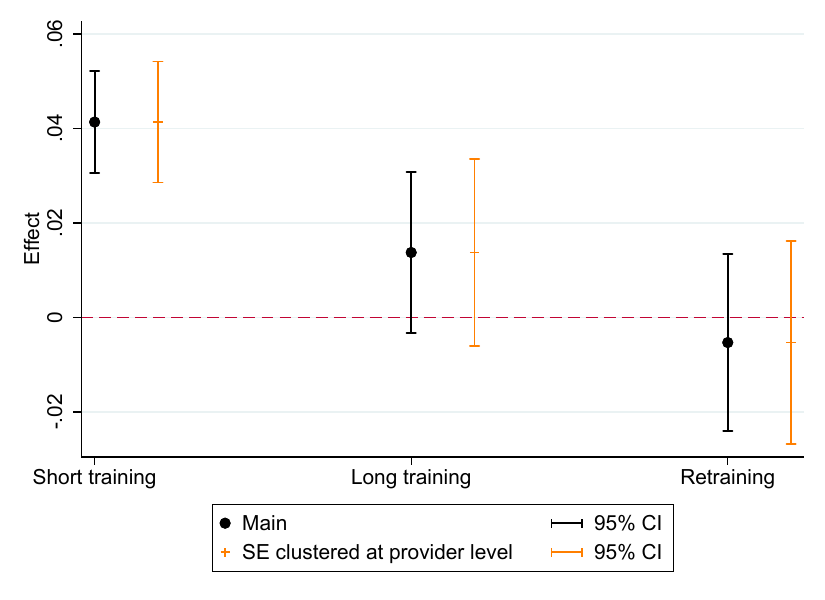}
        \caption*{(c) \small{Log Earnings in First Job}}
    \end{minipage}
    \hfill
    \begin{minipage}{0.49\textwidth}
        \centering
        \includegraphics[width=\linewidth,trim=0 50 0 0,clip]{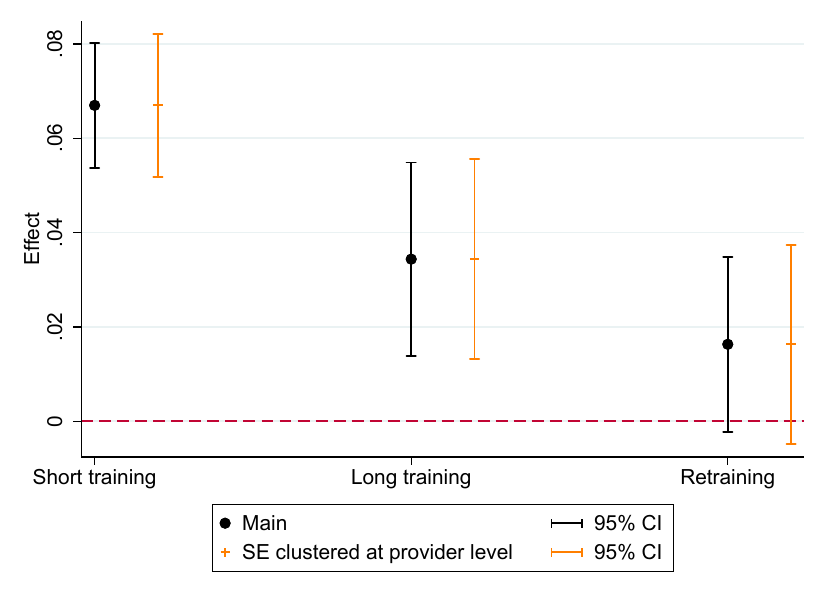}
        \caption*{(d) \small{Log Total Earnings in month 60}}
    \end{minipage}\\[10pt]
        \centering
         \begin{minipage}{0.70\textwidth}
        \includegraphics[width=\linewidth,trim=0 10 0 240,clip]{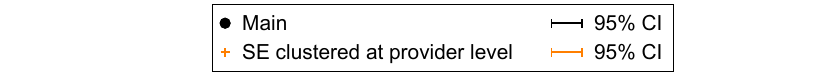}
    \end{minipage}
        \vspace{10pt}
    \begin{minipage}{\textwidth}
          \vspace{10pt}
          \footnotesize \textit{Notes:} This figure summarizes the results of the main specification ("Main") which clusters standard errors at the course level and a specification clustering at the level of providers. It shows the effect sizes by program type with the corresponding 95 percent confidence intervals.
     \end{minipage}
\end{figure}

\begin{table}
\centering
  \caption{Effects of Peer Employability on Earnings (Including Zeros) }
  \label{tab:res_earnings}
  \resizebox{0.8\linewidth}{!}{%
\begin{tabular}[htb]{lccc} 
\hline\hline
 &\multicolumn{1}{c}{Short training}&\multicolumn{1}{c}{Long training}&\multicolumn{1}{c}{Retraining}\\
   &\multicolumn{1}{c}{(1)}&\multicolumn{1}{c}{(2)}&\multicolumn{1}{c}{(3)}\\
   \hline
\multicolumn{4}{l}{\textbf{Panel A - Log earnings first job }}\\ 
$\bar{X}_{(-i)pct}$& 0.035***&   -0.004&   -0.012\\
          &  (0.008)&  (0.012)&  (0.012)\\[4pt]
\hline
\multicolumn{4}{l}{\textbf{Panel B - Earnings in first job (in levels)}}\\ 
$\bar{X}_{(-i)pct}$& 1.037***&    0.207&   -0.067\\
          &  (0.157)&  (0.213)&  (0.212)\\[4pt]
\hline
\multicolumn{4}{l}{\textbf{Panel C - Log total earnings in month 60 }}\\ 
$\bar{X}_{(-i)pct}$& 0.102***& 0.068***&    0.023\\
          &  (0.015)&  (0.024)&  (0.021)\\[4pt]
\hline
\multicolumn{4}{l}{\textbf{Panel D - Total earnings in month 60 (in levels)}}\\ 
$\bar{X}_{(-i)pct}$&2737.298***&1901.221***&980.716***\\
          &(248.515)&(378.582)&(270.453)\\[4pt]  
          \hline
Provider x competence level FE&      \checkmark&      \checkmark&      \checkmark\\
Month x year x occupation FE&      \checkmark&      \checkmark&      \checkmark\\
Additional Controls&      \checkmark&      \checkmark&      \checkmark\\
Observations         &    36661&    12683&    12413\\
\hline\hline
\end{tabular}}
      \begin{minipage}{\textwidth}
          \vspace{6pt}
          \footnotesize \textit{Notes:} $X_{ipct}$ designates the own employability and $\bar{X}_{(-i)pct}$ the leave-one-out mean peer employability. All specifications control for  course-level controls. Standard errors (in round brackets)  are clustered at the course level.  Effects are reported in terms of  SD increases (see Table \ref{tab:peervars_sel}). Earnings are in prices of 2010 and measured in levels in Panels B and D and in log(euro). They are set to 0 in Panels A and C in case individuals are not working.  \sym{*} \(p<0.05\), \sym{**} \(p<0.01\), \sym{***} \(p<0.001\)
     \end{minipage}
\end{table}

\FloatBarrier

\subsubsection{Measurement Error through Unobserved Peers}\label{appendix:validity_checks_me}

There is a possibility that I do not observe all individuals in a training program. The data covers all jobseekers who are participating in publicly sponsored training measures and register at the local employment agencies. I do not observe individuals participating in a training program on their own or on their employer's initiative and do so without registering the employment agencies. There are different groups that could self-select into vocational training programs without being registered. First, there could be individuals who are integrated in the labor market and are either employed or self-employed. At the same time, the likelihood that those individuals would participate in long, full-time courses is very low. If employed, employers would need to grant them leave of absence or in case of self-employment, they would incur major costs of leaving their business. Moreover courses directed at the integration of unemployed individuals into the labor market will not be of great interest for those who already have a job. As for the non-employed individuals, I might not observe women returning to the labor market, non-employed individuals who are not eligible for unemployment assistance and non-registered recipients of social assistance. Nevertheless, all three groups face substantial incentives to register as unemployed if willing to participate in training measures. By registering they could still get access to funding for the course-related costs, for example. By restricting the analysis to full-time courses and excluding courses that are specifically directed at employed individuals the existence of unobserved participants is highly unlikely. A small fraction of unobserved peers would lead to a attenuation bias that is negligible, assuming that there is no selection into courses that specifically relates to quality.  Missing and observed data can come from arbitrarily different distributions (e.g., participants missing in the data may be more skilled than the ones observed) but the distribution needs to be independent of group assignment and $\epsilon_{ipct}$ after controlling for fixed effects and course controls \citep{Ammermueller2009,Sojourner2013}.


\subsubsection{Mechanical Sorting}\label{appendix:mech_sorting}

If individuals select into courses based on preferences that are provider and month-specific, the three-month redemption period of the vouchers might provoke some mechanical sorting which might aggravate the sorting issue. Such sorting is difficult to control for since it is not clear whether it takes place and if there exist specific sorting patterns. What can be controlled for is sorting that follows the same cycle in each quarter.

The intuition behind the identification strategy is illustrated in Figure \ref{fig:identification}:  A jobseeker obtains a voucher in February and decides to participate in a course starting in April at a particular provider. Her course mates may have obtained their vouchers earlier or later than her, i.e., in the months from January to April. Because of the three month redemption period, the jobseeker cannot be grouped with program participants who obtained their vouchers before January nor with participants who obtain their vouchers starting from May. At the same time, no participant in the April course could select into courses starting before January or later than July. As for the August course, no participant could start a course earlier than May or later than November. 

\begin{figure}[h!]
\begin{center}
                 \caption{Identification Strategy Correcting for Mechanical Sorting} 
                  \label{fig:identification}
               \includegraphics[width=1\textwidth, clip=true,  trim={0cm 0cm 4cm 0cm}]{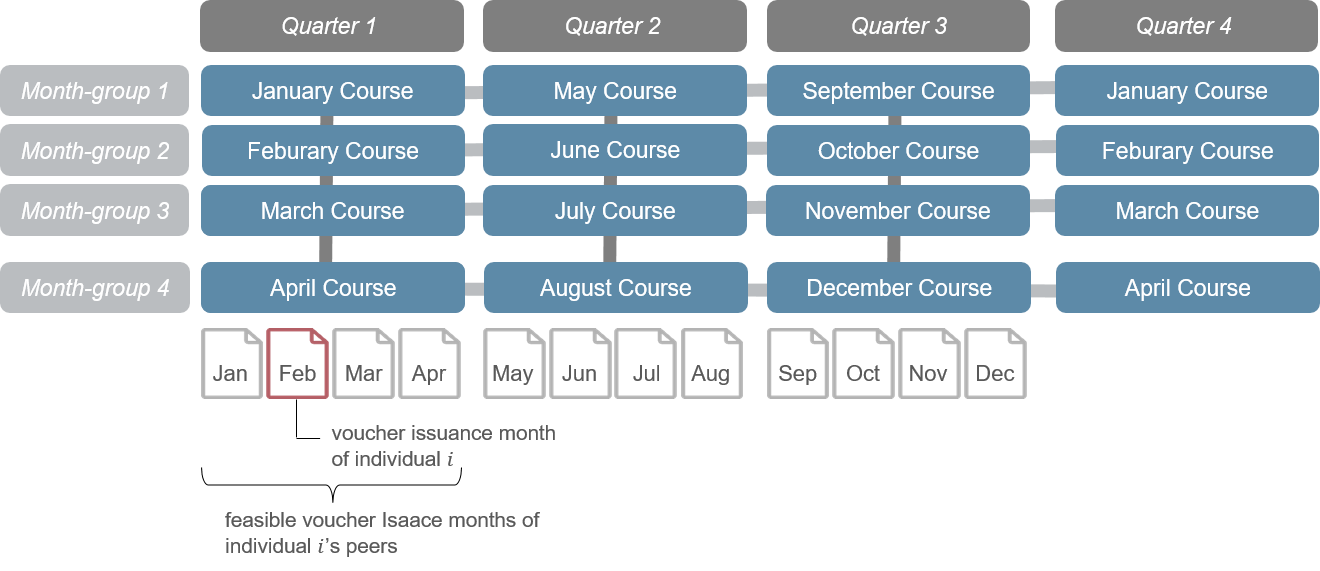}
\end{center}
\footnotesize 
\textit{Notes:} The figure depicts feasible comparison groups depending on the course start month in rows. These groups are provider-specific and contain courses that are each four months apart. The bottom of the figure illustrates how individuals can select into a specific course month depending on their voucher issuance exemplary for the month April. Columns designate year-specific quarters, i.e. 4-month divisions of the calendar year which serve as seasonality controls.
\end{figure}

In line with this reasoning, I can compare all courses starting in the months of April, August and December at a particular provider without participants being able to self-select into the respective other courses. Depending on the start dates of the courses, four groups naturally arise which I label \textit{month groups}: January-May-September, February-June-October, March-July-November and April-August-December. They are listed as rows in Figure \ref{fig:identification}. Conditional on the choice of provider and the month group, the entry into a specific course is driven by the voucher issuance date and the preferences for specific months in a quarter (which are assumed to follow the same pattern for every quarter). The voucher issuance is unlikely to be manipulated. Vouchers are issued in meetings with caseworkers, whose appointment time can hardly be influenced by the jobseekers. It depends on their entry into unemployment and the efficiency of administrative processes. Participants can sort across provider-specific month groups but not within.

\subsection{Compositional Changes and other Mechanisms}\label{appendix:mechanisms_other}
There are other possible mechanisms than knowledge spillovers and competition which could play a role in the context of labor market training. Since peer groups are not perfectly stable throughout the course compositional changes might cause some of the observed effects. To investigate this further, I compare the distribution of peer employability at the beginning and end of a course. As Appendix Figure shows, it remains largely the same. If anything the distribution shifts slightly to the left towards the end of the course in long vocational training courses and retraining courses. This suggests that some highly employable peers leave the course early. Next, I perform a heterogeneity analysis where I compare peer effects in courses with a high and a low share of dropouts. Courses with a high  share of dropouts are courses where the share of dropouts exceeds the median in the sample. The results are shown in Appendix Table \ref{tab:mech_dropouts}. They are qualitatively similar between between both types of courses. One exception is a larger employment effect for courses with many dropouts in retraining. It might be explained by a larger fraction of less employable jobseekers and the effect heterogeneity reported in Section \ref{sec:heterogeneity}. Overall, I can reject compositional changes as a relevant driver behind the effects.

\begin{table}[htbp]\centering
\def\sym#1{\ifmmode^{#1}\else\(^{#1}\)\fi}
\caption{Effects by Compositional Change}  \label{tab:mech_dropouts}
  \resizebox{0.8\linewidth}{!}{%
\begin{tabular}{l*{3}{c}}
\hline\hline
          &\multicolumn{1}{c}{(1)}&\multicolumn{1}{c}{(2)}&\multicolumn{1}{c}{(3)}\\
          &\multicolumn{1}{c}{Short Training}&\multicolumn{1}{c}{Long Training}&\multicolumn{1}{c}{Retraining}\\
          \hline
          \multicolumn{4}{l}{\textbf{Panel A - Search duration}}\\ 
Low share of dropouts&   -1.261&    7.532&    6.133\\
          &  (5.677)&  (8.791)&  (8.767)\\
High share of dropouts&    6.743&   13.534&    4.326\\
          &  (6.643)&  (8.569)&  (7.628)\\
P-value difference&     .278&     .522&     .857\\
\hline
\multicolumn{4}{l}{\textbf{Panel B - Total Employment}}\\ 
Low share of dropouts&19.570***&20.343***&    8.263\\
          &  (3.058)&  (4.739)&  (5.534)\\
High share of dropouts&17.350***&14.706***&19.237***\\
          &  (3.856)&  (4.735)&  (4.502)\\
P-value difference&     .594&     .299&     .083\\
\hline
\multicolumn{4}{l}{\textbf{Panel C - Earnings in First Job (if > 0)}}\\ 
Low share of dropouts& 0.041***&    0.012&    0.010\\
          &  (0.007)&  (0.011)&  (0.013)\\
High share of dropouts& 0.037***&    0.014&   -0.017\\
          &  (0.007)&  (0.011)&  (0.011)\\
P-value difference&     .577&     .896&     .089\\
\hline      
\multicolumn{4}{l}{\textbf{Panel D - Total Earnings (if > 0)}}\\ 
Low share of dropouts& 0.067***&  0.028**&    0.012\\
          &  (0.008)&  (0.013)&  (0.014)\\
High share of dropouts& 0.064***& 0.041***&   0.019*\\
          &  (0.009)&  (0.012)&  (0.011)\\
P-value difference&     .778&     .368&     .667\\
\hline      
Provider x competence level FE&      \checkmark&      \checkmark&      \checkmark\\
Month x year x occupation FE&      \checkmark&      \checkmark&      \checkmark\\
Additional Controls&      \checkmark&      \checkmark&      \checkmark\\
\hline\hline
\end{tabular}}
\begin{minipage}{\textwidth}
          \vspace{6pt}
          \footnotesize \textit{Notes:} All specifications control for  course-level controls and individual employability . Standard errors (in round brackets)  are clustered at the course level.  Peer effects for courses with low and high shares of dropouts are reported in terms of  SD increases (see Table \ref{tab:peervars_sel}).  Earnings are in prices of 2010 and measured in log(euro). \sym{*} \(p<0.05\), \sym{**} \(p<0.01\), \sym{***} \(p<0.001\)
     \end{minipage}
\end{table}

Other  possible mechanisms discussed in the literature are social pressure and so-called teacher-effects. Social pressure might lead jobseekers to respond to the job search behavior of their peers, e.g. look for jobs faster once they observe highly employable peers find a job \citep{Mas2009, Fu2019}. Nevertheless, I find little evidence for a shortening of the job search duration in response to a higher peer group quality, which speaks against this mechanism playing a dominant role. Finally, course instructors might endogenously respond to the average group quality and adapt their teaching style or the course content. While such responses cannot be ruled out, they are unlikely to be the main explanation of the results. Trainings are often based on standardized curricula. If jobseekers would systematically select into specific courses offered by popular and good teachers, my peer effect estimates might pick up these teacher effects. However, such sorting behavior is limited by the voucher validity and capacity constraints.


\end{document}